\documentclass[preprint,sort,compress]{elsarticle}

\usepackage{amssymb}
\usepackage{amsmath}

\begin{document}

\begin{frontmatter}



\title{How effective delays shape oscillatory dynamics in neuronal networks}


\author[upf,columbia]{Alex Roxin}

\author[upf,nyu,crm]{Ernest Montbri\'o}

\address[upf]{Computational Neuroscience, Department of Information and Communication Technologies, Pompeu Fabra University, Barcelona 08018, Spain.}
\address[columbia]{Center for Theoretical Neuroscience, Columbia University, New York, New York, USA.}
\address[nyu]{Center for Neural Science, New York University, New York, 10012 New York, USA.}

\begin{abstract}
Synaptic, dendritic and single-cell kinetics generate significant time
delays that shape the dynamics of large networks of spiking neurons. Previous
work has shown that such effective delays can be taken into account with a
rate model through the addition of an explicit, fixed delay \cite{roxin05,roxin06}.
Here we extend this work to account for
arbitrary symmetric patterns of synaptic connectivity and generic
nonlinear transfer functions.  
Specifically, we conduct a weakly
nonlinear analysis of the dynamical states arising via primary
instabilities of the asynchronous state. In this way we
determine analytically how the nature and stability of these states
depend on the choice of transfer function and connectivity.  
We arrive at two general observations of physiological relevance 
that could not be explained in previous works. These are: 1 - Fast
oscillations are always supercritical for realistic transfer
functions. 2 - Traveling waves are preferred over standing  waves
given plausible patterns of  local connectivity. 
We finally demonstrate that these results 
show a good agreement with those obtained performing
numerical simulations of a network of Hodgkin-Huxley neurons.
\end{abstract}


\end{frontmatter}



\section{\label{intro}Introduction}

When studying the collective dynamics of cortical neurons 
computationally,
networks of large numbers of spiking neurons  have naturally 
been the benchmark model.  Network models incorporate the most fundamental
physiological  properties of neurons:
sub-threshold voltage dynamics, spiking (via spike generation 
dynamics or a fixed threshold), and  discontinuous synaptic interactions.  For this
reason, networks of spiking neurons  are considered to be
biologically realistic.  However, with few exceptions, e.g.  
\cite{amit97,brunel99,brunel00}, network  models of spiking neurons
are not amenable to analytical work and thus constitute above all a
computational tool.  Rather, researchers use reduced or simplified
models which describe  some measure of the mean activity in a
population of cells, oftentimes taken as the firing rate (for reviews, 
see~\cite{ermentrout98,coombes05}).
Firing-rate models are simple, phenomenological models of neuronal activity,
generally in the form of continuous, first-order ordinary differential
equations \cite{wilson72,amari77}.  Such firing-rate models can be analyzed
using  standard techniques for differential equations, allowing one to
understand the qualitative  dependence of the dynamics on parameters.
Nonetheless, firing-rate models do not represent, in general,  proper
mathematical reductions of the original network dynamics but rather
are heuristic, but see \cite{ermentrout94}.  As such, there is in general 
no clear relationship between the
parameters  in the rate model and those in the full network of spiking
neurons, although for at least some specific cases quasi-analytical 
approaches may be of value \cite{shriki03}.  
It therefore  behooves the researcher to study rate models
in conjunction with network simulations  in order to ensure there is
good qualitative agreement between the two.  

Luckily, rate models have proven remarkably accurate in capturing the
main types of qualitative  dynamical states seen in networks of large
numbers of asynchronously spiking neurons. 
For example, it is well known that in such networks the different 
temporal dynamics of excitatory and inhibitory neurons 
can lead  to oscillations. These oscillations can be well captured using rate models \cite{wilson72}. 
When the pattern of
synaptic connectivity depends on the distance between neurons, 
these differences in the temporal dynamics can also lead to the emergence of waves 
\cite{amari77,ermentrout93,pinto01a,pinto01b}. This
is certainly a relevant case for  local circuits in cortical tissue, where the
likelihood of finding a connection between any two neurons  decreases
as a function of the distance between them, e.g.~\cite{holmgren03}.

When considering the spatial dependence of the patterns of synaptic connectivity between neurons, 
one must take into account the presence
of time delays due to the
finite velocity propagation of action potentials along axons.  
Such delays depend depend linearly on the
distance between any two neurons. This has  been the topic of much
theoretical study using rate models with a space-dependent time delay
e.g. \cite{hutt03,coombes03,jirsa96,pinto01b,pinto01a,hutt04,atay05,
laing05,hutt06,venkov07,coombes07}. 
The presence of propagation delays can cause an oscillatory instability of the 
unpatterned state leading to homogeneous oscillations and waves
\cite{coombes03,atay05}. The weakly nonlinear dynamics of waves in
spatially extended rate models, i.e. describing large-scale (on the
order of centimeters) activity, is described  by the coupled
mean-field Ginzburg-Landau equations \cite{venkov07}, and thus
exhibits the phenomenology of small amplitude waves familiar from
other pattern forming systems \cite{cross93}. Also, it is important to 
note that discrete fixed delays have been used 
to model the time delayed interaction between discrete neuronal regions, as well 
as to model neuronal feedback, e.g. \cite{atay06,battaglia07,hutt08,roberts08,coombes09}. 
   
Localized solutions of integro-differential equations describing
neuronal activity, including  fronts and pulses, are also affected by
distance-dependent axonal delays
\cite{pinto01a,pinto01b,coombes03,hutt04,hutt06}.  Specifically,  the velocity
of propagation of the localized solution is proportional to the
conduction velocity along the axon for small conduction velocities,
while for  large conduction velocities it is essentially constant.
This reflects the fact that  the propagation of activity in neuronal
tissue is driven by local integration in which  synaptic and membrane
time constants provide the bottleneck. Also, allowing for different
conduction velocities for separate excitatory and  inhibitory
populations can lead to bifurcations of localized bump states to
breathers and traveling pulses \cite{laing05}.

Although the presence of time delays in the nervous system are most often 
associated with axonal propagation, significant time delays 
are also produced by the synaptic kinetics and single-cell dynamics
(see the next section for a detailed discussion about the origin of such effective 
time delays in networks of spiking neurons). 
As a relevant example for 
the present work, it was shown in \cite{roxin05,roxin06} that the addition 
of an explicit, non-space-dependent delay in a
rate equation was sufficient to explain the emergence of fast
oscillations prevalent in networks of spiking neurons 
with strong inhibition and in the absence of any explicit delays. 

Specifically, in \cite{roxin05,roxin06} the authors
studied  a rate model with a fixed delay on a ring geometry with two
simplifying assumptions.  First they  assumed that the strength of
connection between neurons could be expressed as a constant  plus the
cosine of the distance between the neurons.  Secondly, they  assumed a
linear rectified form for the transfer function which relates inputs
to outputs.   These assumptions allowed them to construct a detailed
phase diagram  of dynamical states, to a large degree analytically.
In addition to the stationary  bump state (SB) which had been studied
previously \cite{benyishai95,hansel98}, the presence of  a delay led
to two new states arising from primary instabilities of the stationary
uniform  state (SU): an oscillatory uniform state (OU) and a traveling
wave state (TW).  Secondary bifurcations  of these three states
(SB,OU,TW) led to yet more complex states including standing waves
(SW) and  oscillatory bump states (OB).  Several regions of
bistability between primary and  secondary states were found,
including OU-TW, OU-SB and OU-OB.   They subsequently confirmed these
results through simulations of networks of Hodgkin-Huxley neurons.
Despite the good agreement between the  rate equation and network
simulations, several important issues remain unresolved:  
\begin{itemize}
\item The rate
equation predicted that the primary instability of the SU
state  to waves should be to traveling waves, while in the network
simulations standing  waves were robustly observed.  
\item The
linear-threshold transfer function, albeit  amenable to analysis,
nonetheless leads to degenerate behavior at a bifurcation point.
Specifically, any perturbations with a positive linear growth rate
will continue to grow until the lower threshold of the  transfer
function is reached.  This means that the amplitude of new solution
branches at a bifurcation is always finite, although the solution
itself may  not be subcritical.  In a practical sense then, this means
that it is not possible to assess whether a  particular solution, for
example oscillations or bumps, will emerge continuously  from the
SU state as a parameter is changed, or if it will appear  at
finite amplitude and therefore be bistable with the SU state
over some range.  
\item The previous work only considered a simplified cosine connectivity.  More 
realistic patterns of synaptic connectivity such as a Gaussian dependence of 
connection strength as a function of distance might lead to different 
dynamical regimes. It remains to be explored the effect of a general 
connectivity kernel in the dynamics of both the rate equation and the 
spiking neurons network with fixed time delays.
\end{itemize}
In order to address these issues, and provide a more
complete  analysis of the role of fixed delays in neuronal tissue, we
here study a  rate equation with delay without imposing any
restrictions on the form of the transfer function beyond smoothness or
on the shape of the connectivity kernel  beyond being symmetric. 
Our approach is similar to that of Curtu and Ermentrout in \cite{curtu04}, who 
extended a simplified rate model with adaptation for orientation 
selectivity \cite{hansel98} to include 
a nonlinear transfer function and general connectivity kernel.  
Here we do the same for a rate model with a fixed time delay.  

Thus in what follows we will study a rate equation with fixed delay
and spatially modulated connectivity.  In conjunction with this
analysis we  will conduct numerical simulations of a network of large
numbers of spiking  neurons in order to assess the qualitative
agreement between the rate model  and the network for the delay-driven 
instabilities, which are the primary focus of this work.  

This article is organized as follows: In section
\ref{sect:delay} we provide an overview of the origin of the
effective delay.  We do this by looking at the dynamics of
synaptically  coupled conductance-based neurons.  This will motivate
the presence of an  explicit fixed delay in a rate-model description
of the dynamics in  recurrently coupled networks of neurons.  In
section \ref{model} we formulate  the rate model and conduct a linear
stability analysis of the SU state.  In section  \ref{sect:codim1} we
conduct a weakly nonlinear analysis for the four possible primary
instabilities of the SU state (asynchronous unpatterned state in a
network  model), thereby deriving amplitude equations for a steady,
Turing (bumps), Hopf (global oscillations), and  Turing-Hopf (waves)
bifurcations.  We will focus on the delay-driven instabilities, i.e.
Hopf and Turing-Hopf.  Finally, in section \ref{sect:codim2} we will
study the  interactions  of pairs of solutions: bumps and global
oscillations, and global  oscillations  and waves, respectively.


\section{The origin of effective time delays}
\label{sect:delay}

This section is intended to provide an intuitive illustration of the origin 
of an effective fixed delay in networks of spiking neurons.  A detailed, 
analytical study of this phenomenon can be found in \cite{brunel99,brunel03,geisler05}.

\begin{figure}
\center
\includegraphics[scale=0.4]{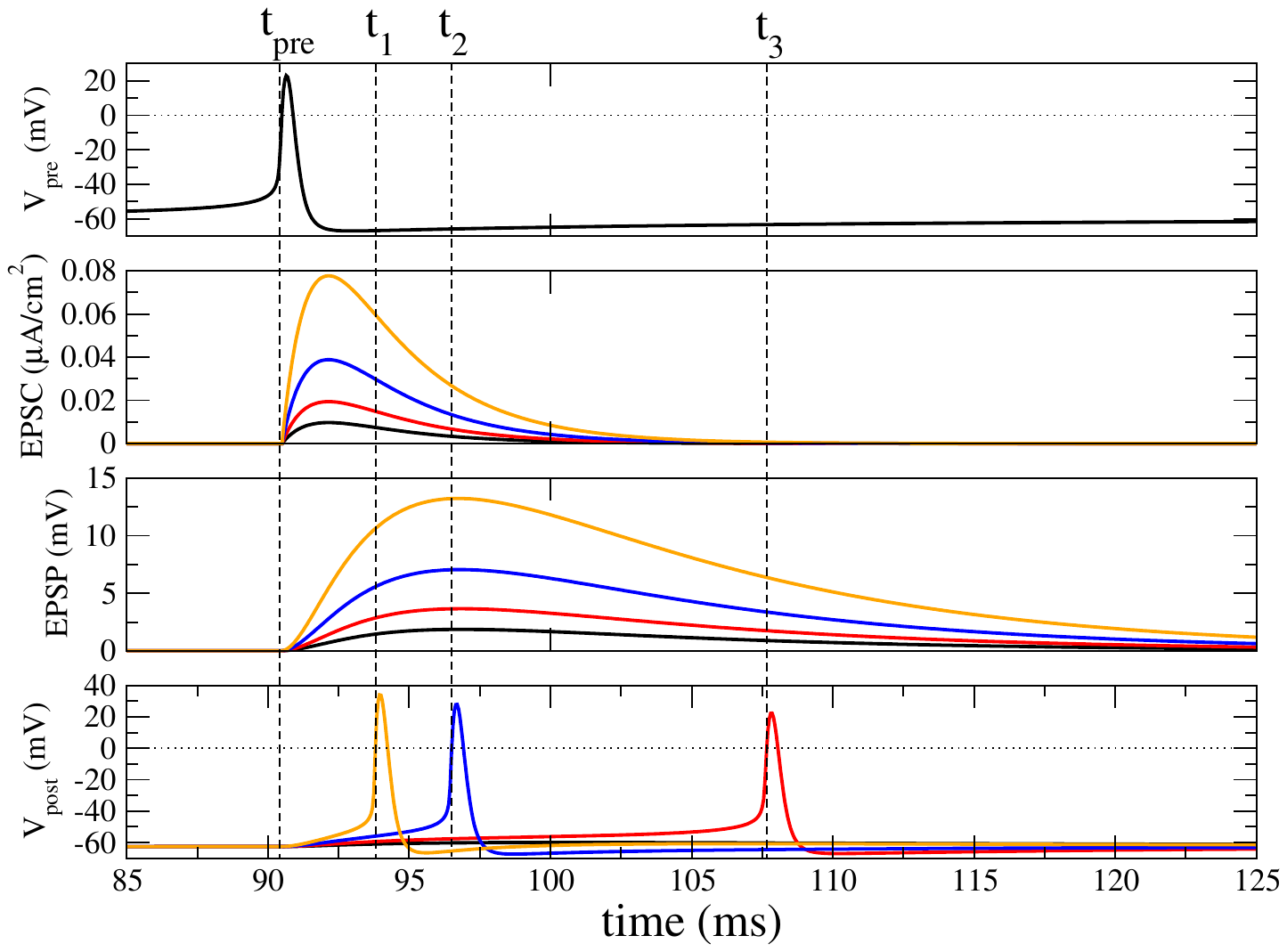}
\caption{(Color online) 
The synaptic time course introduces an effective fixed delay in the interaction between 
model neurons.  Top panel: The membrane voltage of the presynaptic neuron which is driven by a 
steady input current of $I_{app}=0.2\mu$A/cm$^{2}$ causing it to spike periodically.  Here only one 
action potential is shown.  The time of the action potential, defined as the time at which the membrane 
potential crosses zero mV from below, is indicated by the dashed line labeled $t_{pre}$.  Second panel 
from top:  The EPSC in the postsynaptic cell is initiated at time $t_{pre}$ with no delay.  It consists 
of a difference-of-exponential conductance change with rise and decay times of $1$ and $3$ milliseconds 
respectively, times the voltage of the postsynaptic cell (the reversal potential for excitatory synapses is 
$0$mV).  The colors indicate different values of the maximum conductance: black $g_{E}=0.05$, 
red $g_{E}=0.1$, blue $g_{E}=0.2$ and orange $g_{E}=0.4$ mS$\cdot$ms/cm$^{2}$.  See \ref{app:network} for 
details of the model and explanation of units.  
Third panel from top:  The EPSP in the postsynaptic cell.  These curves were generated by eliminating 
the action potential generating currents from the model, i.e. Na and K, and subtracting off the rest 
potential $\sim -64$mV.  Bottom panel: The membrane potential of the postsynaptic cell with $Na$ and 
$K$ currents intact.  Note that the time of the postsynaptic action potential, indicated by the dashed 
lines, approaches $t_{pre}$ with increasing synaptic strength.  The very long latency at time $t_{3}$ is 
due to an input which puts the cell membrane potential very close to threshold and is therefore due 
in part to the action potential generating mechanism of the model.  The postsynaptic cell is driven by 
a steady input current of $I_{app}=0.1 \mu$ A/cm$^{2}$ which is insufficient to cause it to spike.} 
\label{fig:prepost}
\end{figure}

Fig.\ref{fig:prepost} illustrates the origin of the effective delay in
networks of  model neurons.  In this case we look at a single neuron
pair: one presynaptic and one  postsynaptic.  The single-neuron
dynamics are described in detail in \ref{app:network}.   The top panel
of Fig.\ref{fig:prepost} shows the membrane potential of an excitatory
neuron  subjected to a current injection of $I_{app}=0.2
\mu$A/cm$^{2}$ which causes it to fire action  potentials.
Numerically, an action potential is detected whenever the membrane
voltage exceeds  $0$mV from below.  When this occurs, an excitatory
postsynaptic current (EPSC)  is generated in the post-synaptic
neuron, as seen in the second panel.  This current is generated by the
activation of an excitatory  conductance which has the functional form
of a difference of exponentials with rise time $\tau_{1}=1$ms  and
decay time $\tau_{2}=3$ms.  The different colored curves correspond to
different conductance  strengths: black $g_{E}=0.05$, red $g_{E}=0.1$,
blue $g_{E}=0.2$ and orange $g_{E}=0.4$ mS$\cdot$ms/cm$^{2}$.   The
resulting excitatory postsynaptic potential (EPSP) in millivolts is
shown in the third panel.   At this point it is already clear that the
postsynaptic response, although initiated here  simultaneously with
the presynaptic action potential, will nonetheless take a finite
amount of time to  bring the postsynaptic cell to threshold, thereby
altering its firing rate.  This is shown in the  bottom panel.  In
this case the weakest input (black) was insufficient to cause the
neuron to spike,  while the other three inputs were all strong enough
to cause an action potential.  The latency until  action potential
firing is a function of the synaptic strength, with the latency going
to zero as the  synaptic strength goes to infinity.  The very long
latency for $g_{E}=0.1$ mS$\cdot$ms/cm$^{2}$ (red curve) is  due in
part to the intrinsic action potential generating mechanism.  In fact, 
an input  which brings the neuron sufficiently close to the
bifurcation to spiking can generate arbitrarily  long latencies.  

\begin{figure}
\centerline{\includegraphics[scale=0.4]{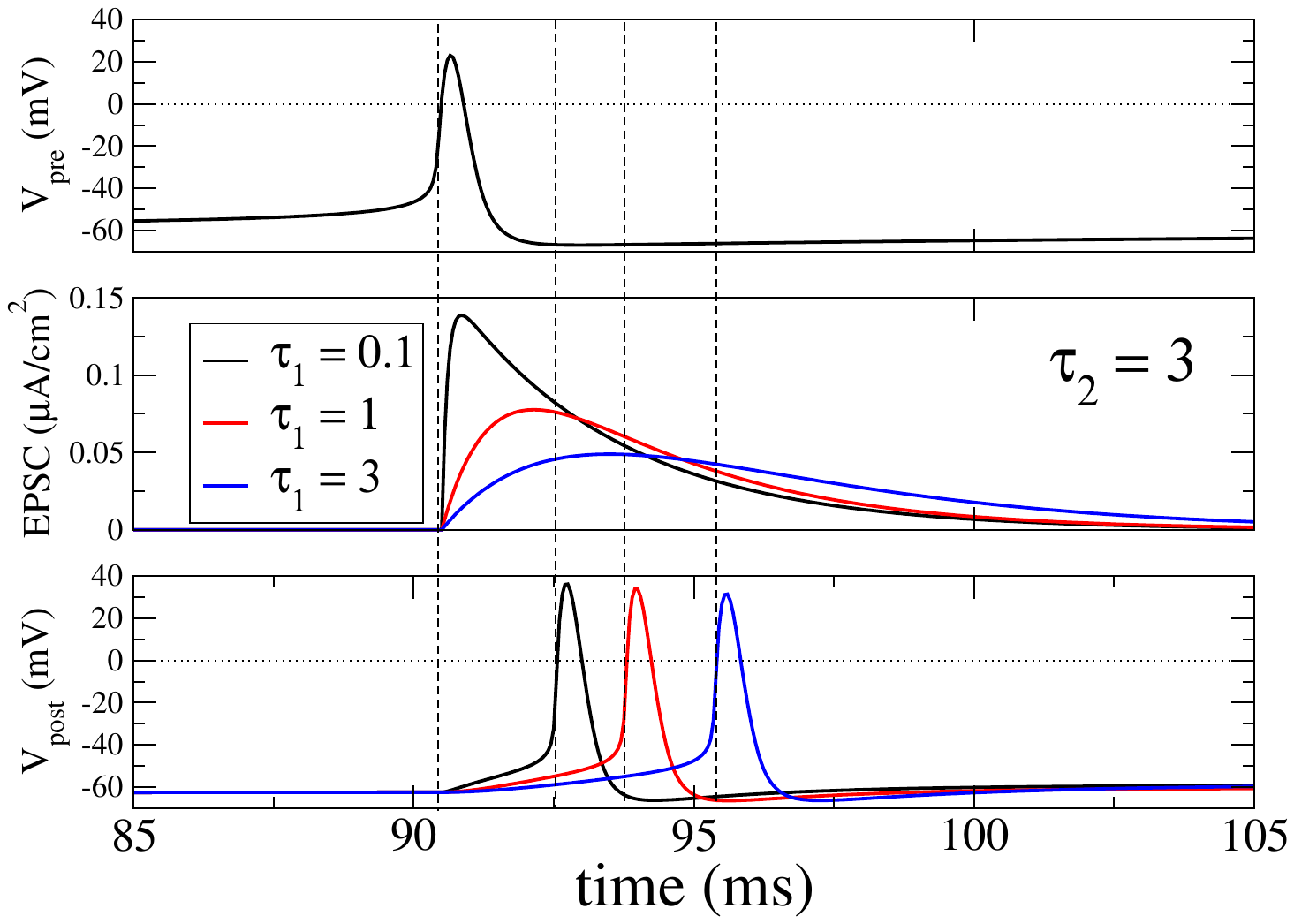}}\vspace{-0.5in}
\centerline{\includegraphics[scale=0.4]{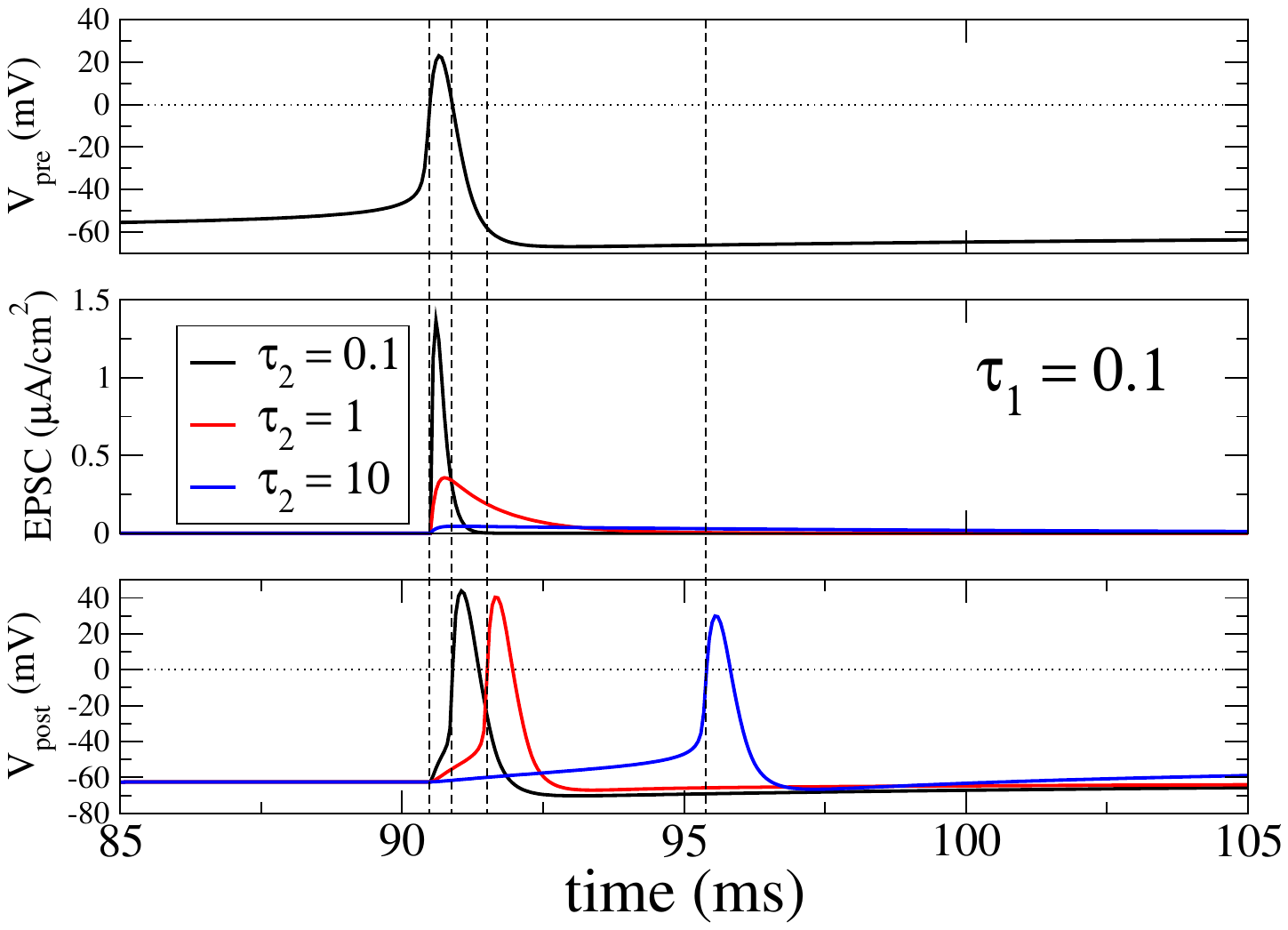}}
\caption{(Color online) 
The effective delay is affected by both the rise and decay time of the EPSC.  Top: The decay is 
kept fixed at $3ms$ and the rise time is varied.  Bottom: The rise time is kept fixed at $0.1$ms and the 
decay time is varied.  The parameter values are the same as in Fig.\ref{fig:prepost} with 
$g_{E}=0.4$ mS$\cdot$ ms/cm$^{2}$.}\label{fig:tau1}
\end{figure}

Fig.\ref{fig:tau1} illustrates how this effective delay is proportional to both the rise and decay time of the 
EPSC.  In the top panel, the decay time is fixed at $3$ms while the rise time is varied, while in the bottom 
panel, the rise time is fixed at $0.1$ms and the decay time is varied.  From these figures 
it is clear that the effective delay is proportional to both the rise and decay times.  Simulations with 
an EPSC modeled by a jump followed by a simple exponential decay reveal that the effective delay is 
proportional to the decay time in this case (not shown).

\begin{figure}
\centerline{\includegraphics[scale=0.40]{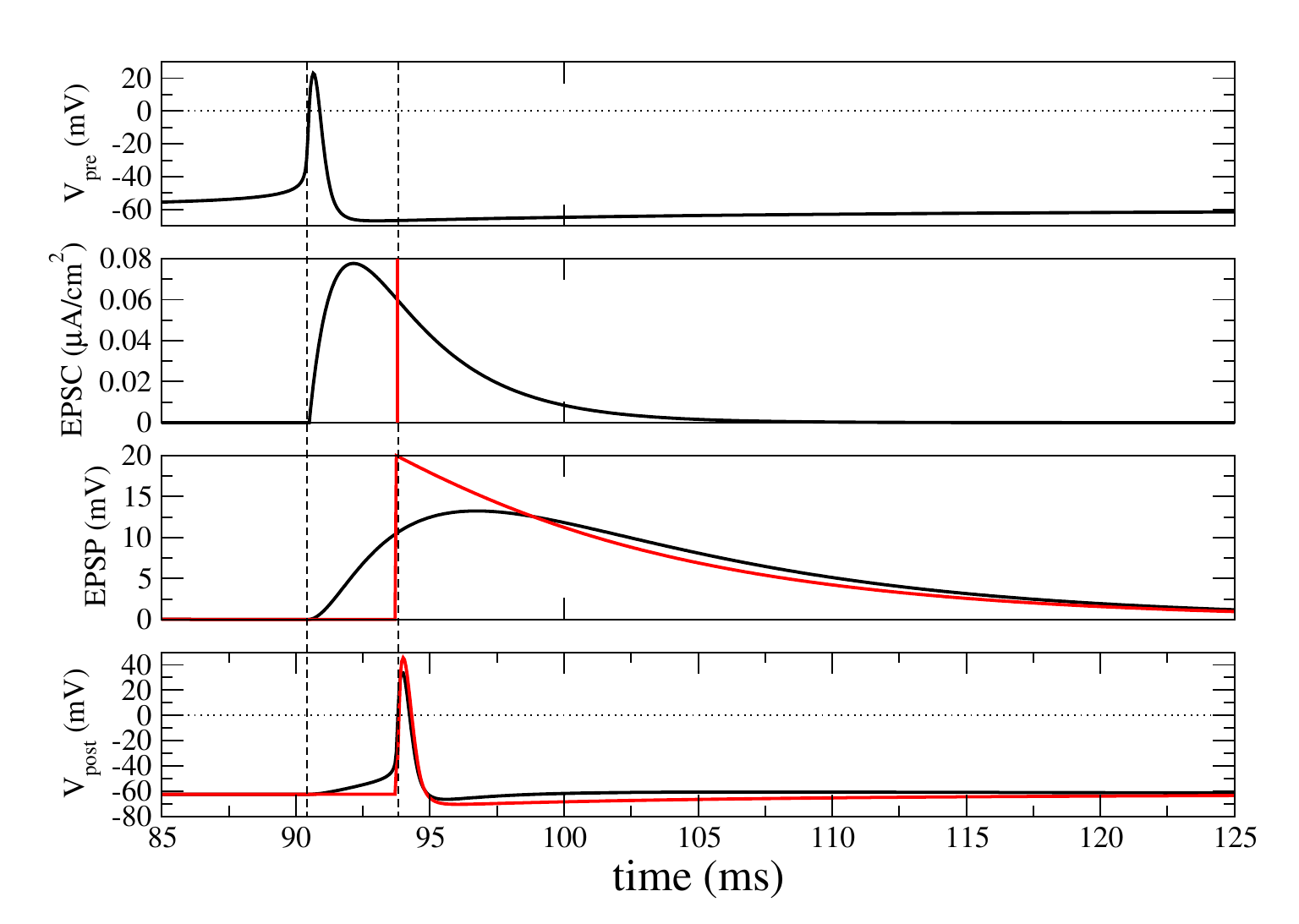}}
\caption{(Color online) 
Continuous synaptic kinetics can be replaced by a jump in the voltage which occurs at a 
fixed delay.  The black curves in the four panels are the same as in Fig.\ref{fig:prepost} with 
$g_{E}=0.4$ mS$\cdot$ms/cm$^{2}$.  The red curves show the effect of an EPSC modeled as a Dirac delta 
function which occurs with a delay of $3.2$ms after the presynaptic action potential.}\label{fig:delta}
\end{figure}

It is instructive to note that the effective delay, due to the time course of the synaptic kinetics in the 
model neuron, can be captured by modeling the EPSC as a Dirac delta function with a fixed delay.  This is 
shown in Fig.\ref{fig:delta}.  In Fig.\ref{fig:delta}, the curves shown in black are the same as in Fig.
\ref{fig:prepost} for $g_{E}=0.4$ mS$\cdot$ms/cm$^{2}$, while the red EPSC is a Dirac delta function which 
arrives with a fixed delay of $3.2$ms.  Note that the decay of the EPSP and the postsynaptic spike time 
are well captured here.  
The fact that a jump in voltage with a fixed delay can capture the effect 
of having continuous synaptic kinetics was described already in \cite{brunel99}.  In that work, the 
authors studied a network of recurrently coupled integrate-and-fire neurons with inhibitory synapses, 
the time course of which was modeled as jump in the voltage at a fixed delay.  They showed that the 
fixed delay led to the emergence of fast oscillations, the period of which was proportional to 
approximately several times the delay.  The advantage of using EPSCs modeled as Dirac delta functions 
is that the input current is delta-correlated in time, allowing one to solve the associated 
Fokker-Planck equation for the distribution of the membrane voltages in a straightforward way. 
 
Subsequent work studied the emergence of fast oscillations in networks of integrate-and-fire neurons 
with continuous synaptic kinetics  \cite{brunel03}.  There the authors determined the frequency of oscillations 
analytically and found that it is proportional to both the rise and decay times of the synaptic response.  
An extension of that work showed that for networks of Hogkin-Huxley conductance-based neurons, the frequency 
of oscillations also depends on the single cell dynamics and specifically the membrane time constant and 
action potential generation mechanism \cite{geisler05}.  This is consistent with the effect of the synaptic response and single-cell dynamics on the response latency that we have illustrated above.   

\begin{figure}
\centerline{\includegraphics[scale=0.5]{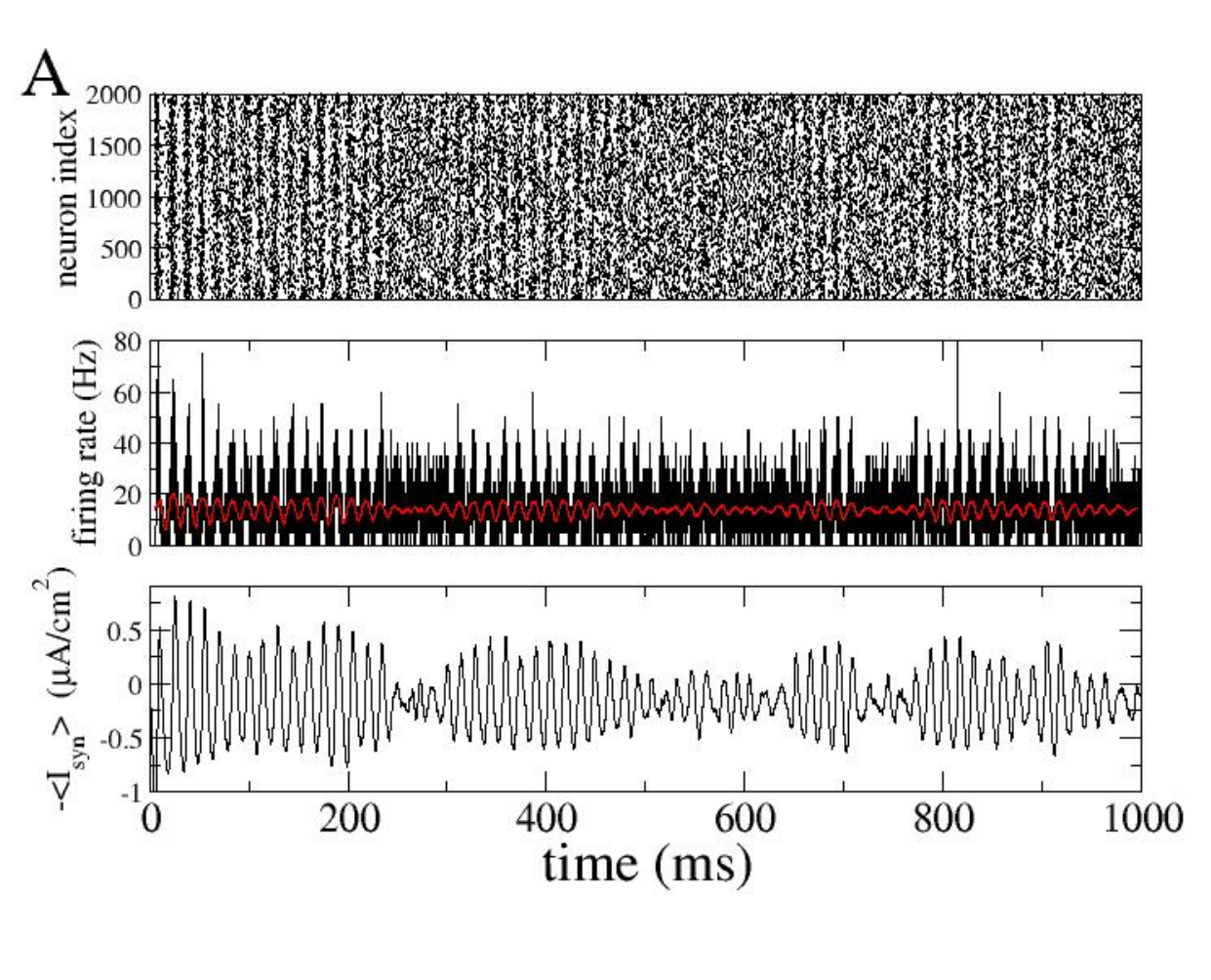}}
\centerline{\includegraphics[scale=0.35]{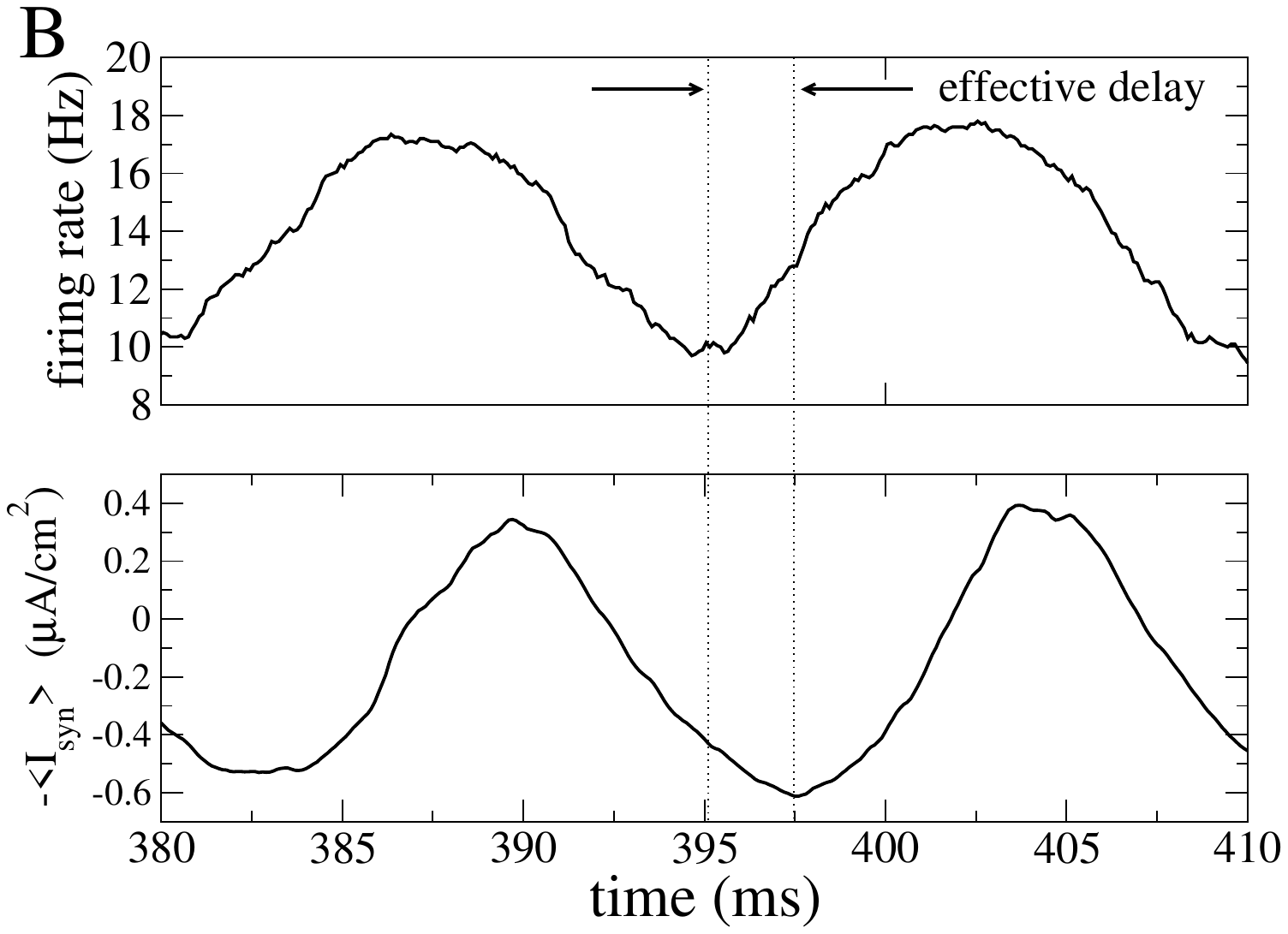}}
\caption{(Color Online) 
The effective delay can lead to oscillations in networks of recurrently coupled 
inhibitory neurons.  A. Raster plot of the spiking activity in a network of 2000 inhibitory cells (top).  
The average firing rate shows large fluctuations (center, black), which when smoothed are clearly 
identifiable as noisy, periodic oscillations (red).  Network oscillations are also clearly 
visible in the average subthreshold input (bottom).  B. A blowup of the average firing rate (top) and 
average input current (bottom) from the simulation in A.  The input current is clearly a time-delayed 
copy of the firing rate.  The delay is between $2$ and $3$ms.}\label{fig:fastosc}
\end{figure}

Thus the same mechanism which generates an effective delay in the response of a postsynaptic neuron to a single excitatory presynaptic input, can also generate coherent oscillations in a network of neurons coupled through inhibitory synapses.  This can be seen in Fig.
\ref{fig:fastosc}, which shows the results of simulations of a network of $2000$ recurrently coupled inhibitory 
cells.  The single-cell model is the same conductance based model used in Figs.\ref{fig:prepost}
-\ref{fig:delta}, see \ref{app:network} for details.  Synaptic connections are made between neurons 
with a probability of $p_{0}=0.2$, leading to a sparse, random connectivity with each cell receiving an 
average of $400$ connections.  Synapses are modeled as the difference of exponentials with a rise time 
and a decay time of $1$ms and $3$ms respectively and $g_{I}=0.01$ mS$\cdot$ms/cm$^{2}$.  
All cells receive uncorrelated Poisson inputs with 
a rate of $12000$Hz and $g_{ext}=0.0019$mS$\cdot$ms/cm$^{2}$, and there is no delay in the interactions.  Fig.
\ref{fig:fastosc}A shows a raster plot of the network activity in the top panel.  The activity is noisy 
although periods of network synchrony are visible.  The middle panel shows the firing rate 
averaged over all neurons in time bins of $0.1$ms (black) and smoothed by averaging 
with a sliding window of $10$ms (red).  The large fluctuations in the firing rate indicate network synchrony, 
while the averaged trace shows clear periodic oscillations.  This is even more evident in the bottom panel 
which shows the subthreshold input current averaged over all cells.  One can clearly see the ongoing 
oscillation, the amplitude of which undergoes slow fluctuations due to the noisy dynamics.  Fig.
\ref{fig:fastosc}B shows the smoothed firing rate (top) and average input current (bottom) from 
the same simulation, but on a shorter time scale.  Note that the sign of the input current has been 
inverted so that downward deflections mean increasing positive currents.  Here it is clear that the input 
current is a \textit{delayed} copy of the firing rate, with a delay on the order of $\sim 2-3$ms which 
matches with the time scale of the synaptic kinetics ($\tau_{1}=1$ms, $\tau_{2}=3$ms).  

Therefore, Fig.~\ref{fig:fastosc} provides 
a clear prescription for developing a rate model description of fast oscillations in networks in the 
asynchronous regime.  The input a neuron receives is not simply a nonlinear function of the instantaneous 
firing rate, rather it is a function of the \textit{delayed} firing rate.  
Thus one should introduce a fixed delay in the rate model description.  This 
was the underlying assumption behind the work in \cite{roxin05}.  

Before presenting the model we would like to emphasize that fixed delays, which are 
primarily due to synaptic and dendritic integration, and conduction delays due to the propagation of
action potentials along the axon, are both present in real neuronal
systems.   Importantly, this means that the delay in neuronal
interactions at zero distance is  not zero.   In fact, \textit{fixed}
delays are always observed in paired intracellular recordings in
cortical slices.  The latency from the  start of the fast rising phase
of the action potential to the start of the post-synaptic current  (or
potential) has been measured for pairs of pyramidal cells in rat
layers 3 to 6 and  is on the order of milliseconds, see
\cite{thomson07} for a recent review.  Recordings from  cat cortex and
between pyramidal cells and other cells including spiny cells and
interneurons in  the rat cortex also reveal fixed delays which are
rarely less than a millisecond.  These delays are seen  when neurons
are spatially adjacent, indicating that axonal propagation is  not an
important contributing factor.   On the other hand the speed of
propagation of action potentials along unmyelinated axons in  mammals
is on the order of $10^{-1}-10^{1}$ m/s, which means a delay of 0.1-10
ms for  neurons separated by 1 millimeter \cite{kandel91,girard01}.  Thus fixed
delays and conduction delays are of similar  magnitude within a local
patch of cortex and both would be expected to shape the dynamics of
non-steady activity, i.e. neither is negligible.  Here we have decided
to focus   on fixed delays, as in previous work
\cite{roxin05,roxin06}, due both to  their physiological relevance and
prevalence in networks of spiking neurons.

\section{The rate model with fixed time delay}
\label{model}

An effective delay roughly proportional to the time scale of the post-synaptic 
currents is always present in networks of spiking neurons as we have illustrated in the
previous section and has been shown extensively elsewhere, e.g. \cite{brunel03,geisler05}.  
In particular, this is true for networks in the asynchronous regime, for 
which a rate-equation description is, in general, appropriate.  Given this, we consider here 
a rate model with fixed delay.
Specifically, we study a heuristic equation describing the activity of  a small
patch of neural tissue consisting of two populations of recurrently
coupled  excitatory and inhibitory neurons respectively.  Our
formulation is equivalent to the  Wilson-Cowan equations without
refractory period \cite{wilson72}, and with spatially dependent
synaptic connectivity which was studied originally in
\cite{ermentrout80}.  Additionally, we  consider a fixed delay in the
neuronal interactions.  Given these assumptions,  the full
Wilson-Cowan equations are 
\begin{subequations}
\begin{equation}
\tau_{e}\dot{r}_{e}=-r_{e}+
\Phi_{e}\bigg[\int_{\Omega}dy[J_{ee}(|x-y|)r_{e}(y,t-d_{e})
-J_{ei}(|x-y|)r_{i}(y,t-d_{i})]+I_{e}\bigg],\label{eq:wc_e}\\ 
\end{equation}
\begin{equation}
\tau_{i}\dot{r}_{i}=-r_{i}+
\Phi_{i}\bigg[\int_{\Omega}dy[J_{ie}(|x-y|)r_{e}(y,t-d_{e})
-J_{ii}(|x-y|)r_{i}(y,t-d_{i})]+I_{i}\bigg]. \label{eq:wc_i}
\end{equation}
\end{subequations}
In the original formulation \cite{wilson72}, $r_{e}(x,t)$ and
$r_{i}(x,t)$  represent the average number of active cells in the
excitatory and inhibitory populations  respectively, in this case at a
position $x$ and at a time $t$.  The time constant  $\tau_{e}$
($\tau_{i}$) is roughly the time it takes for a an excitatory
(inhibitory)  cell receiving ``at least threshold excitation''
\cite{wilson72} to generate a spike.   This can reasonably be taken as
the membrane time constant which is generally on the  order of 10-20
ms.  The functions $\Phi_{a}(x) (a=e,i)$ are usually taken to be
sigmoidal.  Specifically,  if all neurons in the population receive
equal excitatory drive, and there is heterogeneity in some parameter
across neurons, e.g. the threshold to spiking,  which obeys a unimodal
distribution, then the fraction of active neurons is just the integral
over the  distribution, up to the given level of excitation.  The
integral of a unimodal distribution is sigmoidal.  In
Eqs.(\ref{eq:wc_e}-\ref{eq:wc_i}), the functions $J_{ab}(|x|)
(a=e,i)(b=e,i)$ represent the strength of synaptic connection from a
neuron  in population $b$ to a neuron in population $a$ separated by a
distance $x$.  Here the neurons  are arranged in one dimension on a
domain $\Omega $.  Input from excitatory (inhibitory) cells  is
furthermore delayed by a fixed amount $d_{e}$ ($d_{i}$), which, as we
have discussed in the  introduction, is on the order of one
millisecond.  Finally, the excitatory and inhibitory populations  are
subject to an external drive of strength $I_{e}$ and $I_{i}$
respectively. 

A general analysis of Eqs.(\ref{eq:wc_e}-\ref{eq:wc_i}) would be
technically arduous although it is a natural extension of the 
work presented here.  Rather, we  choose to study the dynamics of this
system under the simplifying assumption that the  excitatory and
inhibitory neurons follow the same dynamics, i.e. $\tau_{e}=\tau_{i}=\tau $,
$d_{e}=d_{i}=d$, $J_{ee}=J_{ie}=J_{e}$, $J_{ei}=J_{ii}=J_{i}$,
$\Phi_{e}=\Phi_{i}=\Phi $, $I_{e}=I_{i}=I$.   If this the case, then
$r_{e}=r_{i}=r$ and the variable $r$ follows the dynamics given by   
\begin{equation}
\dot{r}(x,t) = -r(x,t)+\Phi\bigg[\frac{1}{2\pi}\int_{-\pi}^{\pi}dy
J(|x-y|)r(y,t-D)+I\bigg], \label{eq:rate}
\end{equation}
where we have chosen the domain $\Omega$ to be a ring of normalized
length $L=2\pi$. Furthermore, we have re-scaled time by the time
constant $\tau $.  The normalized delay  is therefore $D=d/\tau$,
which is the ratio of the effective delay in neuronal interactions  to
the integration time constant and should be much less than one in
general.  The synaptic  connectivity expressed in terms of the
excitatory and inhibitory contributions is
$J(|x|)=J_{e}(|x|)-J_{i}(|x|)$, and thus represents an effective mixed
coupling which may have  both positive and negative regions.

Eq.(\ref{eq:rate})  with the choice of $\Phi(I)=I$ for $x>0$ and 0
otherwise and with $J(x)=J_{0}+J_{1}\cos{(x)}$ is  precisely the model
studied in \cite{roxin05,roxin06}.  We now wish to study
Eq.(\ref{eq:rate})  for arbitrary choices of $\Phi(I) $ and $J(x)$.   

In presenting Eq.(\ref{eq:rate}) we have relied on  the heuristic
physiological motivation first put forth in \cite{wilson72}.
Nonetheless, as a  phenomenological model, the terms and parameters in
Eq.(\ref{eq:rate}) may have alternative and  equally plausible
interpretations.  Indeed, the variable $r$ is often thought of as the
firing rate as opposed to the fraction of active cells, in which case
the  function $\Phi(I)$ can be thought of as the transfer function or
fI  curve of a cell.  

Experimentally the function $\Phi(I)$ has been found to
be  well approximated by a power-law nonlinearity with a power greater
than one  \cite{miller02,hansel02}.  Modeling studies show that the
same nonlinearity  applies to integrate-and-fire neurons and
conductance based neurons driven by  noisy inputs \cite{hansel02}.
Therefore it may be that such a choice of  $\Phi $ leads to better
agreement of Eq.(\ref{eq:rate}) with networks of spiking  neurons and
hence with actual neuronal activity.  More fundamentally, we may  ask
if choosing $\Phi$ as a sigmoid or a power law qualitatively alters
the  dynamical states arising in Eq.(\ref{eq:rate}).  This is precisely
why we choose  here not to impose restrictions on $\Phi$ but rather
conduct an analysis valid for  any $\Phi$.  How the choice of $\Phi$
affects the generation of oscillations and  waves is an issue we will
return to in the corresponding sections of this paper.

\subsection{Linear stability analysis}

Stationary uniform solutions (SU) of Eq.(\ref{eq:rate}) are given by 
\begin{equation}
R = \Phi\Big[J_{0}R+I\Big], \label{eq:ss}
\end{equation} 
where $R$ is a constant non-zero rate, 
$J_{0}$ is the zeroth order spatial Fourier coefficient of the 
symmetric connectivity which can be expressed as 
\begin{equation}
J(x)=J_0+\left(\sum_{k=1}^{\infty}J_{k}e^{ikx}+c.c.\right)   
\end{equation}
and $k$ is an integer.
Depending on the form of $\Phi$, Eq.(\ref{eq:ss}) may admit one or
several solutions.

We study the linear stability of the SU state with the ansatz
\begin{equation}
r(x,t) = R+\sum_{k=0}^{\infty}\delta r_{k}e^{ikx+\alpha(k)t},
\label{eq:ansatz}
\end{equation}
where $\delta r_{k}\ll 1$ and the spatial wavenumber $k$ is an integer
due to the periodic boundary conditions.  Plugging Eq.(\ref{eq:ansatz})
into  Eq.(\ref{eq:rate}) leads to an equation for the complex eigenvalue $\alpha (k)$ 
\begin{equation}
\alpha(k) = -1+\Phi^{'}J_{k}e^{-\alpha(k)D},\label{eq:growth}
\end{equation}
where the slope $\Phi^{'}$ is evaluated at the fixed point given by
Eq.(\ref{eq:ss}).  The real and imaginary parts of the eigenvalue 
$\alpha(k) = \lambda(k) +i\omega(k)$ represent the linear growth rate 
and frequency of perturbations with spatial wavenumber $k$ respectively.  
At the bifurcation of a single mode, the growth rate will reach zero 
at exactly one point and be negative elsewhere.  That is, $\lambda (k_{cr})=0$ 
for the critical mode $k_{cr}$.  Given this, Eq.(\ref{eq:growth}) yields the 
dispersion relation for the frequency of oscillation of the critical mode
\begin{equation}
i\omega(k_{cr}) = -1+\Phi^{'}J_{k_{cr}}e^{-i\omega (k_{cr})D}.\label{eq:disp}
\end{equation}
From Eq.(\ref{eq:disp}) it is clear that the wavelength of the critical mode 
depends crucially on the synaptic connectivity.  In particular, the spatial 
Fourier coefficients of the connectivity kernel $J(x)$ depend on the 
wavenumber $k$, i.e. $J_{k}=J(k)$.  Thus, the critical wavenumber is, in effect, 
selected by the choice of connectivity kernel.  It is in this way that the 
nature of the instability depends on the synaptic connectivity at the linear level.

Depending on the values of $\omega$ and $k_{cr}$ in Eq.(\ref{eq:disp}) at the bifurcation from
the SU state, four types of instabilities are possible:
\begin{itemize}
\item Steady ($\omega = 0$, $k_{cr}=0$): The instability leads to a global increase in activity.
\item Turing ($\omega = 0$, $k_{cr}\ne 0$): The instability leads to a stationary 
bump state (SU).
\item Hopf ($\omega\ne 0$, $k_{cr}=0$): The instability leads to an oscillatory 
uniform state (OU).
\item Turing-Hopf ($\omega\ne 0$,$k_{cr}\ne 0$): The instability leads to waves (SW, TW).
\end{itemize}
For the non-oscillatory instabilities (i.e. $\omega=0$),
Eq.(\ref{eq:disp}) gives the critical value
\begin{equation}
\bar J_k= 1/\Phi' 
\label{eq:disp0}
\end{equation}
while for the  oscillatory ones Eq.(\ref{eq:disp}) is equivalent to the
system of two transcendental  equations
\begin{subequations}
\begin{equation}
\bar \omega = -\tan{\bar \omega D},   
\label{eq:disp2}
\end{equation}
\begin{equation}
\bar \omega = -\Phi^{'}\bar J_{k}\sin{\bar \omega D}. 
\label{eq:disp3}
\end{equation}
\end{subequations}

Note  that we have defined the critical values as $J_{k_{cr}}\equiv \bar
J_k$ and $\omega_{cr}\equiv\bar \omega$.

\subsubsection{The small delay limit ($D \to 0$)}
\begin{figure}
\center
\includegraphics[scale=0.3]{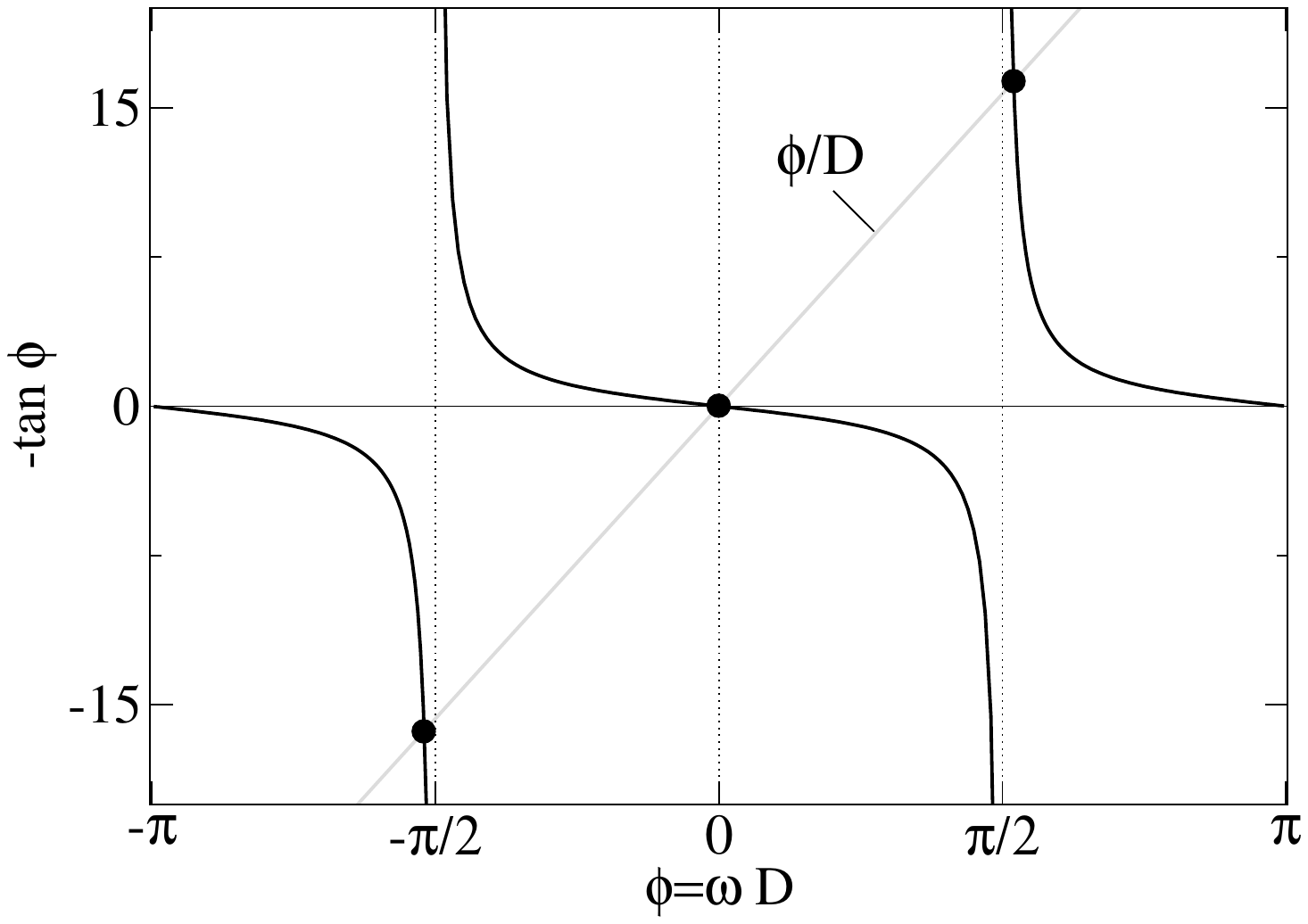}
\caption{The critical frequency at the instability to oscillations is given by the 
intersection of the grey ($D=0.1$) and black curves, the left and right hand sides of Eq.(\ref{eq:disp2}) 
respectively.  As $D\to 0$, solutions clearly approach $\phi=(2n+1)\pi /2$ ($n$ integer). Eq.(\ref{eq:disp3}) shows that 
the first potentially unstable mode corresponds to the solutions $\phi = \pm \pi /2$ (see text). }
\label{fig:roots}
\end{figure}

It is possible to gain some 
intuition regarding the effect of fixed delays on the dynamics, by deriving 
asymptotic results in the limit of small delay. This limit 
is a relevant one physiologically, since 
fixed delays are on the order of a few milliseconds and the integration 
time constant is about an order of magnitude larger.
Therefore throughout this work we will present asymptotic 
results, and compare them to the full analytical formulas 
as well as to numerical simulations. 

\begin{figure}
\center
\includegraphics[scale=0.35]{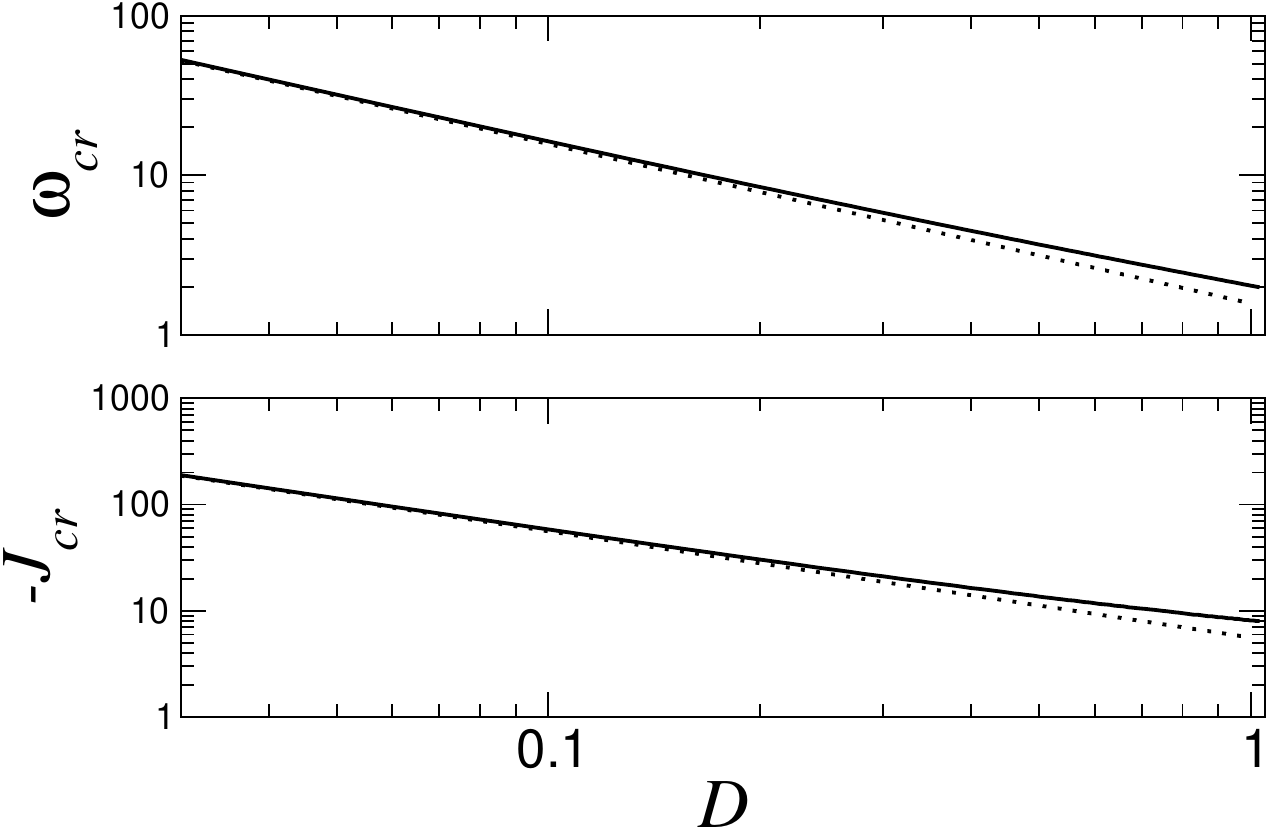}
\caption{Top: The critical frequency of oscillatory instabilities as a
function of the  delay $D$ from the dispersion equation
Eq.(\ref{eq:disp2}) (solid line) and in the small delay  limit (dotted
line).  Bottom: The critical coupling as a function of the delay $D$
from  Eq.(\ref{eq:disp3}) (solid) and in the small delay limit
(dotted).} \label{fig:disp}
\end{figure}

In the limit $D\to 0$, the asymptotic solutions of 
Eq.(\ref{eq:disp2}) can be easily obtained graphically. 
Fig.\ref{fig:roots} shows two curves (black and grey) 
representing the right and left hand sides of Eq.(\ref{eq:disp2}) respectively, 
where we defined $\phi \equiv \bar \omega D$. The intersections of these curves correspond to
the roots of  Eq.(\ref{eq:disp2}). The plot shows three solutions, 
the trivial one $\phi=0$ (corresponding to the non-oscillatory instabilities), 
and an infinite number of solutions that clearly 
approach $\phi= (2n+1) \pi/2$ ($n$ integer) in the small delay limit, since the slope 
of the straight line 
goes to infinity as $D\to 0$. Substituting these solutions into 
Eq.~(\ref{eq:disp3}), we find that the first potentially unstable 
solution of the $k^{th}$ spatial Fourier mode is $\phi=\pm \pi/2$, 
that occurs at the critical value of the coupling
\begin{equation}
\bar J_k = -\frac{\pi}{2D\Phi^{'}},
\end{equation}
with a frequency
\begin{equation}
\bar \omega = \frac{\pi}{2D}.
\label{eq:D0}
\end{equation}
Fig.~(\ref{fig:disp}) shows the critical frequency and coupling as a
function of the delay, up to a delay $D=1$.  The solution obtained
from the dispersion relation Eqs.(\ref{eq:disp2}) and  (\ref{eq:disp3})
are given by solid lines, while the expressions obtained in the small
delay limit  are given by dotted lines.  Thus the expressions in the 
small delay limit agree quite well with the full expressions even 
for $D=1$.  

\subsection{An illustrative Phase Diagram}

Throughout the analysis which follows we will illustrate our results with 
a phase diagram of dynamical states.
Specifically, we will follow the 
analysis in \cite{roxin05,roxin06} in constructing a phase diagram of 
dynamical states as a function of $J_{0}$ and $J_{1}$, the first two 
Fourier coefficients of the synaptic connectivity.  
We will set the higher order 
coefficients to zero for this particular phase diagram, although we will discuss 
the effect of additional modes in the text.  Furthermore, unless otherwise 
noted, for simulations we choose a sigmoidal 
transfer function $\Phi(I)=\frac{\alpha}{1+e^{-\beta I}}$ with $\alpha = 1.5$ and 
$\beta = 3$.  As we vary the connectivity in the phase diagram, we also vary the 
constant input $I$ in order to maintain the same level of mean activity, i.e. 
we keep $R=0.1$ fixed.  For the values of the parameters we have chosen here 
this results in $I\sim -0.1J_{0}-0.88$.  We also take $D=0.1$ unless noted otherwise

\begin{figure}[tp]
\center
\includegraphics[scale=0.35]{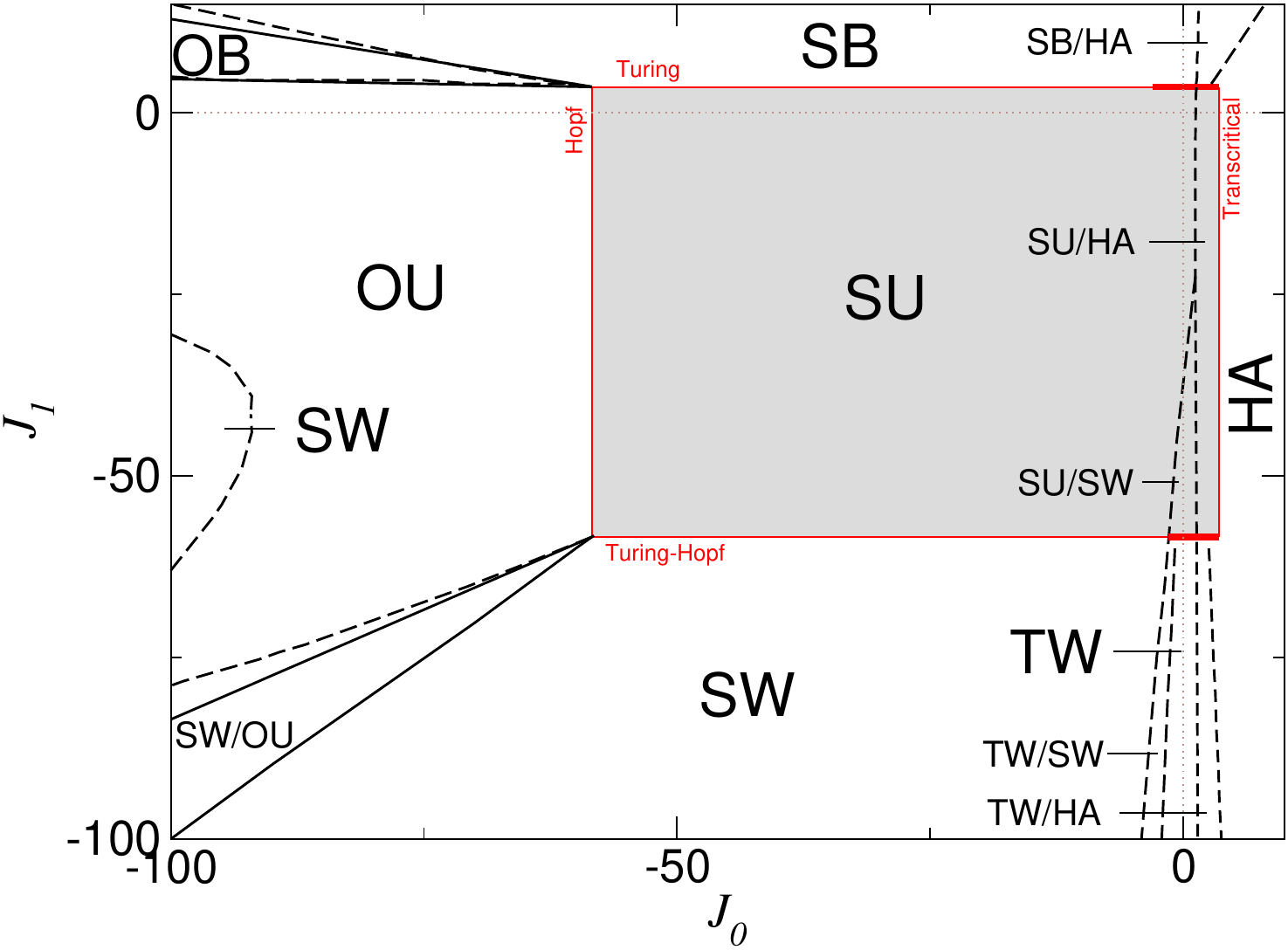}
\caption{(Color online) Phase diagram of the rate model Eq.(\ref{eq:rate}). 
In each region, the type of
solution seen in numerical  simulations is indicated by a letter
code: SU - stationary uniform (grey region), 
HA - high activity,  SB - stationary
bump, OB - oscillatory bump, SW - standing waves, TW - traveling
waves.   Solid lines indicate analytical expressions.  In
particular, the four possible instabilities 
of the  SU state are depicted in red     
(thick lines correspond to subcritical bifurcations) and are given by
the linear stability criteria Eqs.(\ref{eq:disp0}-\ref{eq:disp3}). 
The four  lines emanating from the upper and lower left corners of
the SU region were determined from a  weakly nonlinear analysis at
the two corners (codimension two points) [see section IV].  The
region  marked OB corresponds to a mixed mode solution of SB-OU,
while in the lower left-hand region the OU  and SW solutions are
bistable. Parameters: $\Phi (x) = \frac{\alpha}{1+e^{-\beta x}}$ where
$\alpha=1.5$ and $\beta = 3$. We consider the coupling function
$J(x) = J_{0}+2J_{1}\cos{x}$.  The time delay is $D=0.1$ and the
input current $I$ is varied so as to keep the uniform  stationary
solution fixed at $R=0.1$.} \label{fig:diagram}
\end{figure}

The primary instability lines for the SU state can be seen in the phase diagram, 
Fig.~(\ref{fig:diagram}).
The region in 
$(J_{0},J_{1})$ space where the SU state is stable is shown in gray, while the 
primary instabilities, listed above, are shown as red lines.  
In Section \ref{sect:codim1} we will provide a detailed analysis of
the bumps, global oscillations and waves (SB, OU and SW/TW) which
arise due to the Turing, Hopf and Turing-Hopf instabilities
respectively. The derivation of the  amplitude equations is given in \ref{app:ampeq}, 
as well as a brief discussion of the steady,
transcritical bifurcation which occurs for strong excitatory coupling
and is  not of primary interest for this study.    
Finally, in Section \ref{sect:codim2} we will analyze the codimension
2 bifurcations: Hopf and Turing-Hopf (OU and waves), and Turing and
Hopf (SU and OU).  This analysis will allow us to understand the dynamical 
states which appear near the upper and lower left hand corners of the 
grey shaded region in Fig.~(\ref{fig:diagram}), i.e. the SW/OU and OB states.

\section{\label{sect:codim1}Bifurcations of codimension 1}

As we are interested in creating a phase diagram as a function of the 
connectivity, we will take changes in the connectivity as the
bifurcation parameter.  The small parameter  $\epsilon$ is therefore
defined by the expansion
\begin{equation}
J_k = \bar{J}_k+\epsilon^{2}\Delta J_k, 
\label{eq:Jkexp}
\end{equation}
%
The perturbative method we apply,  which makes use of this small parameter, is 
called the multiple-scales method and is a standard approach for determining the 
weakly nonlinear behavior of pattern-forming instabilities \cite{cross93}.  
We choose 
the particular scaling of $\epsilon^{2}$ in the foreknowledge that if the amplitudes 
of the patterns of interest are 
scaled as $\epsilon$, a solvability condition will arise at order $\epsilon^{3}$.  
This solvability condition yields a dynamical equation governing the temporal 
evolution of the pattern (see Appendix A for details).
Without loss of generality we will assume that an instability of  a
nonzero spatial wavenumber is for $k=1$.  We will furthermore
co-expand the constant input $I$ so as to  maintain a fixed value for
the spatially homogeneous steady state solution $R$
\begin{eqnarray}
I &=& \bar{I}+\epsilon^{2}\Delta I, \label{eq:Iexp}\\   
r &=& R+\epsilon r_{1}+\epsilon^{2}r_{2}+\dots,\label{eq:rexp}
\end{eqnarray}
where the small parameter $\epsilon$ is defined by
Eq.(\ref{eq:Jkexp}). Additionally we define the slow time
\begin{equation}
T = \epsilon^{2}t.
\label{eq:T}
\end{equation}

\subsection{\label{subsect:T}Turing Bifurcation}

The emergence and nature of stationary bumps in rate equations have
been  extensively studied elsewhere, e.g. \cite{ermentrout80}.  We
briefly describe this state here for  completeness.  The $k^{th}$
spatial Fourier mode of the connectivity is given by the critical
value Eq.~(\ref{eq:disp0}), while  we assume that all other Fourier
modes are sufficiently below their critical values to avoid additional
instabilities.  Without loss of generality we  assume $k=1$ here.

We expand the parameters $J_1$, $I$ and $r$ as in
Eqs.~(\ref{eq:Jkexp},\ref{eq:Iexp},\ref{eq:rexp}), and define the slow
time Eq.~(\ref{eq:T}). The solution of Eq.(\ref{eq:rate}) linearized about the 
SU state $R$ is a spatially periodic
amplitude which  we allow to vary slowly in time, i.e. $r_{1} =
A(T)e^{ix}+c.c.$.  Carrying out a weakly nonlinear analysis to third
order in $\epsilon$ leads to the amplitude equation
\begin{equation}
\partial_{T}A = \eta\Delta J_{1}A+\Gamma |A|^{2}A,
\label{eq:ampeq_turing}
\end{equation}
with the coefficients
\begin{subequations}
\begin{equation}
\eta = \frac{\Phi^{'}}{1+D},
\label{eq:turing_lc} 
\end{equation}
\begin{equation}
\Gamma =
\frac{\bar{J}_{1}^{3}}{1+D}\Bigg(\frac{J_{0}(\Phi^{''})^{2}}{1-J_{0}\Phi^{'}}
+\frac{J_{2}(\Phi{''})^{2}}{2(1-J_{2}\Phi^{'})}+\frac{\Phi^{'''}}{2}\Bigg). \label{eq:turing_cc}
\end{equation}
\end{subequations}

\begin{figure}
\center
\includegraphics[scale=0.35]{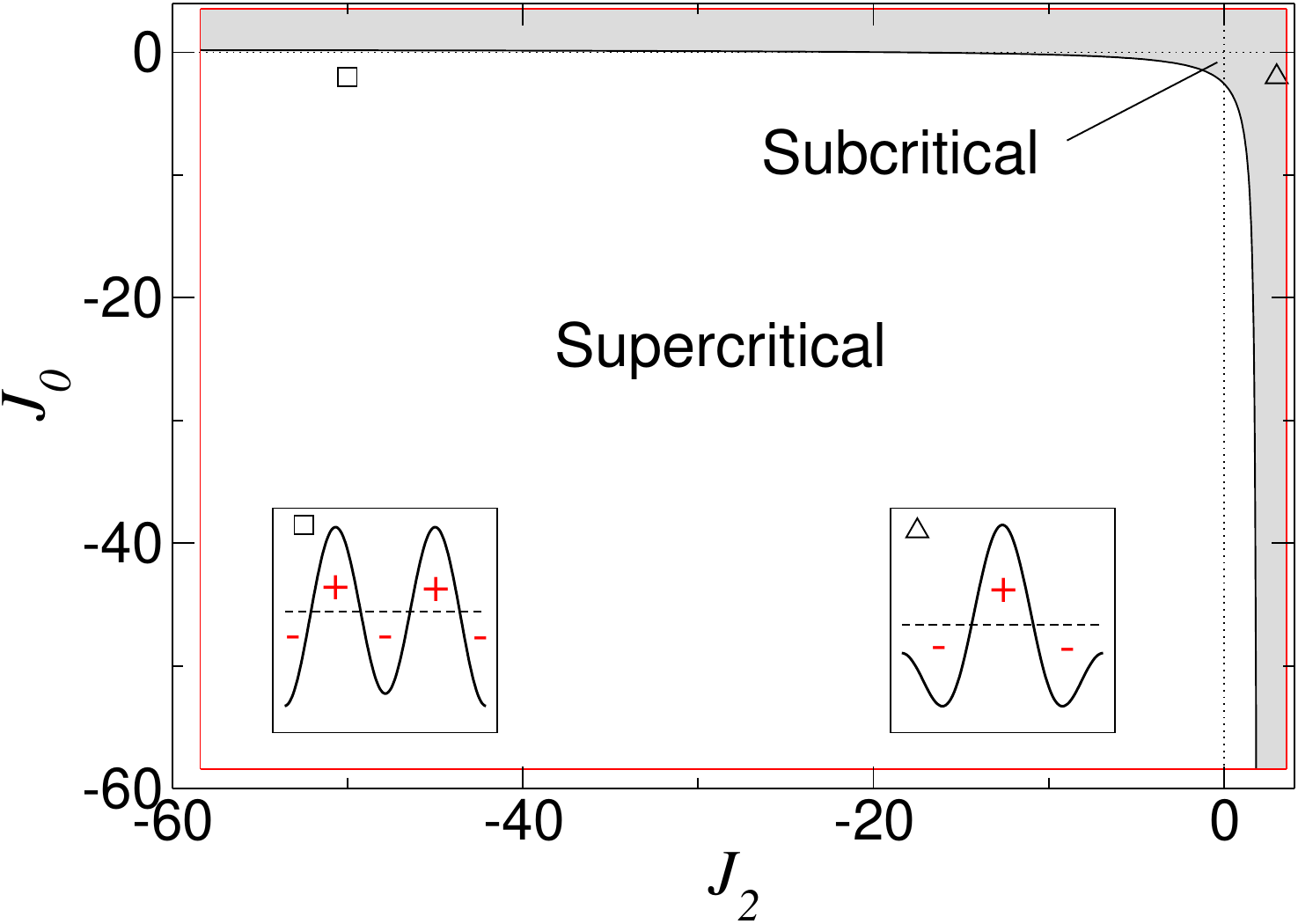}
\caption{(Color online) The phase diagram for stationary bumps as a function of the
zeroth  and second spatial Fourier modes of the connectivity kernel.  
The region of bistability between the unpatterned and the bump state is 
shaded.
Here the critical  spatial Fourier coefficient $\bar J_{1}=3.54$.
Red lines
indicate the boundaries of the SU state (obtained via 
Eqs.(\ref{eq:disp0}-\ref{eq:disp3}).
The functions $\Phi$ and $J(x)$ as well as the input current $I$ and
the delay $D$ are taken as in Fig.~(\ref{fig:diagram}).  Insets: example 
connectivity patterns corresponding to the values of $J_{0}$ and $J_{2}$ 
marked by the square and triangle respectively.  Note that standard 
Mexican Hat connectivity tends to favor bistability.}
\label{fig:turing_cubic}
\end{figure}

The nature of the bifurcation (sub- or supercritical) clearly depends
strongly on the  sign and magnitude of mean connectivity $J_{0}$ and
the second spatial Fourier mode $J_{2}$.  Figure~\ref{fig:turing_cubic}
shows a phase diagram of the bump state  at the critical value of
$\bar{J}_{1}=3.54$.  The red lines indicate oscillatory and steady instability 
boundaries for the modes $J_{0}$ and $J_{2}$.  Clearly $J_{0}<0$ and $J_{2}<0$ over 
most of the region of allowable values, and the bump is therefore supercritical.  
There is only a narrow region of predominantly positive values (shaded region 
in Fig.~\ref{fig:turing_cubic} for which the cubic coefficient is positive.  This 
indicates that the bifurcating solution branch is unstable.  However, neuronal 
activity is bounded, which is captured in Eq.(\ref{eq:rate}) by a saturating transfer 
function $\Phi$. Thus the instability will not grow without bound but rather will 
saturate, producing a finite amplitude bump solution.  This stable, large amplitude 
branch and the unstable branch annihilate in a saddle-node bifurcation for values of 
$J_{1}$ below the critical value for the Turing instability.  Such finite-amplitude 
bumps are therefore bistable with the SU state.  In Fig.~\ref{fig:turing_cubic}, 
the two insets show the connectivity kernel $J(x)$ for parameter values given by the 
placement of the open triangle (subcritical bump) and the open square (supercritical bump).  

In the phase diagram (\ref{fig:diagram}), the Turing instability line (upper horizontal 
red line) is shown thin for supercritical, and thick for subcritical bumps (here 
$J_{2}=0$). 

\subsection{\label{subsect:H}Hopf Bifurcation}

\subsubsection{Network simulations}

\begin{figure}
\centerline{\includegraphics[scale=0.5]{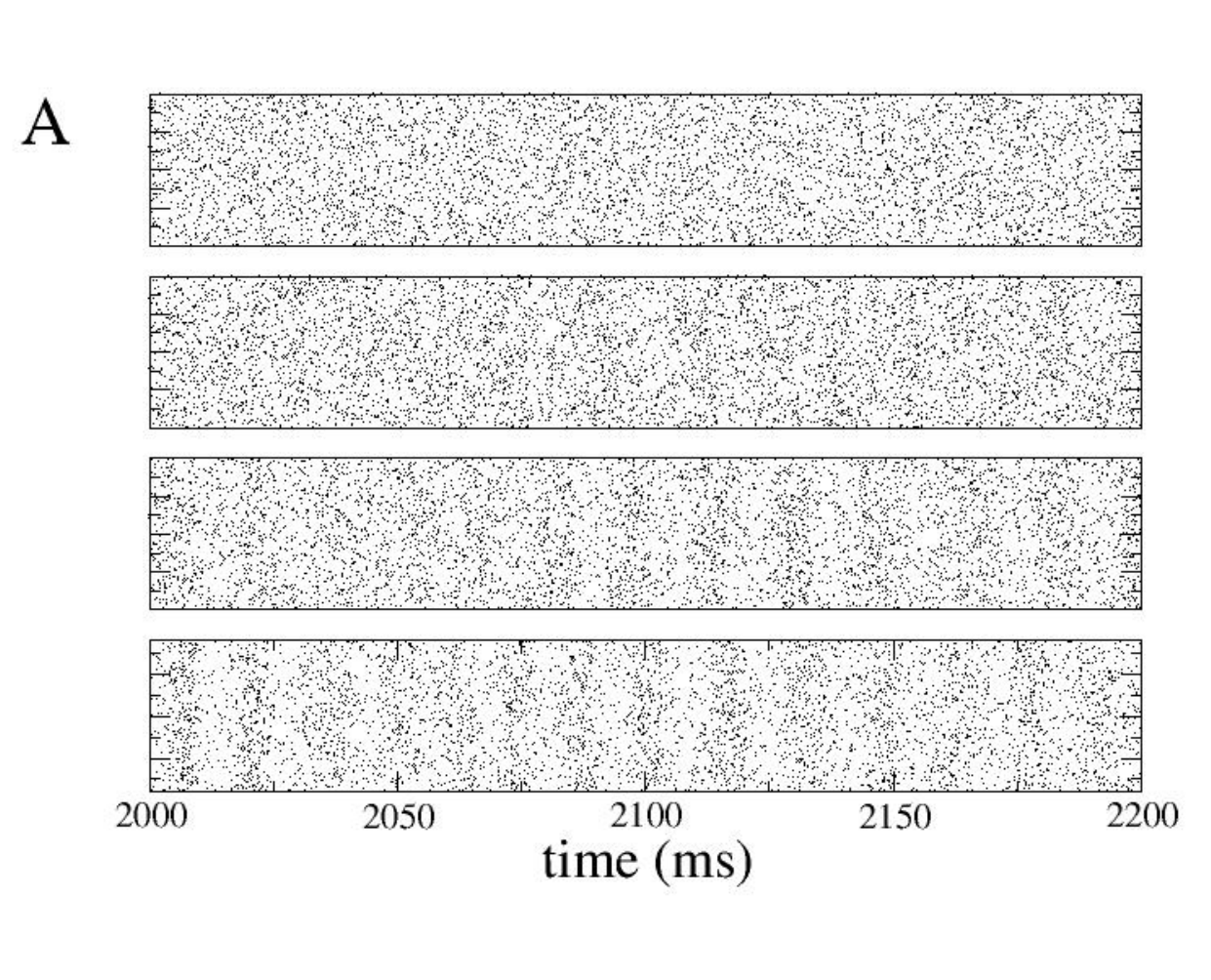}}
\centerline{\includegraphics[scale=0.4]{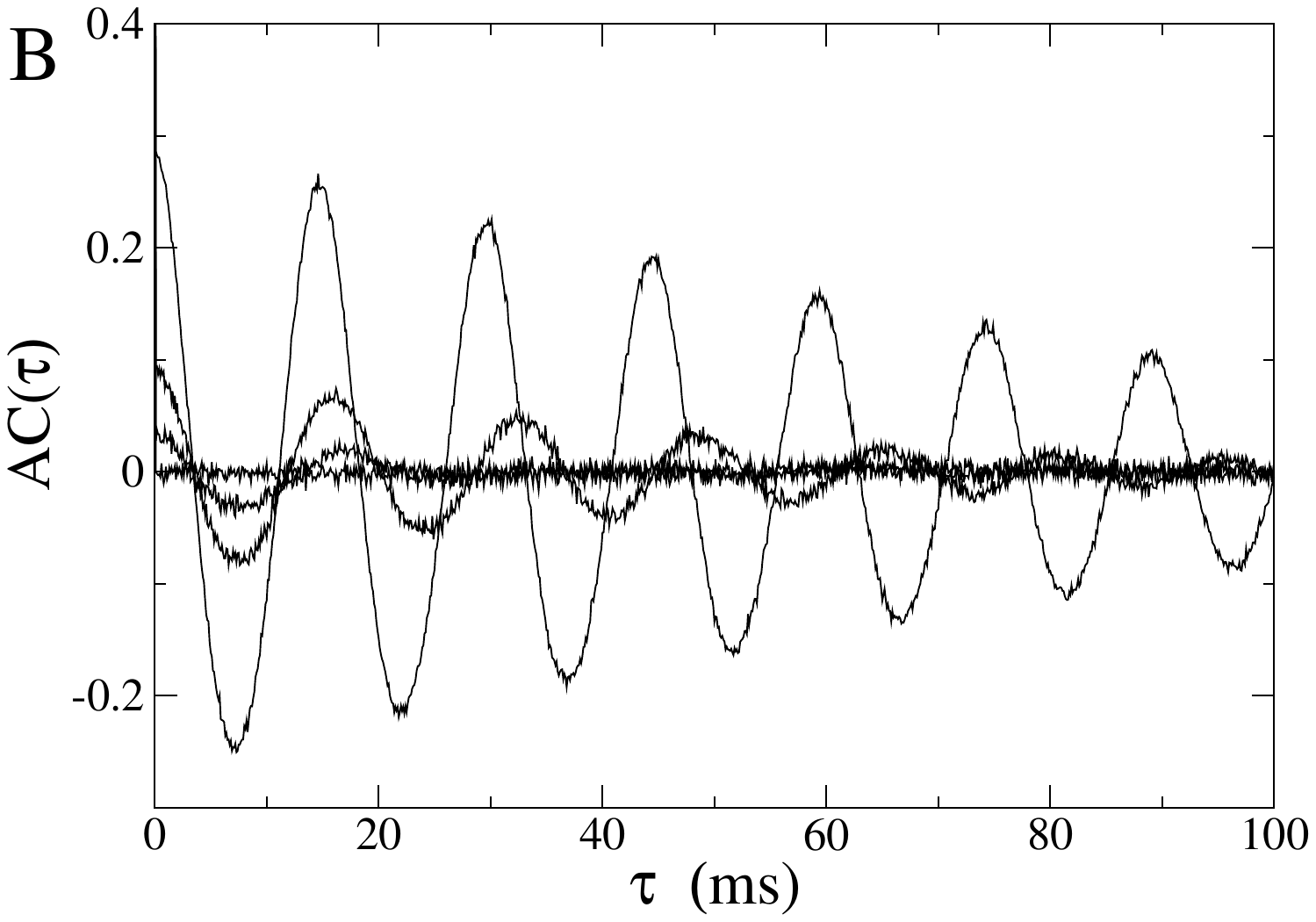}}
\caption{Fast oscillations in a sparse, randomly connected network of 2000 
randomly connected inhibitory 
neurons without delay ($p_{0}^{I}=0.2$).  A: Raster plots of spiking activity in four networks with 
increasing values of the synaptic weight. From top to bottom: $g_{I} = 0$, $0.005$, 
$0.008$, $0.01$mS$\cdot$ms/cm$^{2}$. The mean firing 
rate was kept fixed at approximately $14$Hz  by varying the input current. From top to bottom:  
$\nu_{ext} = 2150$, $7700$, $10000$ and $12000$Hz. See \ref{app:network} for additional 
model details.  B: The autocorrelation function of the network-averaged firing rate for the 
four networks.}
\label{fig:network_hopf}
\end{figure}
As shown elsewhere previously \cite{brunel99,brunel00,roxin05}, a network of 
recurrently coupled inhibitory neurons can generate fast oscillations due to 
the effective delay in the synaptic interactions.  
Fig.\ref{fig:network_hopf}A shows 
raster plots of the spiking activity in a network randomly connected 
inhibitory neurons  
as the synaptic weight $g_{I}$ is increased (see \ref{app:network} for a description of 
the network).
The raster plots clearly show oscillations emerging as the parameter 
$g_{I}$ increases in strength.  Fig.\ref{fig:network_hopf}B shows the autocorrelation 
function of the 
network-averaged firing rate from four seconds
 of simulation time.  The curve is 
completely flat for the uncoupled case as expected, while periodic peaks appear and grow as 
the synaptic weights are increased, indicating the presence of coherent network 
oscillations.  Note that the oscillations 
appear to emerge continuously, which would indicate a supercritical bifurcation.  This is 
consistent with the finding in \cite{brunel99}, where fast oscillations in a network of 
integrate-and-fire neurons were shown to be supercritical analytically.

We may ask if fast oscillations are in general expected to bifurcate supercritically, 
or if a subcritical bifurcation is also possible. 
To this end we study the rate equation which allows us to 
detemine the nature of the bifurcation analytically as a function of the transfer function 
and connectivity.

\subsubsection{Rate model}

In the rate model there is a spatially homogeneous oscillatory instability with
frequency $\omega$ given  by Eq.(\ref{eq:disp2}).  This occurs for a
value of the $0^{th}$ spatial Fourier mode of  the connectivity given
by Eq.(\ref{eq:disp3}), while  we assume that all other Fourier modes
are sufficiently below their critical values to avoid additional
instabilities.   We expand the parameters $J_0$, $I$ and $r$ as in
Eqs.~(\ref{eq:Jkexp},\ref{eq:Iexp},\ref{eq:rexp}), and define the slow
time Eq.~(\ref{eq:T}).  The linear solution has an   amplitude which  we
allow to vary slowly in time, i.e. $r_{1} = H(T)e^{i\omega t}+c.c.$.
Carrying out a weakly nonlinear analysis to third order in $\epsilon$
leads to the amplitude equation
\begin{equation}
\partial_{T}H = (\mu +i\Omega )\Delta J_{0}H +(\alpha
+i\beta)|H|^{2}H, \label{eq:hopf}
\end{equation}
where the coefficients $(\mu + i\Omega)$ and $(\alpha+i\beta)$ are
specified by the Eqs.~(\ref{eq:hopf_lc}) and (\ref{eq:hopf_cc}) in
the Appendix B.

\begin{figure}
\center
\includegraphics[scale=0.3]{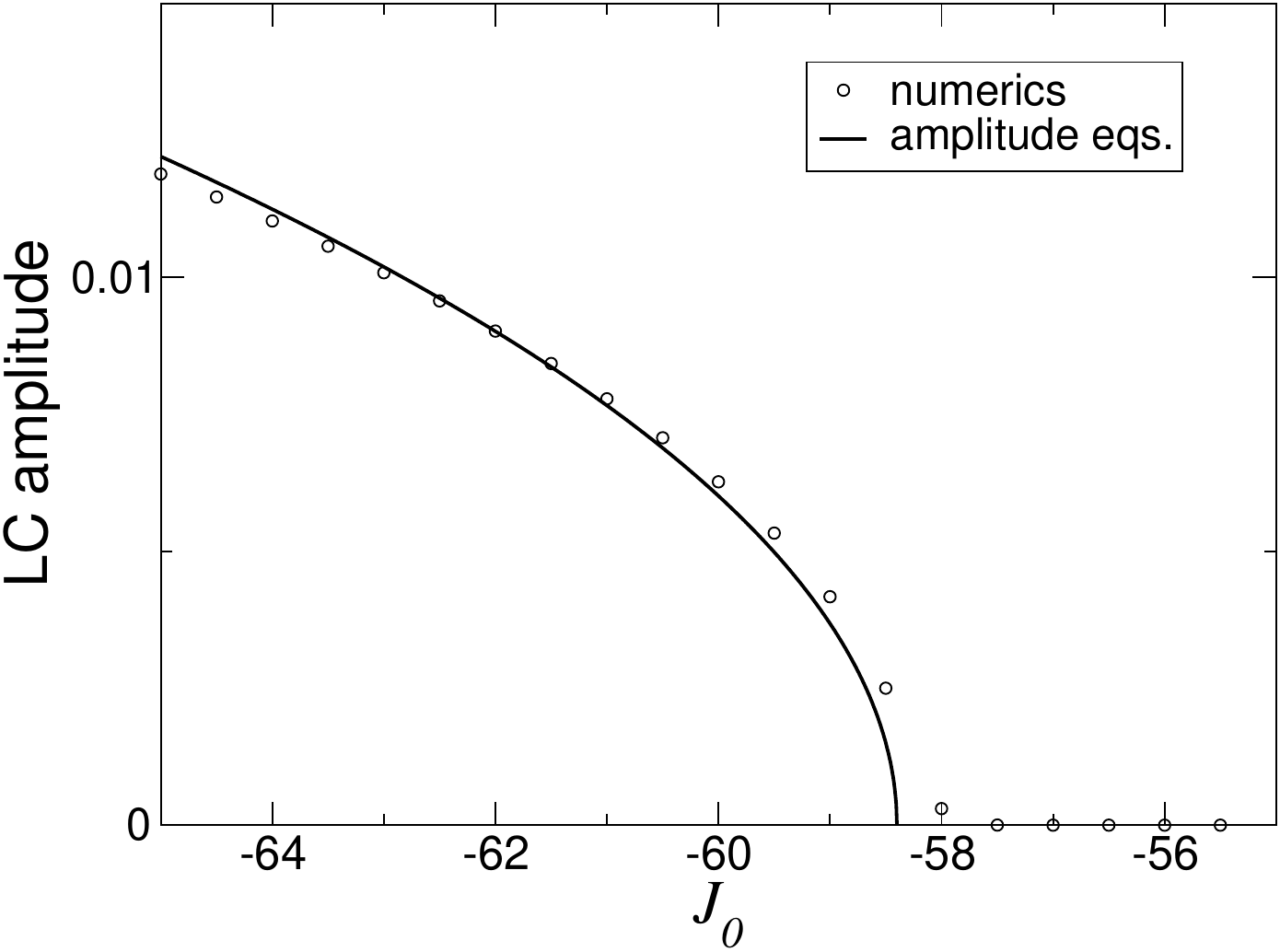}
\caption{The bifurcation diagram for a supercritical Hopf bifurcation.
Shown is the amplitude of the limit cycle  as a function of the
$0^{th}$ order spatial Fourier coefficient of the coupling $J(x)$.
Open circles are from  numerical simulation of Eq.(\ref{eq:rate}) and
solid lines show the solution from the amplitude equation,
Eq.(\ref{eq:ampeq_turing}). The functions $\Phi$ and $J(x)$ as well as
the input current $I$ and the delay $D$ are taken as in
Fig.~\ref{fig:diagram}.} \label{fig:hopf_bifdiag}
\end{figure}

Figure~\ref{fig:hopf_bifdiag} shows a typical bifurcation diagram (in
this case $J_{1}=0$) for the Hopf bifurcation.   Plotted is the
amplitude of the limit cycle as a function of $J_{0}$ where symbols
are from numerical simulation  of Eq.(\ref{eq:rate}) and the lines are
from the amplitude equation, Eq.(\ref{eq:hopf}).

\begin{figure}
\center
\includegraphics[scale=0.35]{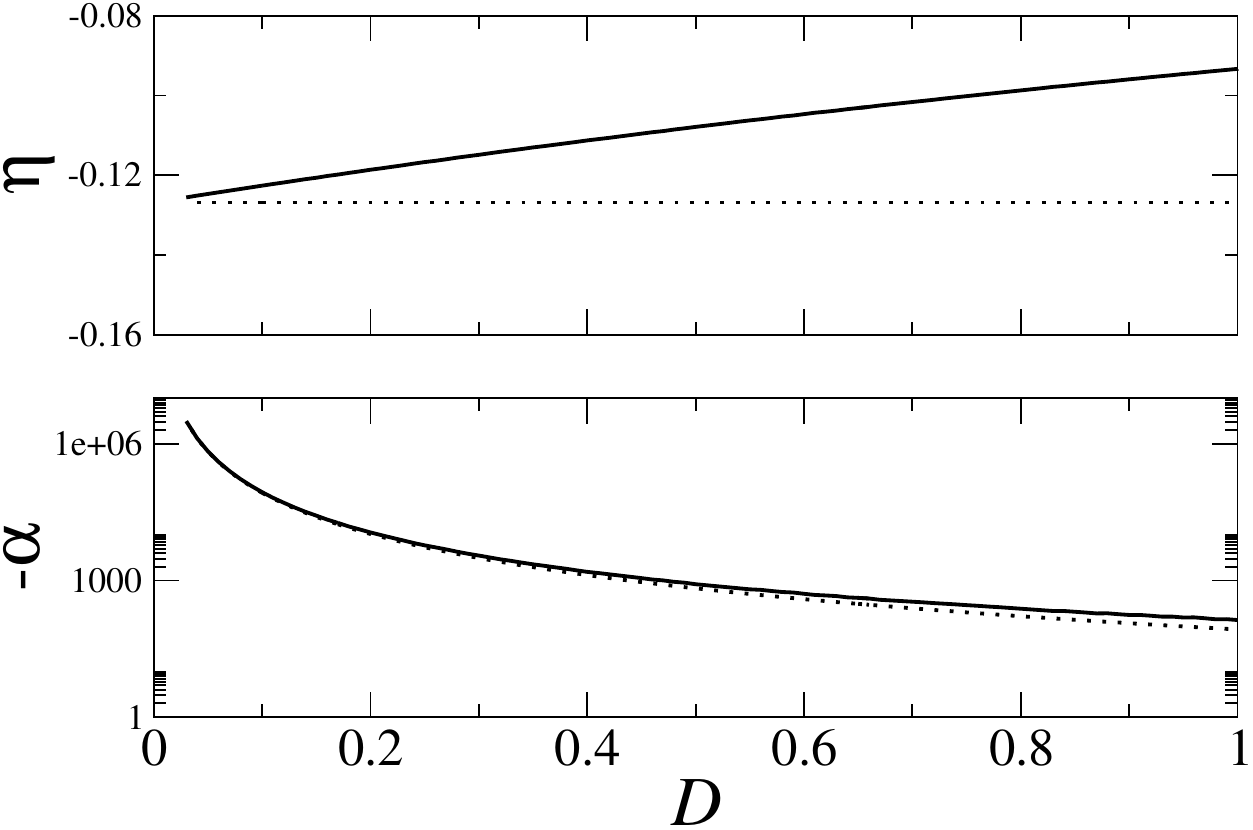}
\caption{Top: The real part of the linear coefficient $\mu $.  Bottom:
Minus the real part of the  cubic coefficient $-\alpha$.  Solid lines
are from the full expressions Eq.(\ref{eq:hopf_lc}-\ref{eq:hopf_cc}) and
dotted lines are the leading order terms in the small delay limit,
Eqs.~(\ref{eq:hopf_lc2}-\ref{eq:hopf_cc2}). The functions $\Phi$ and
$J(x)$ as well as the input current $I$ are taken as in
Fig.~(\ref{fig:diagram}).}
\label{fig:hopf_coefs}
\end{figure}

In the small delay limit ($D\to 0$) we can use the asymptotic
values (\ref{eq:D0}) to obtain, to leading order,
\begin{subequations}
\begin{equation}
\mu +i\Omega =
-\frac{(\frac{\pi}{2}+i)\Phi^{'}}{1+\frac{\pi^{2}}{4}},
\label{eq:hopf_lc2}
\end{equation}
\begin{eqnarray} 
\alpha +i\beta &=& -\frac{\chi}{(D \Phi') ^3} \Bigg(\frac{11\pi-4}{20}
\frac{(\Phi^{''})^{2}} {\Phi^{'}} -
\frac{\pi\Phi^{'''}}{4}-i\bigg[\frac{11+\pi}{10}-\frac{\Phi^{'''}}{2}\bigg]\Bigg),
\label{eq:hopf_cc2}
\end{eqnarray} 
\end{subequations}
where we have defined the quantity $\chi \equiv \pi^{3}/(8+2 \pi^{2})$.
Figure~\ref{fig:hopf_coefs} shows a comparison of the full expressions
for the coefficients of the amplitude equation,
Eqs.~(\ref{eq:hopf_lc}-\ref{eq:hopf_cc}) with the expressions obtained in
the limit $D\to 0$, Eqs.~(\ref{eq:hopf_lc2}-\ref{eq:hopf_cc2}).  Again, the 
agreement is quite good, even up to $D=1$, especially for the real part of the 
cubic coefficient $\alpha$, which is of primary interest here.

The asymptotic expression for the cubic coefficient $\alpha$,
Eq.(\ref{eq:hopf_cc2}), indicates that  a subcritical limit cycle should
occur for $\Phi^{'''}\Phi^{'}/(\Phi^{''})^{2}>(11\pi-4)/(5\pi )$.  This 
provides a simple criterion for determining whether or not a particular 
choice of the transfer function can generate oscillations which are 
bistable with the SU state.  In fact, it 
is a difficult condition to fulfill given a sigmoidal-like
input-output function.  For example, given a sigmoid of the form $\Phi
(x)=\alpha/(1+e^{-\beta x})$, one finds that
\begin{equation}
\frac{\Phi^{'''}\Phi^{'}}{(\Phi^{''})^{2}} = 1-2\frac{e^{-3\beta
x}}{(e^{-4\beta x}-2e^{-3\beta x}+e^{-2\beta x})}. \label{eq:bound}
\end{equation}

It is straightforward to show that the expression of the right hand side 
of Eq.(\ref{eq:bound}) is bounded above
by $1$.  In fact, $-\infty\le\Phi^{'''}\Phi^{'} /(\Phi^{''})^{2} < 1 <
(11-4\pi)/(5\pi)\sim 1.95$.  Such a nonlinear transfer function will
therefore always generate  supercritical oscillations.

If the nonlinear transfer function is interpreted as the single-cell
fI curve, which is  common in the literature, then we can use the
fact that cortical cells operate in the fluctuation-driven regime.  In
particular,  the mean input current to cortical cells is too low to
cause spiking.  Rather, this occurs at very low rates  due to
fluctuations in the membrane voltage.  Although the fI curve for
spiking neurons in the supra-threshold regime is  concave down and
saturates, in the fluctuation-driven, sub-threshold regime the fI curve
exhibits a smoothed out tail which  is concave up.  It has been shown
that the sub-threshold portion of the fI curve of actual cells can be
well fit by  a function of the form $\Phi(x)=Ax^{\gamma}$, where
$\gamma >1$ (see e.g. \cite{miller02,hansel02}).  In this case
\begin{equation}
\frac{\Phi^{'''}\Phi^{'}}{(\Phi^{''})^{2}} = 1-\frac{1}{\gamma -1},
\nonumber
\end{equation}
which again is bounded between $-\infty$ and $1$.  This again rules
out subcritical oscillations in the small delay limit.  

\begin{figure}
\center
\includegraphics[scale=0.27]{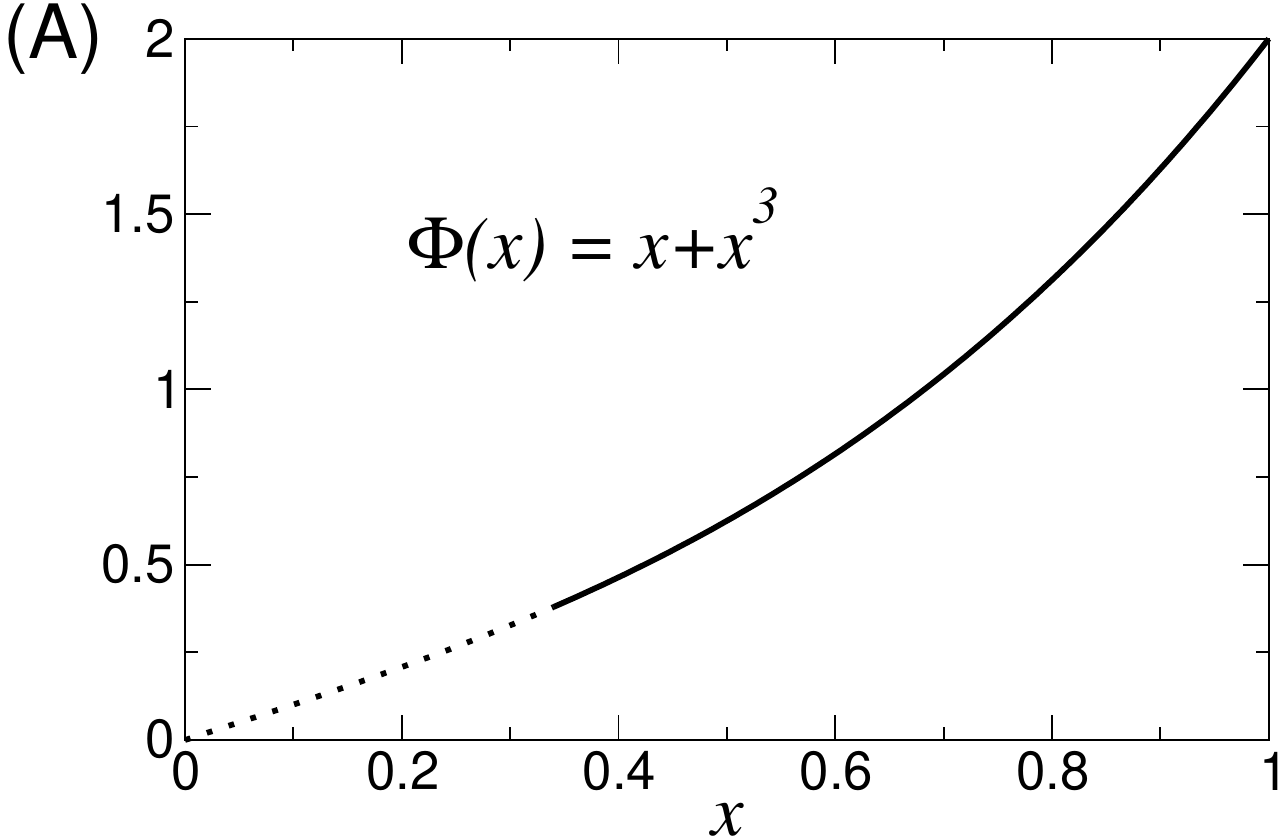}
\includegraphics[scale=0.27]{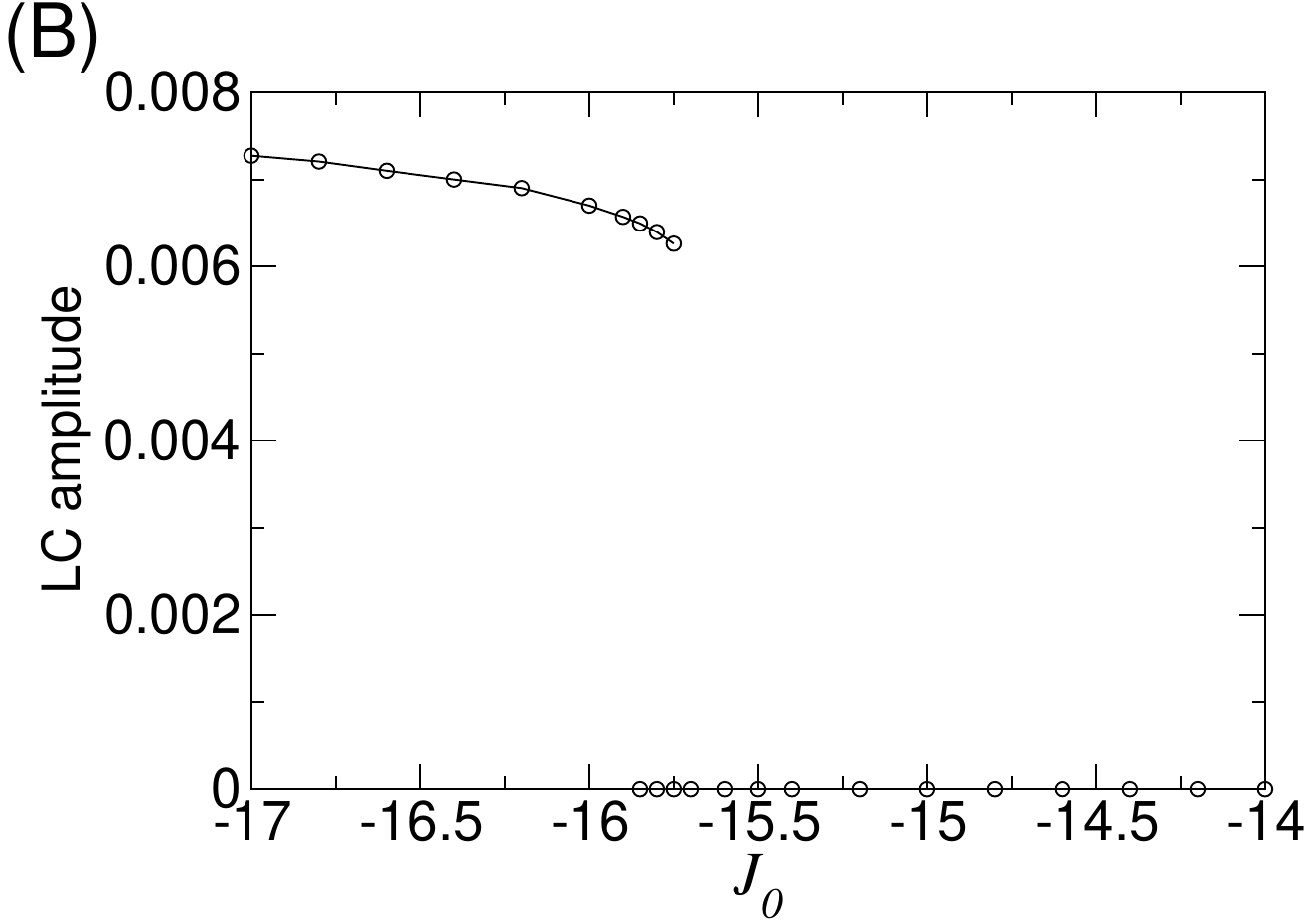}
\caption{A. An example of a function $\Phi (x)$ for which subcritical
oscillations are possible.   The dotted curve indicates the range of
the  function $\Phi$over which oscillations are subcritical. B.  A
bifurcation diagram for subcritical oscillations when the function
$\Phi (x)$ is the same as in panel A. Open circles: the limit cycle
amplitude computed numerically as  a function of $J_{0}$.  
Here $D=0.1$ and the critical coupling is
$\bar{J}_{0}=-15.89$.  The fixed point is held at $R=0.1$ and thus the
value of $x$  in panel A is close to 0.1 ($x+x^{3}=0.1$).}
\label{fig:hopf_sub}
\end{figure}

Nonetheless,
suitable  functions $\Phi$ for generating subcritical oscillations
can be contrived, as shown in Fig.~\ref{fig:hopf_sub}~A.  Numerical
simulation of Eq.(\ref{eq:rate})  indeed reveals a subcritical
bifurcation in this case (see Fig.~\ref{fig:hopf_sub}~B).  However, this type 
of transfer function does not seem consistent with the interpretation of 
$\Phi$ as a single-cell fI curve, nor with that of $\Phi$ as a 
cumulative distribution of activation, i.e. a sigmoid.  This strongly suggests 
that delay-driven oscillations in networks of spiking neurons will be 
generically supercritical.  
%

\subsection{\label{subsect:TH}Turing-Hopf Bifurcation}

\subsubsection{Network simulations}

\begin{figure}
\centerline{\includegraphics[scale=0.45]{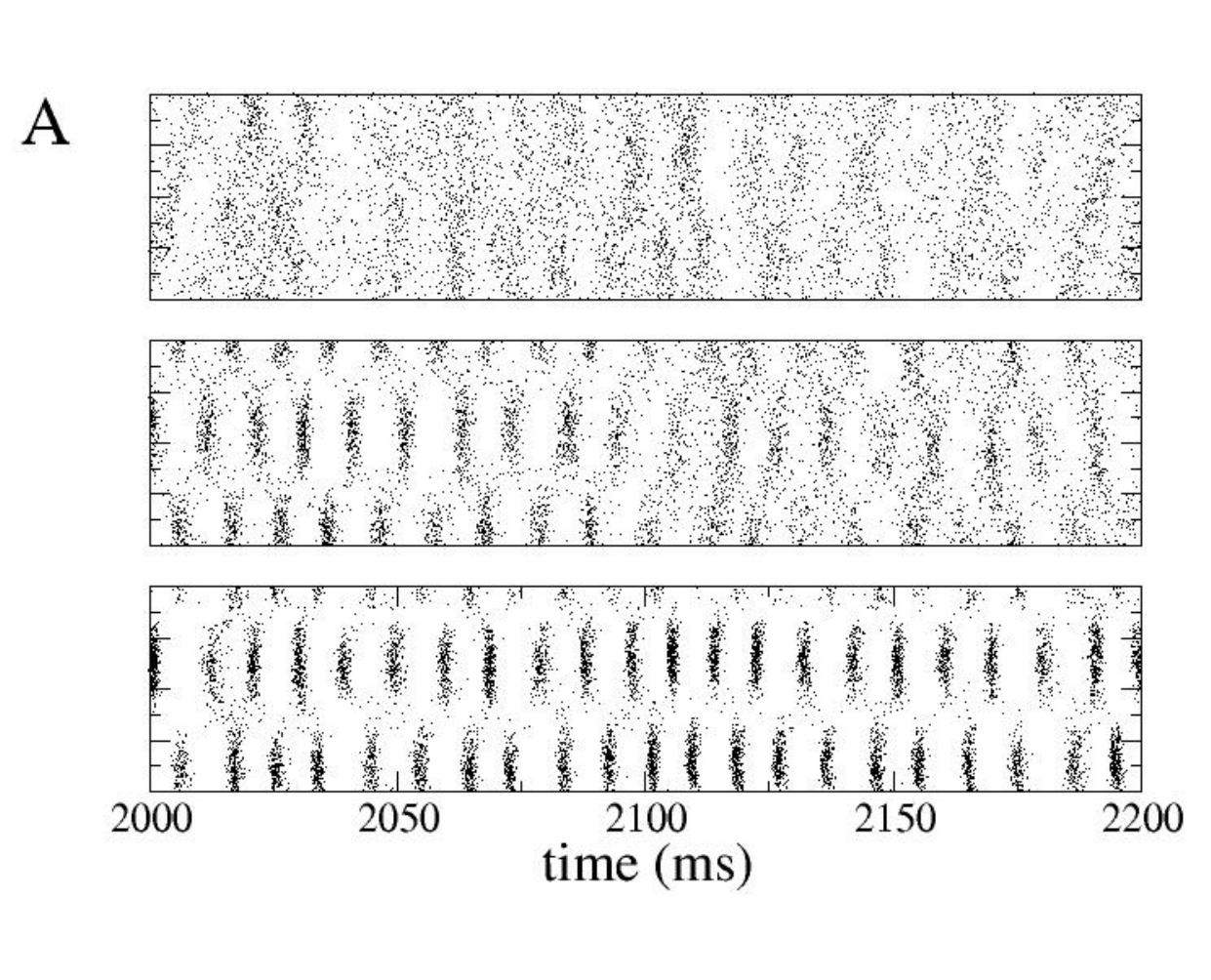}}
\centerline{\includegraphics[scale=0.35]{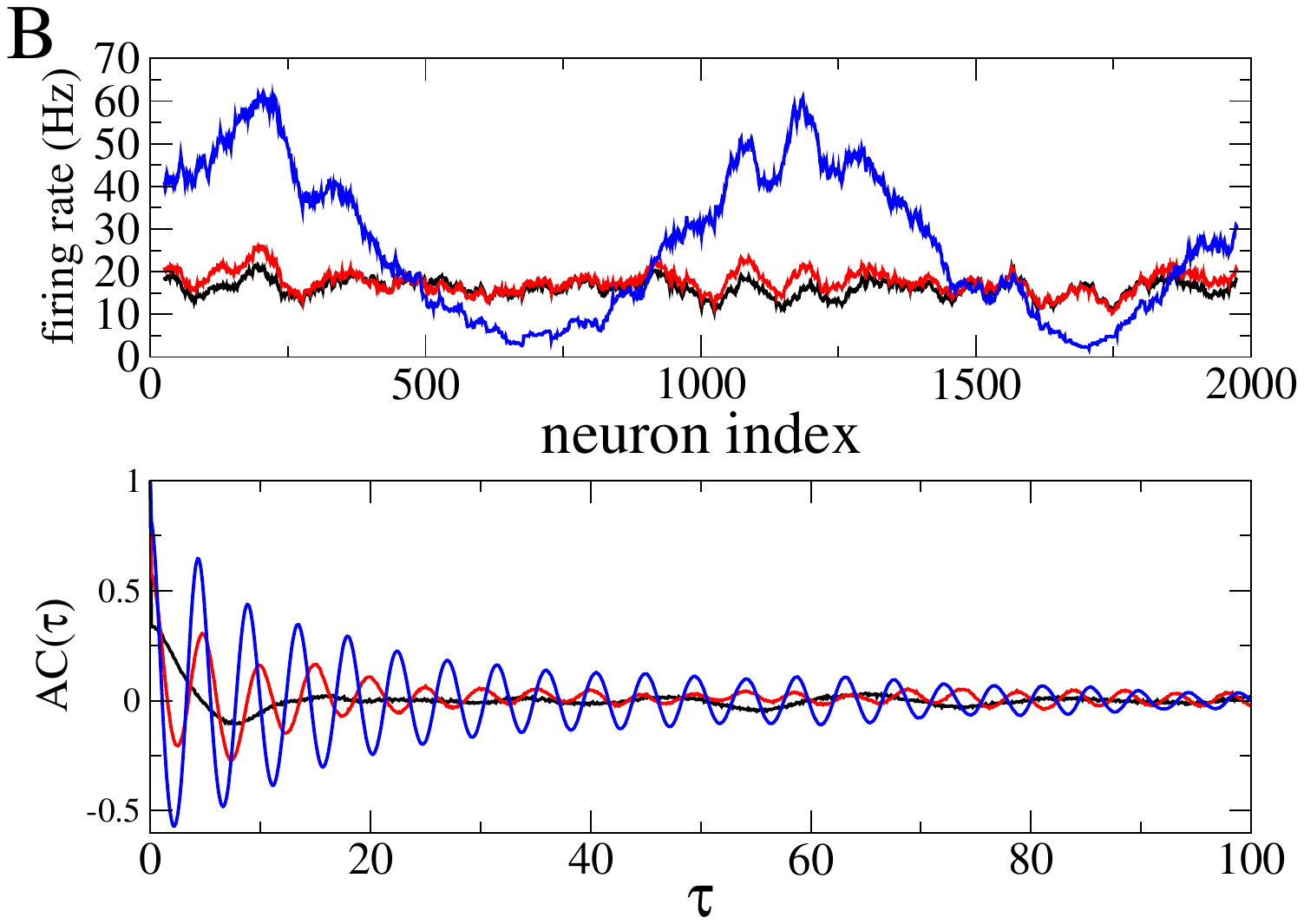}}
\caption{(Color online) Standing waves in a spiking network with inverted Mexican-hat 
connectivity with $p_{0}^{E}=p_{0}^{I}=0.2$, $p_{1}^{E}=0$ and $p_{2}^{E,I}=0$   
(see Eq.(\ref{conn}) in the Appendix A) and $g_{E}=0.01$, $g_{I}=0.028$. 
The rate of the external Poisson inputs is $\nu_{ext}=5000$Hz and $g_{ext}=0.001$.  
A: Raster plots of spiking activity for three simulations with 
increasing spatial modulation of the connection probability between neurons. From 
top to bottom: $p_{1}^{I}=0.15$, $0.17$ and $0.20$.  
B: Top: The profile of activity in the three simulations averaged over 2500ms.  The color code 
is black $p_{1}^{I}=0.15$, red $0.17$ and blue $0.2$.  Bottom: The autocorrelation function 
of the firing rate averaged over all neurons in each simulation.  Note that the instantaneous firing 
rate itself is not shown here.}
\label{fig:network_waves}
\end{figure}

As shown previously in \cite{roxin05}, given an inverted Mexican-hat
connectivity  for which inhibition dominates locally, fast waves may
emerge in networks of  spiking neurons.  This is illustrated in
Fig.\ref{fig:network_waves}A, where  raster plots of three networks
are shown with the degree of spatial modulation  increasing from top
to bottom.  

Additionally, Fig.~\ref{fig:network_waves}B (top) shows a
spatial profile of the network activity averaged over 2500 ms for  the
three networks,
while the bottom panel  shows the autocorrelation function (AC) of the
network firing rate for the three cases (same color scheme).  Note that 
for small inhibitory spatial modulation $p_{1}^{I}$ (black curve)
the profile is essentially flat while the AC exhibits
an initial peak and dip, but an absence of multiple peaks which would
indicate fast oscillations. As $p_{1}^{I}$ is increased (red curve), 
the profile remains flat but the AC clearly exhibits periodic peaks 
indicating that fast oscillations are present in the  firing rate.  
The corresponding raster plot in the middle panel of 
Fig.~\ref{fig:network_waves} shows
intermittent standing wave patterns which emerge and later disappear 
giving rise to a new pattern with a different spatial orientation (not shown).  
This explains why the time average of this spatial profile becomes flat.  
Finally, for strong enough spatial
modulation, a stationary standing  wave pattern is seen in
Fig.~\ref{fig:network_waves}A (bottom). In this case the time-averaged spatial
profile shown in Fig.~\ref{fig:network_waves}B (top, blue) shows two
distinct maxima, whereas the AC indicates fast oscillations in the firing rate.

Extensive simulations with such a cosine connectivity always yielded standing 
wave patterns for various choices of synpatic weights and input rates (not shown).  
We seek to understand why this is so, and if delay-driven traveling wave patterns 
can also be seen in network numerical simulations.  To this end we study the 
emergence of fast oscillations in the rate equation.

\subsubsection{Rate equation}

There is a spatially inhomogeneous oscillatory instability with
frequency $\omega$ given  by Eq.(\ref{eq:disp2}).  This occurs for a
value of the $k^{th}$ spatial Fourier mode of  the connectivity given
by Eq.(\ref{eq:disp3}), while  we assume that all other Fourier modes
are sufficiently below their critical values to avoid additional
instabilities.  Without loss of generality we assume $k=1$.

We expand the parameters $J_1$, $I$ and $r$ as in
Eqs.~(\ref{eq:Jkexp},\ref{eq:Iexp},\ref{eq:rexp}) and define the slow
time Eq.~(\ref{eq:T}).  The linear solution consists of leftwards and
rightwards traveling waves with  an   amplitude which  we allow to
vary slowly in time, i.e. $r_{1} = A(T)e^{i\omega
t+ix}+B(T)e^{-i\omega t+ix}+c.c.$.  Carrying out a weakly nonlinear
analysis to third order in $\epsilon$ leads to the coupled amplitude
equations
\begin{subequations}
\begin{equation}
\partial_{T}A = (\mu+i\Omega)\Delta J_{1}A+(a+ib)|A|^{2}A
+(c+id)|B|^{2}A, \label{eq:ampeq_THA}
\end{equation}
\begin{equation}
\partial_{T}B =(\mu-i\Omega)\Delta J_{1}B+(a-ib)|B|^{2}B
+(c-id)|A|^{2}B, \label{eq:ampeq_THB}
\end{equation}
\end{subequations}
where the coefficients $(a+ib)$, $(c+id)$ and $(\mu +i\Omega)$ are
given by the Eqs.~(\ref{eq:TH_cc1},\ref{eq:TH_ccc1},\ref{eq:hopf_lc}), respectively.

\begin{figure}
\center
\includegraphics[scale=0.3]{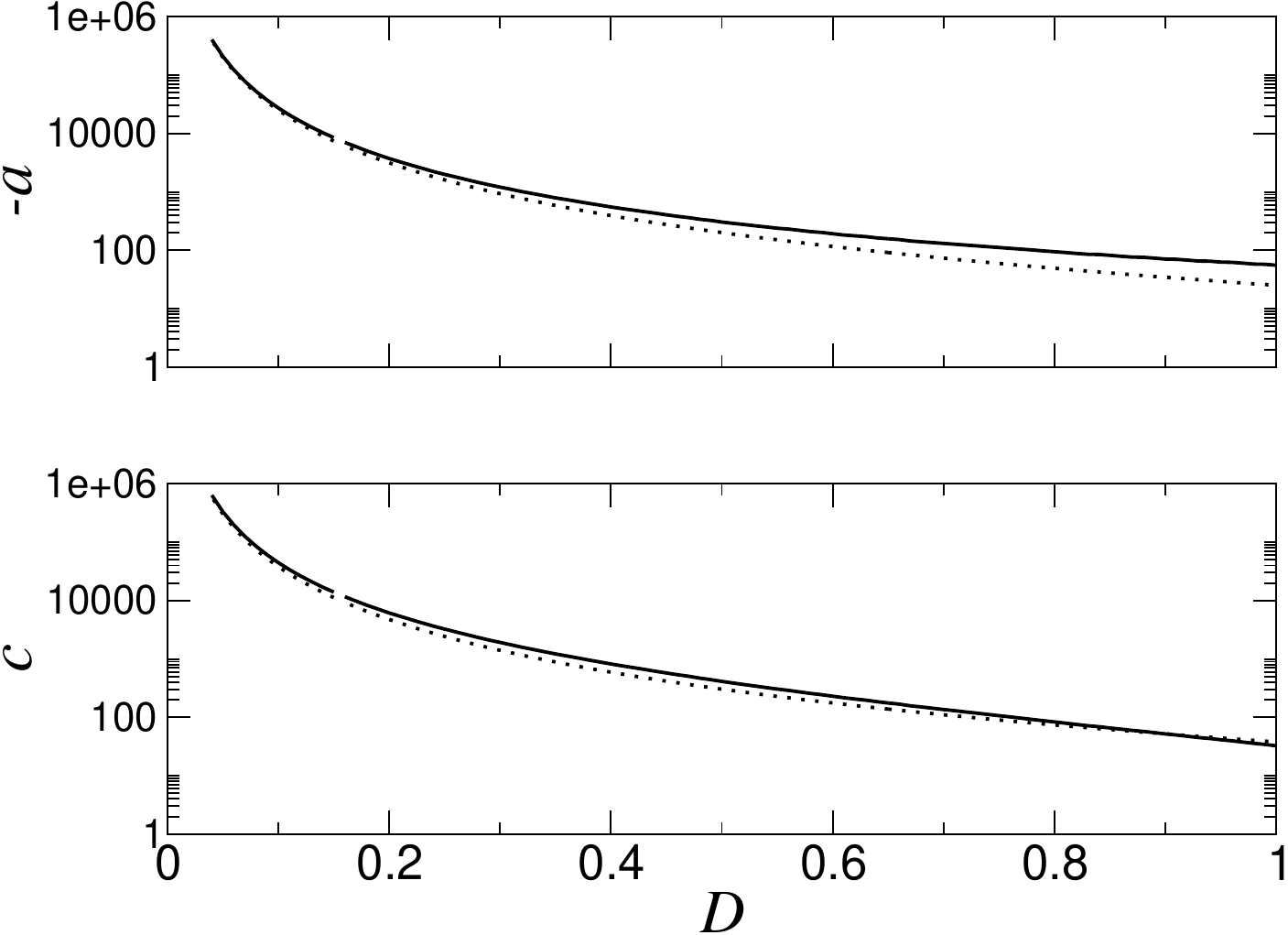}
\caption{Top: The real part of the cubic coefficient $a$.  Bottom: The
real part of the  cross-coupling coefficient $c$.  Solid lines are
from the full expressions Eq.(\ref{eq:TH_cc1}-\ref{eq:TH_ccc1}) and
dotted lines are the leading order terms in the small delay limit,
Eqs.~(\ref{eq:TH_cc2}-\ref{eq:TH_ccc2}).  The functions $\Phi$ and $J(x)$
as well as the  input current $I$  are taken as in
Fig.~(\ref{fig:diagram}).} 
\label{fig:TH_a_and_c}
\end{figure}

In the small delay limit ($D\to 0$) we can use the asymptotic
values~(\ref{eq:D0}) to obtain, to leading order,
\begin{subequations}
\begin{equation}
a+ib = \frac{\chi  (\frac{\pi}{2}+i)}{(D \Phi^{'}) ^3}  \Bigg(
 \frac{J_{0}(\Phi^{''})^{2}}{1-\Phi^{'}J_{0}}+\frac{\Phi^{'''}}{2}\Bigg),
\label{eq:TH_cc2} 
\end{equation}
\begin{equation}
c+id =
\frac{\chi (\frac{\pi}{2}+i)}{(D \Phi^{'})^3} \Bigg(
\frac{J_{0}(\Phi^{''})^{2}}{1-\Phi^{'}J_{0}}+\Phi^{'''}
+\frac{J_{2}(\Phi^{''})^{2}}{1-\Phi^{'}J_{2}}\Bigg), \label{eq:TH_ccc2}
\end{equation}
\end{subequations}
where $\chi \equiv \pi^{3}/(8+2 \pi^{2})$. Figure~\ref{fig:TH_a_and_c} 
shows a comparison of the full expressions
(solid lines) for the real parts of the cubic and cross-coupling
coefficients $a$ and $c$ with the asymptotic expressions above (dotted
lines).
\\

\noindent
\subsubsection{Wave solutions and their stability.}

The equations (\ref{eq:ampeq_THA}) and (\ref{eq:ampeq_THB}) admit solutions of the
form $(A,B)=(\mathcal{A}e^{i\theta_{A}}, \mathcal{B}e^{i\theta_{B}})$,
where the amplitudes $\mathcal{A}$ and $\mathcal{B}$ obey
\begin{subequations}
\begin{equation}
\dot{\mathcal{A}} = \mu\Delta
J_{1}\mathcal{A}+a\mathcal{A}^{3}+c\mathcal{B}^{2}\mathcal{A},
\label{eq:ampeq_THA2}
\end{equation} 
\begin{equation}
\dot{\mathcal{B}} = \mu\Delta
J_{1}\mathcal{B}+a\mathcal{B}^{3}+c\mathcal{A}^{2}\mathcal{B}. \label{eq:ampeq_THB2}
\end{equation}
\end{subequations}

\textit{Traveling waves:}  Leftward and rightward traveling waves in
Eqs.~(\ref{eq:ampeq_THA2}) and  (\ref{eq:ampeq_THB2}) are given by
$(\mathcal{A}_{TW},0)$ and $(0,\mathcal{A}_{TW})$ respectively, where
$\mathcal{A}_{TW} = -\mu\Delta J_{1}/a$.  The stability of
traveling waves can be determined with the  ansatz
$(\mathcal{A},\mathcal{B})=(\mathcal{A}_{TW},0)+(\delta\mathcal{A},\delta\mathcal{B})e^{\lambda
t}.$   The resulting eigenvalues are $\lambda_{1} = -2\mu\Delta J_{1}$
and $\lambda_{2} = -\mu\Delta J_{1}(c/a-1)$.\\

\textit{Standing waves:} Standing waves in Eqs.~(\ref{eq:ampeq_THA2}) and
(\ref{eq:ampeq_THB2}) are given by
$(\mathcal{A}_{SW},\mathcal{A}_{SW})$, where $\mathcal{A}_{TW} =
-\mu\Delta J_{1}/(a+c)$.  The stability of  standing waves can
be determined with the ansatz
$(\mathcal{A},\mathcal{B})=(\mathcal{A}_{SW},\mathcal{A}_{SW})
+(\delta\mathcal{A},\delta\mathcal{B})e^{\lambda t}.$  The resulting
eigenvalues are  $\lambda_{1} = -2\mu\Delta J_{1}$ and $\lambda_{2} =
-2\mu\Delta J_{1}(a-c)/(a+c)$.
\begin{figure}
\center
\includegraphics[scale=0.4]{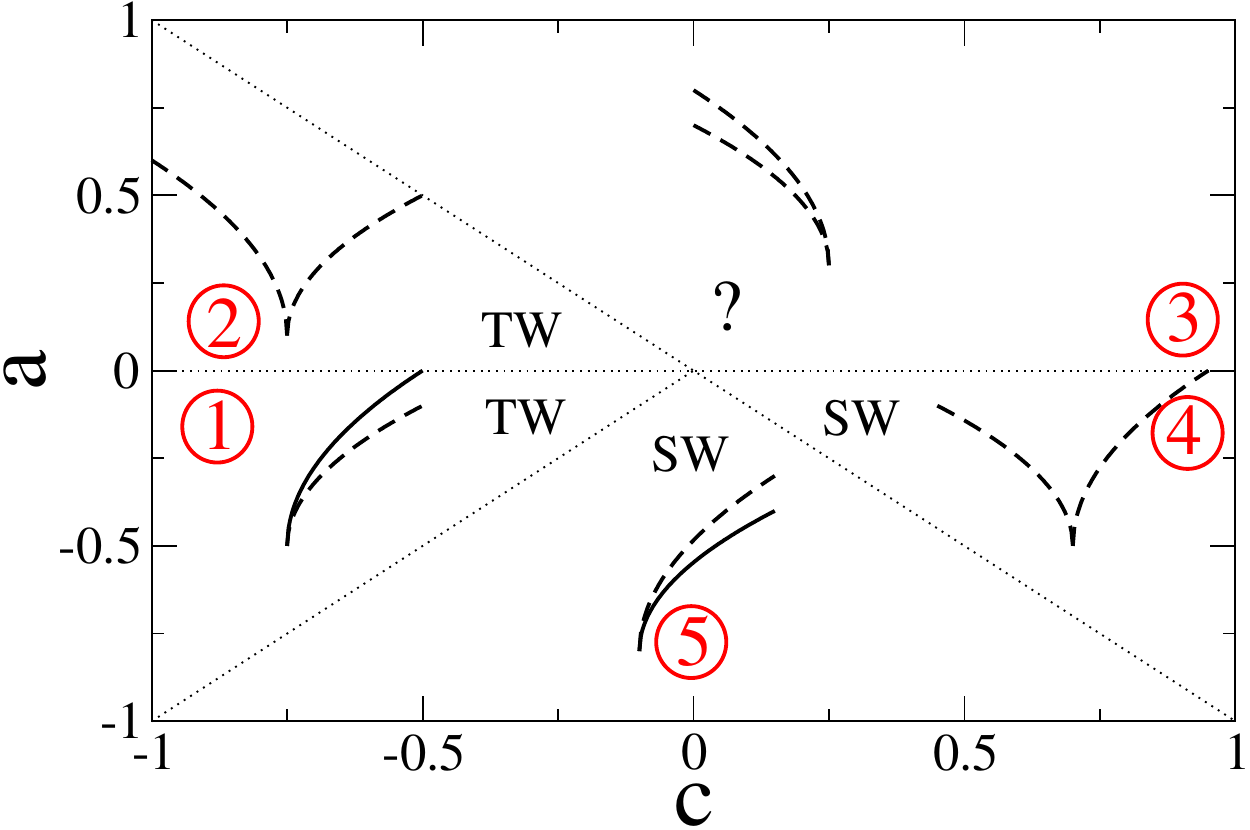}
\caption{The existence and stability of traveling and standing waves
  as a function of the cubic and cross-coupling  coefficients $a$ and
  $c$ given by Eqs.~(\ref{eq:TH_cc2},\ref{eq:TH_cc2}) and, in the small 
delay limit, by Eq.~(\ref{eq:a_vs_c}).  
In each sector of parameter space a representative bifurcation
  diagram is shown.  Supercritical  (subcritical) solutions are shown
  growing from left to right (right to left).  Stable (unstable)
  solutions are given  by solid (dashed) lines.  Also indicated in
  each sector is the type of solution which will be seen numerically.
  A question  mark is placed wherever the type of stable solution
  cannot be determined through a weakly nonlinear analysis.}
\label{fig:a_vs_c}
\end{figure}

The existence and stability of small-amplitude waves as described
above is completely determined by the values of the  cubic and
cross-coupling coefficients $a$ and $c$.  This is illustrated in
Fig.~\ref{fig:a_vs_c}, where the parameter  space is divided into five
sectors.  In each sector the type of solution which will be observed
numerically is indicated  where known, and otherwise a question mark
is placed.  Illustrative bifurcation diagrams are also given.
Specifically,  in the region labeled 1 (red online), the SW solution
is supercritical and  unstable while the TW solution is supercritical
and stable.  TW will therefore  be observed.  In the region labeled 2,
the SW solution is supercritical and  unstable while the TW solution
is subcritical.  Finite-amplitude TW are therefore  expected to occur
past the bifurcation point.  In the region labeled 3, both solution
branches are subcritical, indicating that the analysis up to cubic
order is not  sufficient to identify the type of wave which will be
observed.  In the region  labeled 4, TW are supercritical and unstable
while SW are subcritical.  Finite amplitude  SW are therefore expected
past the bifurcation point.  In the region labeled 5, the  TW solution
is supercritical and unstable while the SW solution is supercritical
and stable.  SW will therefore be observed.

Performing the small delay limit we find, from Eqs.~(\ref{eq:TH_cc2},\ref{eq:TH_ccc2}), 
that
\begin{equation}
a =c-\frac{\pi}{2} \frac{\chi}{(D \Phi)^3} \Bigg(\frac{\Phi^{'''}}{2}
+\frac{J_{2}(\Phi^{''})^{2}}{1-\Phi^{'}J_{2}}\Bigg). \label{eq:a_vs_c}
\end{equation}
From Fig.~\ref{fig:a_vs_c} we can see that the nature of the solution
seen will depend crucially on  the sign of the second term of the
right-hand side of Eq.(\ref{eq:a_vs_c}).  In particular, the diagonal
$a=c$ divides the  the parameter space into two qualitatively
different regions.  Above this line TWs are  favored while below it
SWs are favored.  In the small delay limit, Eq.(\ref{eq:a_vs_c})
indicates  that the balance between the third derivative of the
transfer function $\Phi^{'''}$ and the second spatial  Fourier mode of
the connectivity kernel will determine whether TW or SW are favored.

For sigmoidal transfer functions, the third derivative changes sign
from positive to  negative already below the inflection  point, while
for expansive power-law nonlinearities, which fit cortical neuronal
responses quite well in the  fluctuation-driven regime, the third
derivative is positive if the power is greater than 2 and is negative
otherwise.  The contribution of this term therefore will depend on the
details of the neuronal response.  In simulations of large networks of
conductance-based neurons in the fluctuation-driven regime  in which
$J_{2}$ was zero, the standing wave state was always observed,
indicating a $\Phi^{'''}>0$ \cite{roxin05,roxin06}.

\begin{figure}
\center
\begin{tabular}{cccc}
A & \includegraphics[scale=0.25]{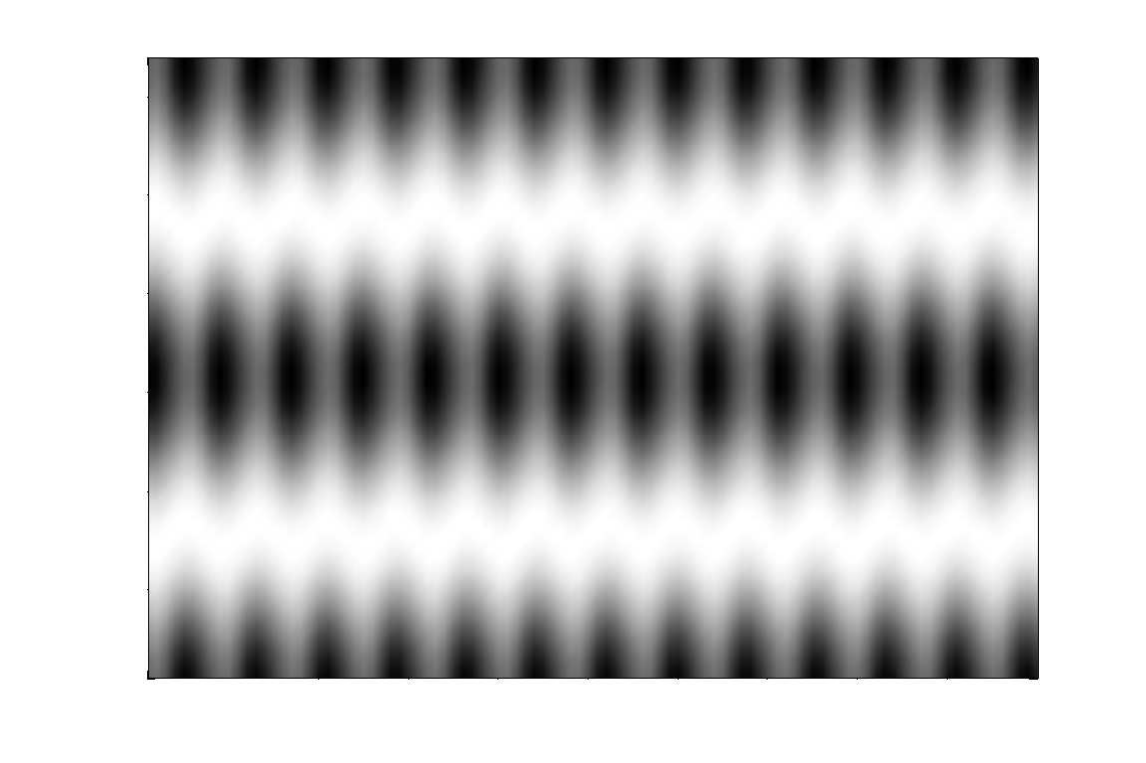} & B &
\includegraphics[scale=0.25]{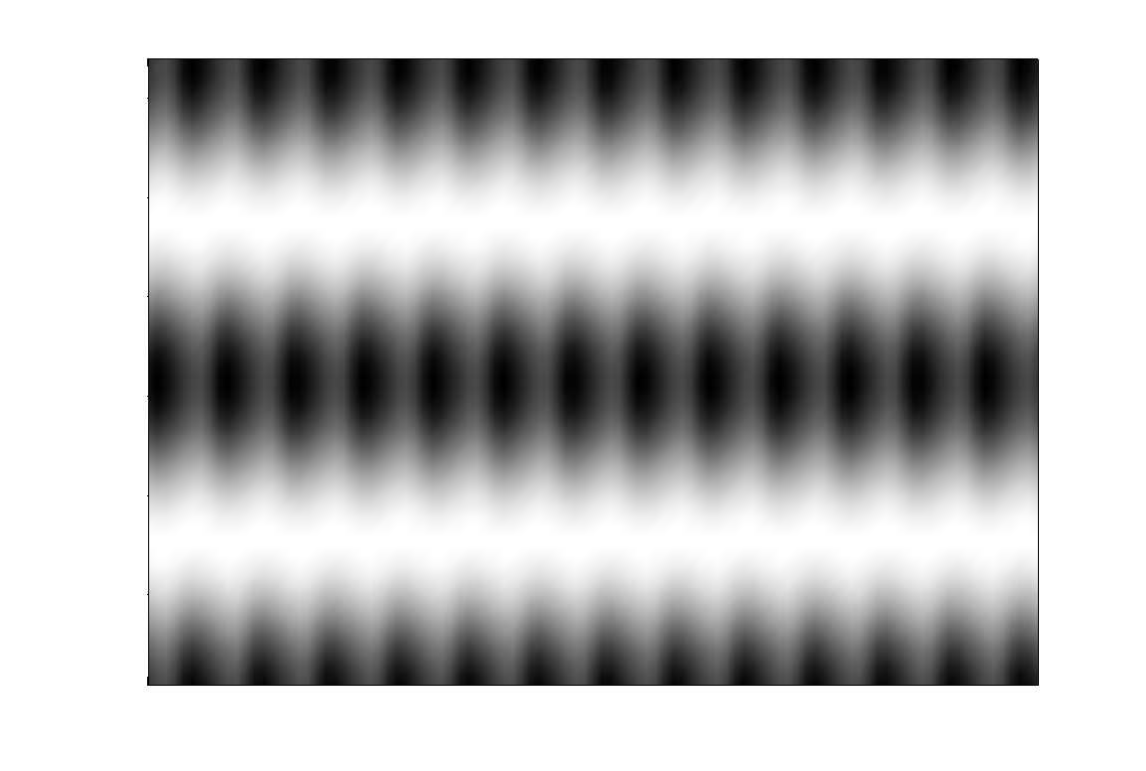}\vspace{-0.05in}\\ &
\textbf{time} & & \textbf{time} \\ C &
\includegraphics[scale=0.25]{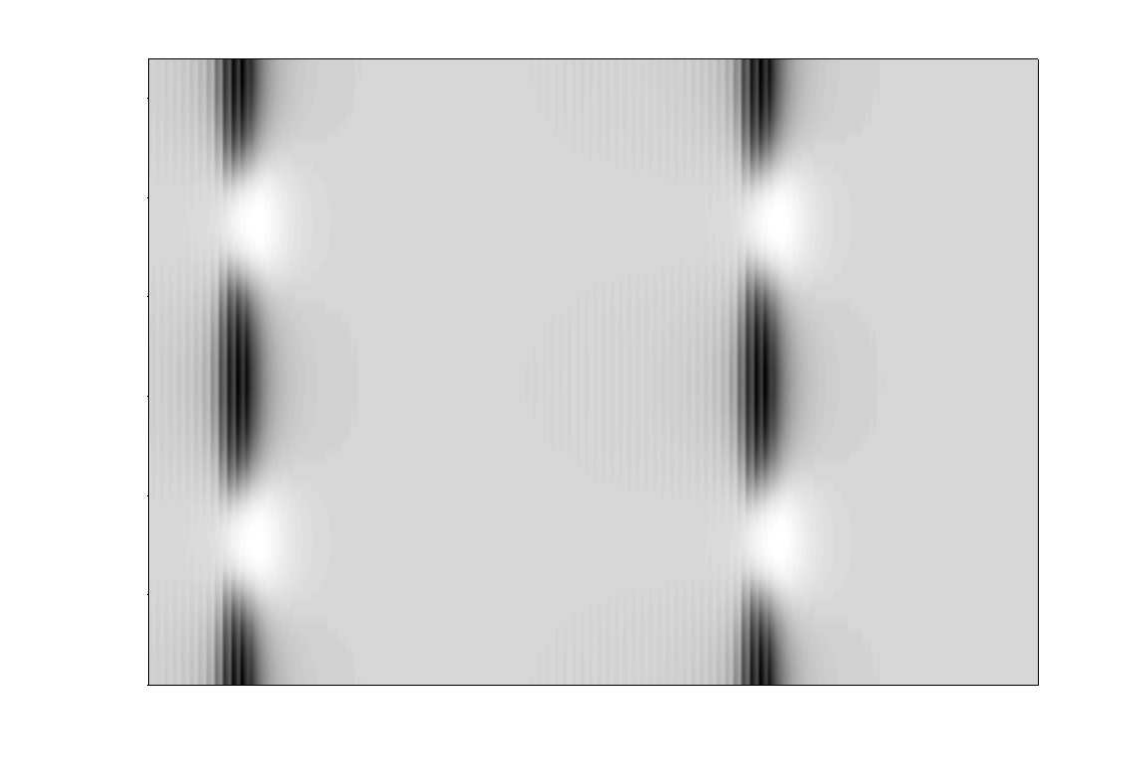} & D &
\includegraphics[scale=0.25]{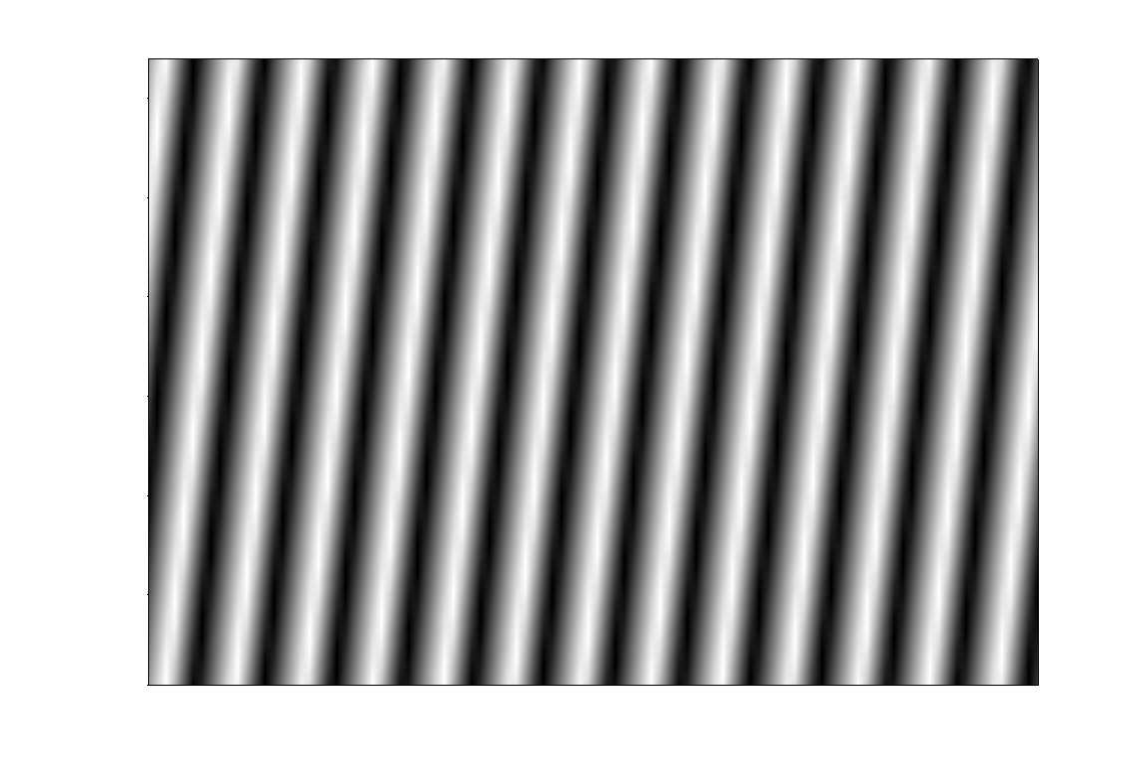}\vspace{-0.05in} \\ &
\textbf{time} & & \textbf{time} \\
\end{tabular}
\caption{Examples of wave solutions from numerical simulation of
Eq.(\ref{eq:rate}).   The functions $\Phi$ and $J(x)$ as well as the
input current $I$  and the delay $D$ are taken as in
Fig.~\ref{fig:diagram}, with $J_{1}=-120$.  A. Supercritical standing
waves:  $J_{0}=-40$ and 5 units of time are shown.   B. Supercritical
standing waves: $J_{0}=-9$ and 5 units of time are shown.
C. Subcritical standing  waves: $J_{0}=-5$ and 40 units of time are
shown.   D. Subcritical traveling waves: $J_{0}=0$ and 5 units of time
are shown.} \label{fig:spacetime_SW}
\end{figure}

The phase diagram for $J_{2}=0$, Fig.~\ref{fig:diagram}, clearly shows
the  dominance of the SW solution, indicating $\Phi^{'''}>0$ for the
parameter values  chosen.  Specifically, for values of $J_{0}<-6.3$, 
supercritical standing waves are stable (see region 5 in Fig.~\ref{fig:a_vs_c}).  
Figures~\ref{fig:spacetime_SW}~A and \ref{fig:spacetime_SW}~B show supercritical 
SW patterns for $J_{0}=-40$ and $J_{0}=-9$, respectively.  For $-6.3<J_{0}<-2.6$, 
TW are supercritical and unstable while SW are subcritical [see region 4 in 
Fig.~\ref{fig:a_vs_c}].  An example of subcritical SW is shown in 
Fig.~\ref{fig:spacetime_SW}~C.  For $-2.6<J_{0}<3.58$, both SW and TW are 
subcritical (see region 3 in Fig.~\ref{fig:a_vs_c}).  
Numerical simulations reveal TW patterns in this region (see an example in 
Fig.~\ref{fig:spacetime_SW}~D).  In the region where SW are subcritical there is a small 
sliver in $(J_{0},J_{1})$ where the SW state is bistable with a TW state (TW/SW in 
the phase diagram).  This TW branch most likely arises in a secondary bifurcation 
slightly below the Turing-Hopf bifurcation line.  There is also a small region of 
bistability between large amplitude TW and the spatially uniform high activity state 
(TW/HA in the phase diagram Fig.~\ref{fig:diagram}).

\begin{figure}
\center
\includegraphics[scale=0.35]{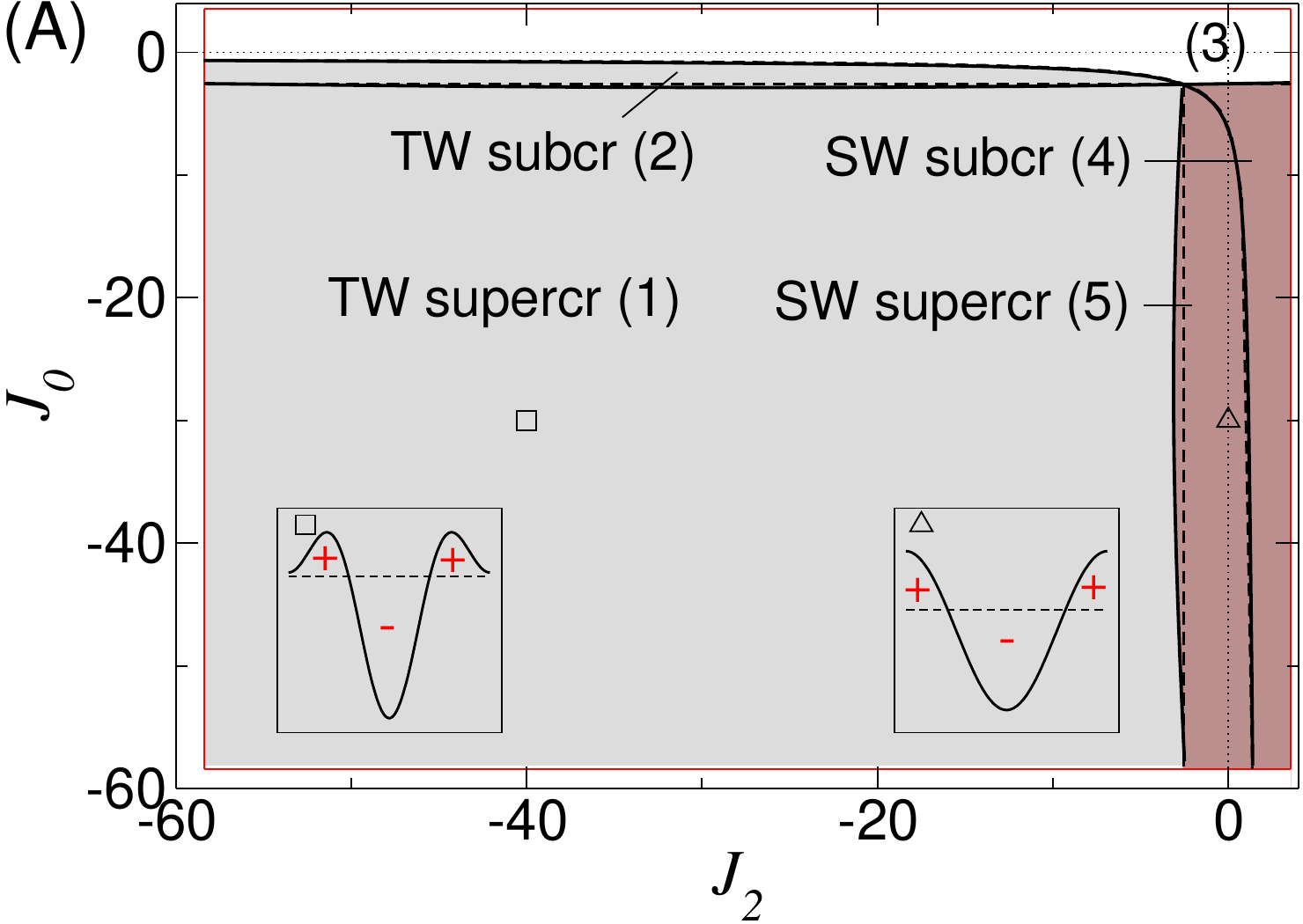}
\includegraphics[scale=0.35]{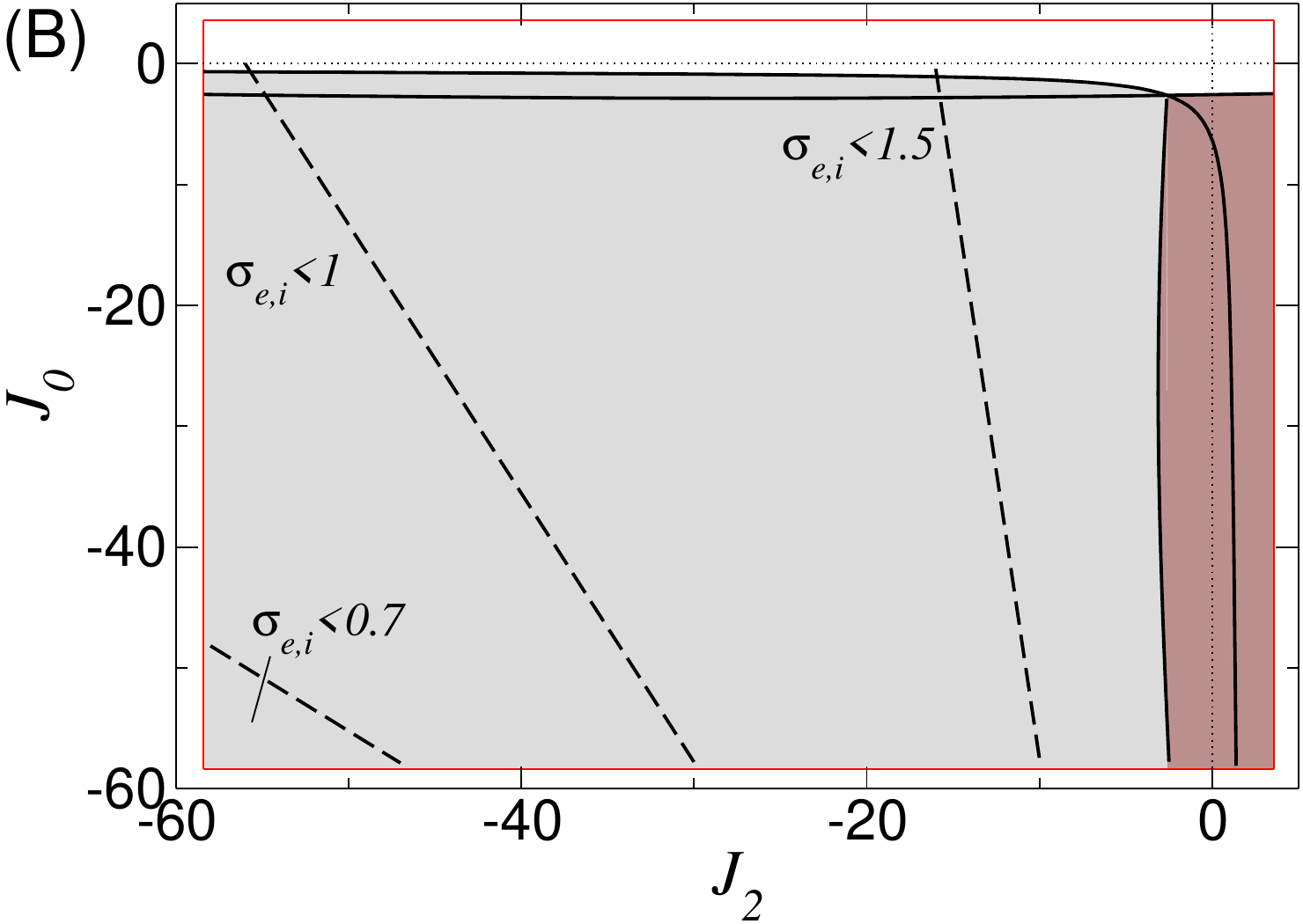}
\caption{(Color online) A. Phase diagram for waves as a function of
  the zeroth and second spatial Fourier  coefficients of the
  connectivity kernel.  The dark-shaded region indicates SW, whereas the light shaded
  region indicates TW.
  Red lines indicate boundaries for primary instabilities of $J_{0}$ and $J_{2}$
  given by  Eqs.~(\ref{eq:disp0}-\ref{eq:disp3}).  Solid stability lines for waves are
  from  Eqs.~(\ref{eq:TH_cc1}-\ref{eq:TH_ccc1}) while the dashed line are
  from the asymptotic expressions (\ref{eq:TH_cc2}-\ref{eq:TH_ccc2}).   
Here $\bar J_{1}=-58.4$.  The function
  $\Phi$ as well as the input current $I$ and the delay $D$ are taken
  as in Fig.~\ref{fig:diagram}.  Insets: example
  connectivity patterns corresponding to the values of $J_{0}$ and
  $J_{2}$  marked by the square and triangle respectively.  B. The
  same phase diagram  as in A, now showing where various types of
  'difference-of-Gaussian' connectivities, Eq.~(\ref{eq:diff_gauss}),
  would lie.  Each dotted  line indicates the border of a region in
  which the standard  deviations of the excitatory and inhibitory
  connectivities are below a certain threshold  (0.7, 1.0 and 1.5,
  respectively).  Relatively narrow connectivities  compared
  to the system size will always generate TW solutions. See text for
  details.}
\label{diagram_tw_sw}
\end{figure}

Thus for $\Phi^{'''}>0$ and with a simple cosine connectivity, SW 
arise for most values of $J_{0}$.
However, adding a non-zero $J_{2}$ can lead to the
TW solution winning out.  The phase diagram of wave states
as a function of  $J_{0}$ and $J_{2}$ is shown in
Fig.~\ref{diagram_tw_sw}~A.  In Fig.~\ref{diagram_tw_sw}~A, the light shaded 
region indicates values of $J_{0}$ and $J_{2}$ for which TW are expected, whereas 
SW are expected in the dark shaded region.  In the unshaded region, both TW and 
SW are subcritical and the solution type is therefore not determined by the 
analysis up to cubic order.  These regions, delimited by the solid lines, 
 were determined by numerically 
evaluating the real parts of the full expressions for the 
cubic and cross-coupling coefficients, Eqs.~(\ref{eq:TH_cc1}-\ref{eq:TH_ccc1}).  
Each region is furthermore numbered according to the existence and 
stability of the TW and SW solution branches as shown in Fig.~\ref{fig:a_vs_c}.  
The dashed lines show 
the approximation to the solid lines given by the asymptotic formulas 
(\ref{eq:TH_cc2}-\ref{eq:TH_ccc2}).  The set of allowable values for 
$J_{0}$ and $J_{2}$ is bounded by the conditions (\ref{eq:disp0}-\ref{eq:disp3}) 
corresponding to steady or oscillatory linear 
instabilities. These stability conditions are shown by the 
horizontal and vertical bounding lines (red online).  All parameter values are as 
in Fig.~\ref{fig:diagram}.

From Fig.~\ref{diagram_tw_sw} we can now understand the discrepancy between the 
analytical results in \cite{roxin05} using a rate equation with a linear threshold 
transfer function, which predicted TW, and network simulations, which showed SW.  
Specifically, given a nonlinear transfer function with $\Phi^{'''}>0$, then 
with a simple cosine coupling SW are predicted over almost the entire range of 
allowable $J_{0}$ (dark shaded region for $J_{2}=0$).  The nonlinear transformation of 
inputs into outputs is thus crucial in determining the type of wave solution.  The 
choice of a threshold linear transfer function results in the second and all 
higher order derivatives being zero.  In this sense it produces degenerate behavior at 
a bifurcation point, and by continuation of the solution branches, in a finite 
region of the phase diagram.  

\subsubsection{'Difference-of-Gaussian' connectivities}

We have shown that varying $J_{0}$ can change the nature of the bifurcation,
e.g. supercritical  to subcritical, while varying $J_{2}$ can switch
the solution type, e.g. from SW to TW.  As an example of a functional
form of connectivity motivated by anatomical findings,
e.g. \cite{hellwig00}, we consider a difference of Gaussians, written
as
\begin{equation}
J(x)=\frac{J_{e}}{\sqrt{2\pi}\sigma_{e}}e^{-\frac{x^{2}}{2\sigma_{e}^{2}}}
-\frac{J_{i}}{\sqrt{2\pi}\sigma_{i}}e^{-\frac{x^{2}}{2\sigma_{i}^{2}}}.
\label{eq:diff_gauss}
\end{equation}
In this case, one finds that the Fourier coefficients are
\begin{equation}
J_{k}=J_{e} e^{-k^{2}\sigma_{e}^{2}/2} f(k,\sigma_e) -J_{i}
e^{-k^{2}\sigma_{i}^{2}/2} f(k,\sigma_i). 
\label{eq:fourrier_gauss}
\end{equation}
where $f(k,\sigma_{e,i})=\text{Re}[\mbox{Erf}((\pi/\sigma_{e,i}+i
  k^2)/\sqrt{2})]/\pi$. Once $J_{1}$ has been fixed at the critical
value for the onset of waves, from  Eq.(\ref{eq:fourrier_gauss}) it is
straightforward to show that $J_{0}=-pJ_{2}+q$ where  both $p$ and $q$
are constants which depend on $\sigma_{e}$ and $\sigma_{i}$, the widths of the excitatory
and  inhibitory axonal projections respectively. 
Thus a difference-of-Gaussian connectivity,
constrains the possible values of $J_{0}$ and $J_{2}$ to lie along a
straight line for fixed connectivity widths. 
This is illustrated in Fig.~\ref{diagram_tw_sw}B where three dashed
lines are superimposed on the  phase diagram, corresponding to 
the values $\sigma_{e,i}=(1.5,1.49)$; $\sigma_{e,i}=(1,0.99)$; and 
$\sigma_{e,i}=(0.7,0.69)$. Each of these lines is bounding a
region to the left where $\sigma_{e}$ and $\sigma_{i}$ are less than
$0.7$, $1.0$ and $1.5$ respectively.  Given periodic boundary
conditions with a system size of $2\pi$,  a Gaussian with $\sigma =
1.5$ is already significantly larger than zero for $x=\pi$ or $-\pi$.
Thus,  restricting ourselves to Gaussians which essentially decay to
zero at the boundaries means that TW will  always be observed.  The
same holds true for qualitatively similar types of connectivity.

\subsubsection{Classes of waves in Network Simulations}

Our analytical results concerning waves from the rate equation Eq.(\ref{eq:rate}) 
predict that a connectivity with a sufficiently
strong second Fourier component with a negative amplitude will lead to
traveling waves (see the phase diagram in Fig.~\ref{diagram_tw_sw}(A)). 

Here we have confirmed this prediction performing numerical simulations 
of the network of spiking neurons described in the Appendix A.  Indeed, 
Fig.~\ref{fig:rasters} shows that
the addition of the second spatial Fourier
component to the inhibitory connections converts standing waves (SW) 
into travelling waves (TW).  

\begin{figure}
\center
\includegraphics[scale=0.24]{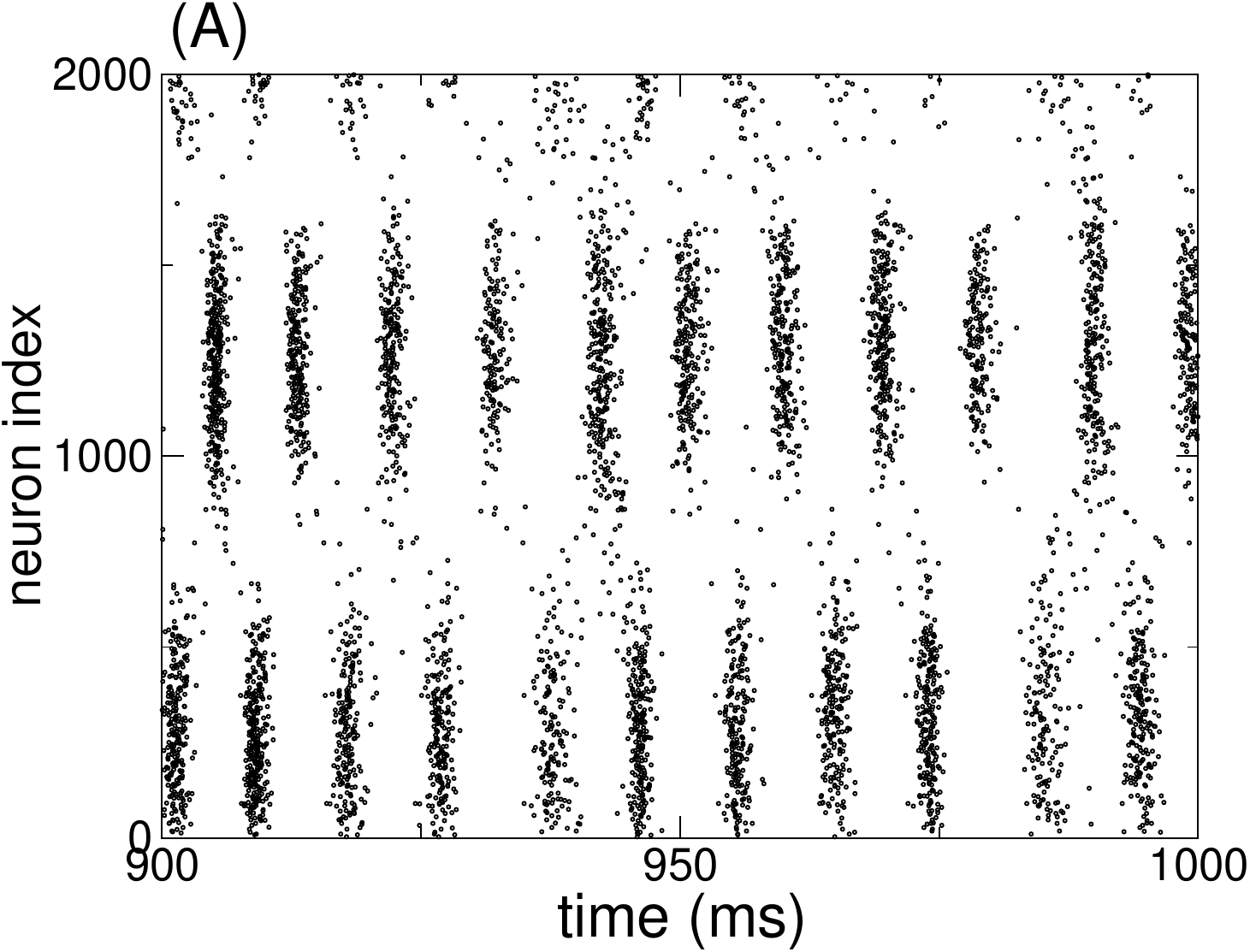}
\includegraphics[scale=0.24]{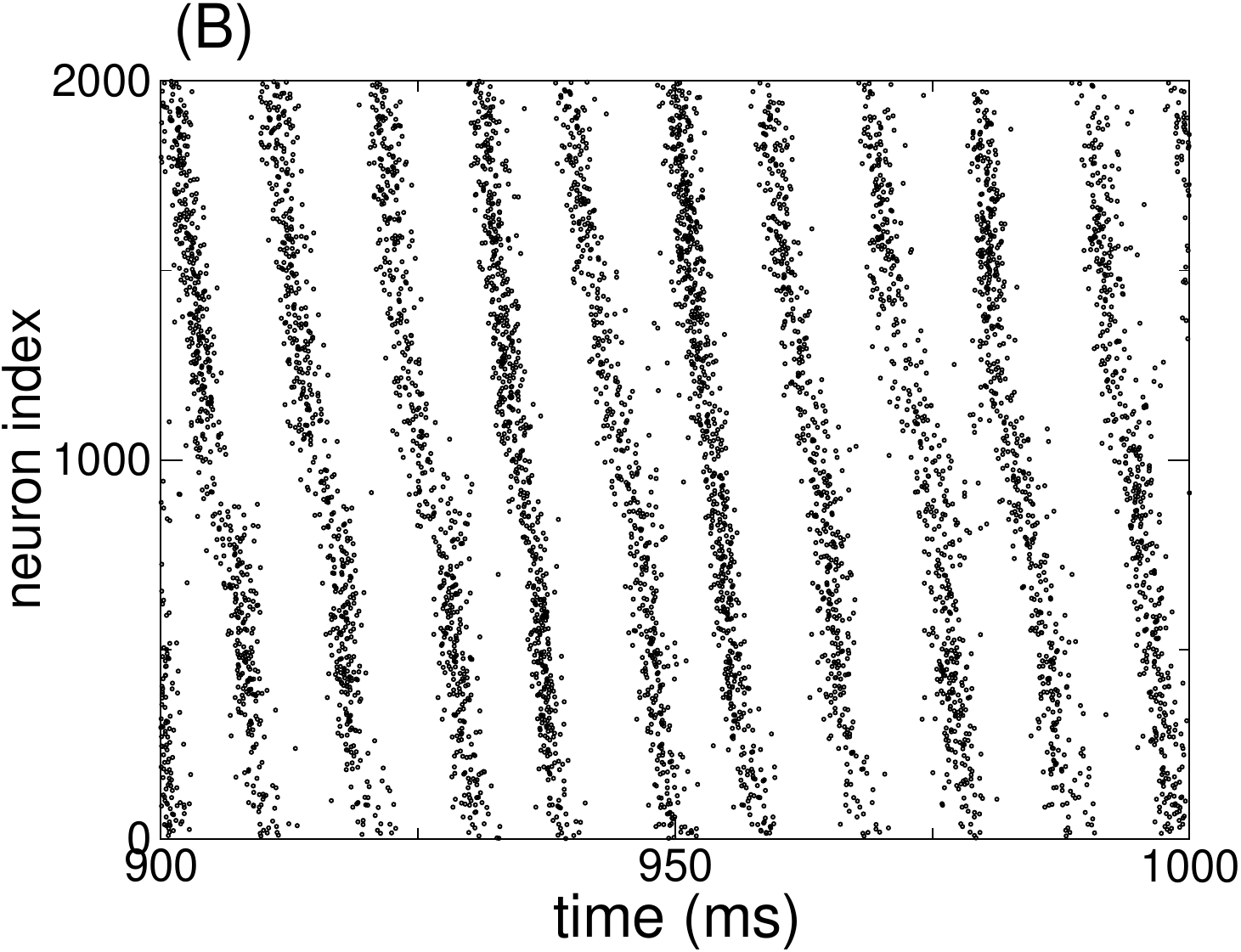}
\caption{The transition from (A) standing to (B) travelling waves 
in a network of conductance-based neurons takes place by 
increasing the second Fourier mode of the synaptic connectivity 
$p_2^{I}$.  
(A) $p_2^{I}=0$, (B) $p_2^{I}=0.1$. Remaining parameters: 
$p_0^{E,I}=p_1^{I}=0.2$, $p_{1,2}^E=0$, $g_E=0.01$, $g_I=0.028$, 
$g_{ext}=0.001$mS~ms/cm$^2$, and $\nu_{ext}=5000$Hz. }
\label{fig:rasters}
\end{figure}

\section{\label{sect:codim2}Bifurcations of codimension 2}

For certain connectivity kernels we may be in the vicinity of two
distinct instabilities.  This is the case for  certain Mexican hat
connectivities (OU and SB) and certain inverted Mexican hat
connectivities (OU and SW/TW).   Although two instabilities will
co-occur only at a single point in the phase diagram
Fig.~\ref{fig:diagram}, i.e. $J_{0}$ and  $J_{1}$ are both at their
critical values, the competition between these instabilities may lead
to solutions which  persist over a broad range of connectivities.
This is the case here.  We can investigate this competition once again
using a  weakly nonlinear approach.  The main results of this section are 
summarized in table \ref{table1}.

\subsection{\label{subsect:H-TH}Hopf and Turing-Hopf bifurcations}

Here we consider the co-occurrence a spatially homogeneous oscillation
and a spatially inhomogeneous oscillation (OU and SW/TW), both with frequency
$\omega$ given  by Eq.(\ref{eq:disp2}).  This instability occurs when
the zeroth and $k^{th}$ spatial Fourier mode of  the connectivity both
satisfy the relation, Eq.(\ref{eq:disp3}), while  we assume that all
other Fourier modes are sufficiently below their critical values to
avoid additional instabilities.  Without loss of generality we take $k=1$ for the 
SW/TW state.
  
We expand the parameters $J_0$, $J_1$, $I$ and $r$ as in
Eqs.~(\ref{eq:Jkexp},\ref{eq:Iexp},\ref{eq:rexp}), and define the slow
time (\ref{eq:T}).  The linear solution consists of homogeneous,
global oscillations, leftwards and rightwards traveling waves with
amplitudes which  we allow to vary slowly in time, i.e. $r_{1} =
H(T)e^{i\omega t}+A(T)e^{i\omega t+ix}+B(T)e^{-i\omega t+ix}+c.c.$.
Carrying out a weakly nonlinear analysis to third order in $\epsilon$
leads to the coupled amplitude equations
\begin{subequations}
\begin{eqnarray}
\partial_{T}H &=& (\mu+i\Omega)\Delta J_{0}H  \label{eq:ampeq_HTHH} \\ &&
+2(\alpha+i\beta)[(\frac{|H|^{2}}{2}+|A|^{2}+|B|^{2})H  
+H^* A B^*], \nonumber
\end{eqnarray}
\begin{eqnarray} 
\partial_{T}A &=&
(\mu+i\Omega)\Delta J_{1}A+(a+ib)|A|^{2}A+(c+id)|B|^{2}A \nonumber \\ &&
+(\alpha+i\beta ) [2 |H|^{2}A +H^{2}B],
\label{eq:ampeq_HTHA}
\end{eqnarray}
\begin{eqnarray}
\partial_{T}B &=& (\mu-i\Omega)\Delta
J_{1}B+(a-ib)|B|^{2}B+(c-id)|A|^{2}B  \nonumber \\ &&
+(\alpha -i\beta) [2 |H|^{2}B +H^{*2}A],
\label{eq:ampeq_HTHB}
\end{eqnarray}
\end{subequations}
where $\alpha +i\beta$, $a+ib$ and $c+id$ are given by
Eqs.~(\ref{eq:hopf_cc},\ref{eq:TH_cc1},\ref{eq:TH_ccc1}), respectively.  
The overbar in $H^*$ represents the complex conjugate.

\subsubsection{Solution types and their stability}

Eqs.~(\ref{eq:ampeq_HTHH}-\ref{eq:ampeq_HTHB}) admit several types of
steady state solutions including oscillatory  uniform solutions (OU),
traveling waves (TW), standing  waves (SW) and mixed mode
oscillations/standing waves (OU-SW).   The stability of these
solutions depends on the  values of the coefficients in
Eqs.~(\ref{eq:ampeq_HTHH}-\ref{eq:ampeq_HTHB}).  In addition,
non-stationary solutions are  also possible.  Here we describe briefly 
some stationary solutions and their stability.  For details see \ref{app:ampeq}.\\

\noindent\textit{Oscillatory Uniform (OU)}: The oscillatory uniform 
solution has the form $(H,A,B)= (\mathcal{H}e^{i\omega t},0,0)$ where
\begin{eqnarray}
\mathcal{H}&=&\sqrt{\frac{-\mu\Delta J_{0}}{\alpha}},\nonumber \\
\omega &=&\Big(\Omega-\frac{\beta\mu}{\alpha}\Big)\Delta J_{0}.
\nonumber
\end{eqnarray}

The OU state undergoes a steady instability along the line
\begin{equation}
\Delta J_{1} = \Delta J_{0}.
\label{eq:OU_steady_inst}
\end{equation}
This stability line agrees very well with the results of
numerical simulations of Eq.(\ref{eq:rate}) [see the phase diagram
Fig.~\ref{fig:diagram}].
\vspace{0.1in}

\noindent\textit{Traveling Waves (TW)}: The traveling wave solution has 
the form $(H,A,B)$ = $(0,\mathcal{A}_{TW}e^{i\omega t},0)$ or 
$(0,0,\mathcal{A}_{TW}e^{-i\omega t})$, where
\begin{eqnarray}
\mathcal{A}_{TW}&=&\sqrt{\frac{-\mu\Delta J_{1}}{a}},\nonumber \\
\omega &=&\Big(\Omega-\frac{b\mu}{a}\Big)\Delta J_{1}.
\nonumber
\end{eqnarray}

The TW state undergoes an oscillatory instability along the line
\begin{equation}
\Delta J_{1} = \frac{a}{2\alpha }\Delta J_{0},
\label{eq:TW_inst}
\end{equation}
with a frequency
\begin{equation}
\bar{\omega} = \Big(\Omega (1-\frac{a}{2\alpha })+(b-2\beta )\frac{\mu }{2\alpha }\Big)\Delta J_{0}. \nonumber
\end{equation}

\noindent\textit{Standing Waves (SW)}:
The standing wave solution has the form $(H,A,B)$ =
$(0,\mathcal{A}_{SW}e^{i\omega t},\mathcal{A}_{SW}e^{-i\omega t})$,
where 
\begin{eqnarray}
\mathcal{A}_{SW}&=&\sqrt{\frac{-\mu\Delta J_{1}}{(a+c)}},\label{eq:HTH_SW_amp}\\
\omega  &=& \Big(\Omega-\frac{(b+d)}{(a+c)}\mu \Big)\Delta J_{1}. \label{eq:HTH_SW_freq}
\end{eqnarray}

An oscillatory instability occurs along the line 
\begin{equation}
\Delta J_{1} = \frac{(a+c)}{4\alpha} \Delta J_{0}, 
\label{eq:HTH_SW_osc}
\end{equation}
with a frequency 
\begin{equation}
\bar{\omega} = \sqrt{ [\Omega
(\frac{a+c}{4\alpha}-1)-\mu\frac{b+d-4\beta}{4\alpha}
]^{2}-\mu^{2}\frac{\alpha^{2}+\beta^{2}}{4\alpha^{2}}}\Delta J_{0}.
\nonumber
\end{equation}   
A stationary instability  occurs along the line 
\begin{equation}
\Delta J_{1} = \Psi \Delta J_{0}, \label{eq:HTH_SW_stead}
\end{equation}
where 
\begin{subequations}
\begin{equation}
\Psi = \frac{-k_{2}+\sqrt{k_{2}^{2}-4 k_{1}k_{3}}}{2k_{1}},\\
\nonumber
\end{equation}
\begin{equation}
k_{1} = \Big[
\Omega-\mu\frac{(b+d-4\beta)}{(a+c)}\Big]^{2}+\mu^{2}\frac{(12\alpha^{2}-4\beta^{2})}{(a+c)^{2}},
\nonumber
\end{equation}
\begin{equation} 
k_{2} =
-8\mu^{2}\frac{\alpha}{(a+c)}-2\Omega^{2}+2\Omega\mu\frac{(b+d-4\beta
)}{(a+c)}, 
\nonumber
\end{equation}
\begin{equation}
k_{3} = \mu^{2}+\Omega^{2}.
\nonumber
\end{equation}
\end{subequations}
For Eq.~(\ref{eq:rate}) with the parameters used to generate the
phase diagram Fig.~\ref{fig:diagram}, we find that the stationary
instability precedes the oscillatory one and that $\Psi \sim
0.6$. This  agrees well with the numerically determined stability
line near the co-dimension 2 point in the diagram~\ref{fig:diagram}.
\vspace{0.1in}

\textit{Mixed Mode}:
We can study the mixed mode solutions in
Eqs.~(\ref{eq:ampeq_HTHH}-\ref{eq:ampeq_HTHB}) by assuming  an ansatz
\begin{equation}
(H,A,B) =
(\mathcal{H}e^{i\theta},\mathcal{A}e^{i\psi_{A}},\mathcal{B}e^{-i\psi_{B}}), \label{eq:HTH_mmansatz}
\end{equation}
which leads to four coupled equations [see (\ref{app:ampeq})].  We 
do not study the stability of mixed mode solutions in this work.

\subsubsection{A simple example}

We now turn to a simple example in order to illustrate the two main
types bifurcation scenarios that  can arise when small amplitude waves
and oscillations interact in harmonic resonance.\\

\begin{figure}
\center
\includegraphics[scale=0.27]{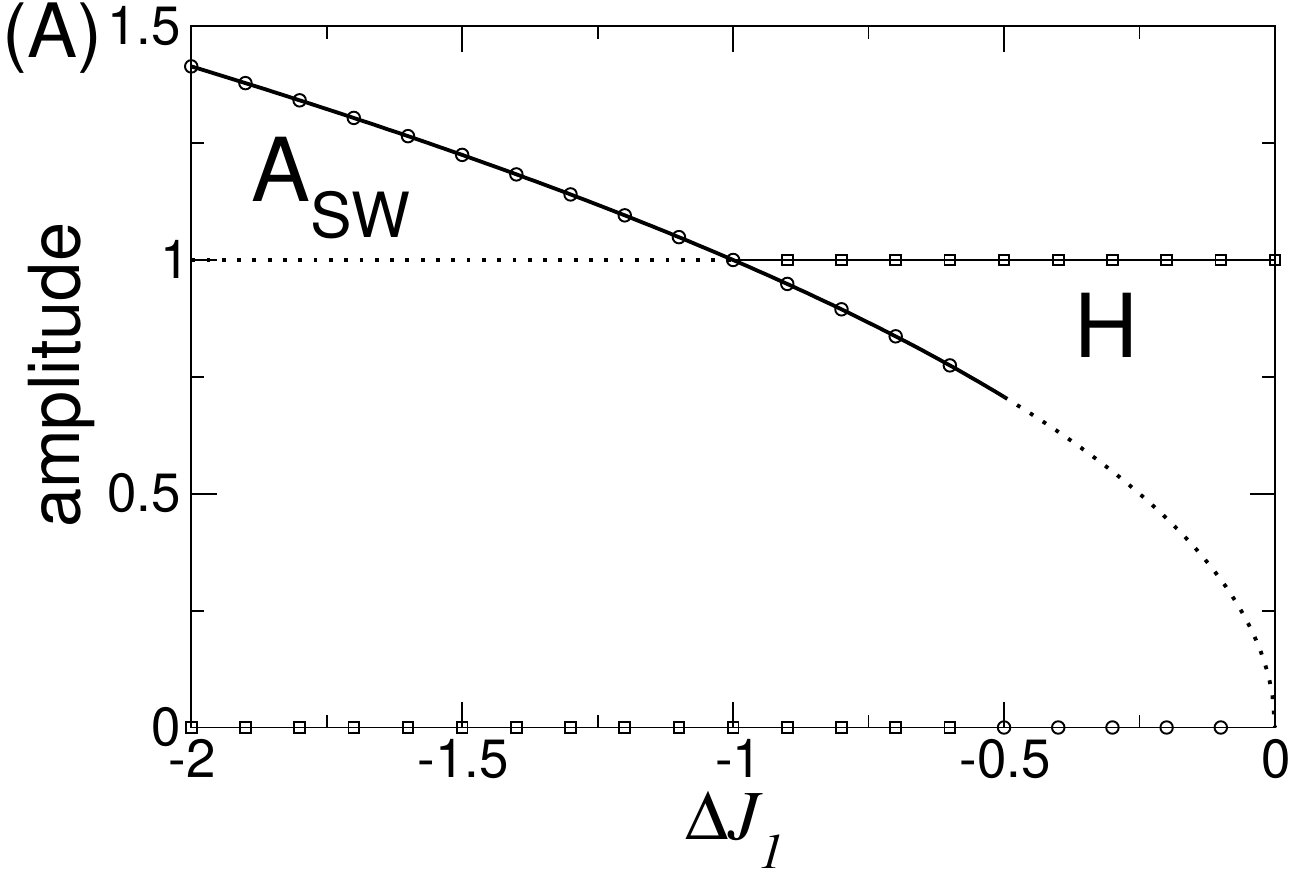}
\includegraphics[scale=0.27]{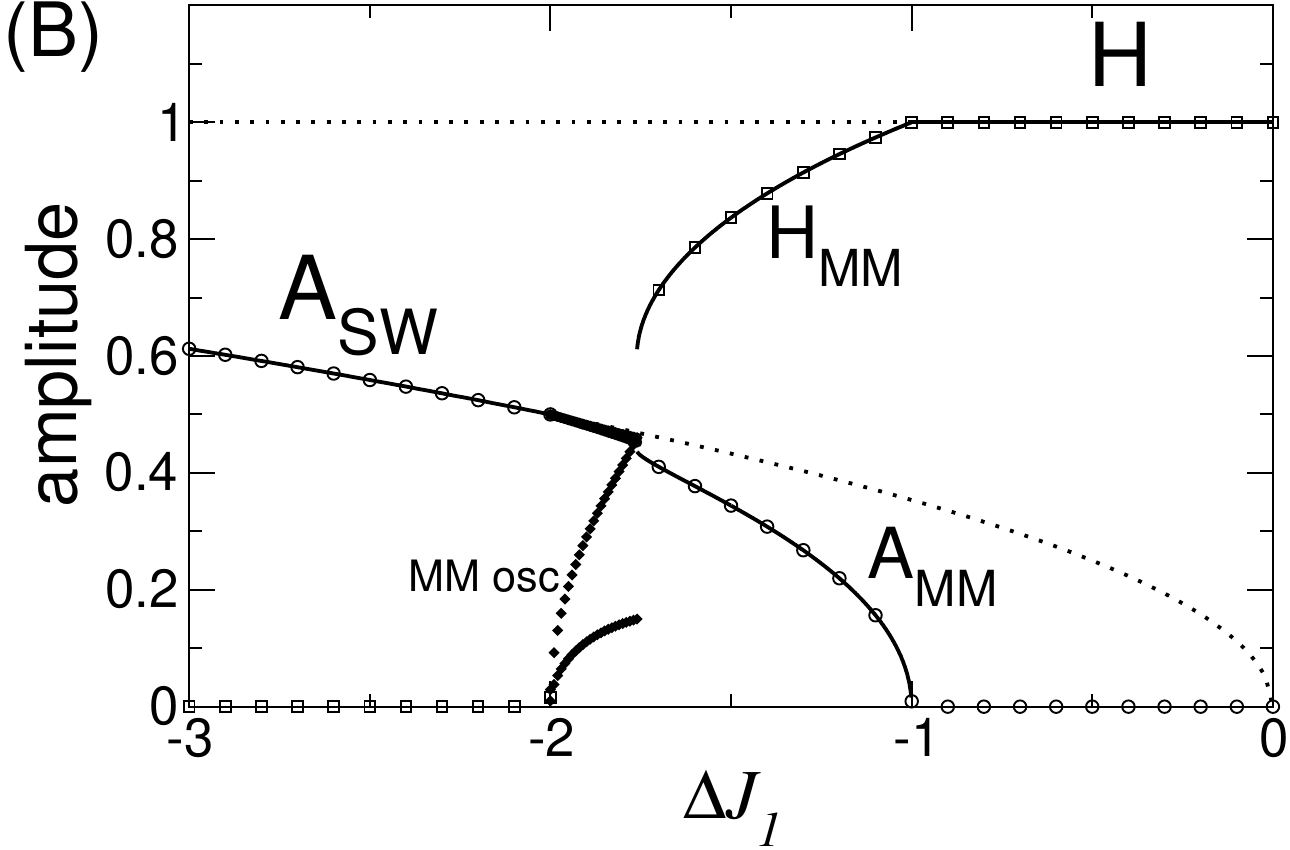}
\caption{Two typical bifurcation diagrams for the case of harmonic
resonance between small-amplitude oscillations  and small-amplitude
standing waves.  A: Here oscillations and standing waves are bistable
for $-1<\Delta J_{1}<-1/2$.   $\Delta J_{0}=-1$, $\alpha =-1$, $a=-1$,
$b=c=d=\beta=\Omega=0$, $\mu=-1$. B: Here the standing wave solution
loses stability to an oscillatory mixed-mode solution at $\Delta
J_{1}=-2$.  At $\Delta J_{1}\sim -1.75$  a steady mixed-mode solution
arises in a saddle-node bifurcation and continuously approaches the
oscillatory pure-mode solution at $\Delta J_{1}=-1$.  Parameters are
$a=-8$, $\beta=1$, $\alpha=-1$, $\mu=-1$,  $b=c=d=\Omega=0$.  The
phase $\phi$ of the mixed-mode solution is not shown.}
\label{fig:exampleHTH}
\end{figure}

\textit{i. Bistability:} Here we take the parameters 
\footnote{$\mu=-1$, 
$\Delta J_{0}=-1$,
$\alpha=-1$, $a=-1$, $b=c=d=\beta = \omega = \Omega = 0$}.   Given
these parameter values one finds, from the analysis above, that the
oscillatory uniform state  has an amplitude $\mathcal{H} = 1$ and
destabilizes along the line $\Delta J_{1} = -1$.  The standing  wave
solution (traveling waves are unstable [see Fig.~\ref{fig:a_vs_c}] has
an amplitude $\mathcal{A}_{SW}  = \sqrt{-\Delta J_{1}}$ which
undergoes a steady bifurcation to the oscillatory uniform state at
$\Delta J_{1} = -1/2$.  Both solutions are therefore stable in the
region $-1<\Delta J_{1}<-1/2$.  This analysis is borne  out by
numerical simulation of Eqs.~(\ref{eq:ampeq_HTHH}-\ref{eq:ampeq_HTHB})
[see Fig.~\ref{fig:exampleHTH}a].   Solid and dotted lines are the
analytical expressions for the stable and unstable solution branches
respectively  (red is OU and black is SW).  Circles are from numerical
simulation of the amplitude equations~(\ref{eq:ampeq_HTHH}-\ref{eq:ampeq_HTHB}).  

Note that this scenario is the relevant one for the phase diagram shown in
Fig.~\ref{fig:diagram}.  That is, we find there is a region of
bistability between the OU and SW solutions, bounded between two
lines with slope $\sim 0.6$ and $1$ respectively.  \\

\textit{ii. Mixed Mode:}  Here we consider the parameters 
\footnote{
$\mu=-1$, $\Delta J_{0}-1$, $\alpha=-1$, $a=-8$,  $\beta=1$,
$b=c=d=\omega = \Omega = 0$}.  Given these parameter values one finds
that the oscillatory  uniform state has an amplitude $\mathcal{H} = 1$
and destabilizes along the line $\Delta J_{1}=-1$.  The  standing
waves solution has an amplitude $\mathcal{A}_{SW}=\sqrt{-\Delta
J_{1}/8}$ (traveling waves are  again unstable) which undergoes an
oscillatory instability at $\Delta J_{1}=-2$.  The mixed-mode solution
is  given by
\begin{subequations}
\begin{eqnarray}
\mathcal{H} &=& \frac{4+\Delta J_{1}
(2-\cos{\phi}-\sin{\phi})}{4-(2-\cos{\phi}-\sin{\phi})^{2}},
\label{eq:MM_H}\\ \mathcal{A}_{SW} &=& \frac{\Delta
J_{1}+(2-\cos{\phi}-\sin{\phi})}{8-2(2-\cos{\phi}-\sin{\phi})^{2}}, 
\end{eqnarray}
\begin{eqnarray}
1 &=& \Delta
J_{1}(1-4\cos{\phi}-2\sin{\phi}+2\sin{\phi}\cos{\phi})
-4\cos{\phi}+8\cos{\phi}^{2}. \label{eq:MM_phase}
\end{eqnarray}
\end{subequations}
It is easy to show that for $\Delta J_{1}\to -1$ the mixed mode
amplitudes approach $(\mathcal{H},\mathcal{A}_{SW})=(1,0)$  and the
phase $\phi\to 0$.  The mixed-mode solution thus bifurcates
continuously from the oscillatory pure mode.
Figure~\ref{fig:exampleHTH}b shows the corresponding bifurcation diagram
where solid and dotted lines are  the analytical expressions for the
solution branches and symbols are from numerical simulation of
Eqs.~(\ref{eq:ampeq_HTHH}-\ref{eq:ampeq_HTHB}).  As $\Delta J_{1}$
increase from the left we see that the SW solution indeed undergoes an
oscillatory instability at $\Delta J_{1}=-2$ leading to an oscillatory
mixed-mode solution indicated by small  circles (the maximum and
minimum amplitude achieved on each cycle is shown).  This oscillatory
solution disappears  in a saddle-node bifurcation, giving rise to a
steady mixed-mode solution whose amplitude is given by
Eq.(\ref{eq:MM_H}- \ref{eq:MM_phase}).  This steady mixed-mode solution
bifurcates from the pure oscillatory mode at $\Delta J_{1}=-1$  as
predicted.

\subsubsection{Summary}

The interaction between the oscillatory uniform state and 
waves may lead to mixed mode solutions or bistability.  The OU state 
always destabilizes along the line $J_{1}=J_{0}$, irrespective of parameter 
values or the choice of $\Phi $ or $J(x)$.  This result from the weakly 
nonlinear analysis, agrees with numerical simulations of Eq.(\ref{eq:rate}) over 
the entire range of values of $J_{0}$ and $J_{1}$ used in the phase diagram, 
Fig.~\ref{fig:diagram} and appears to be exact.  Depending on the value of 
$J_{2}$, supercritical TW or supercritical SW will be stable near the codimension 
2 point.  In the case of TW, the slope of the stability line is one half the ratio of the 
cubic coefficient of waves to that of oscillations.  In the small delay limit 
this expression can be simplified to 
\begin{equation}
\frac{a}{2\alpha }\sim \frac{\pi}{4}\frac{\Big(\frac{(\Phi^{''})^{2}}{\Phi^{'}}
-\frac{\Phi^{'''}}{2}\Big)}{\Big(\frac{(11\pi -4)}{20}\frac{(\Phi^{''})^{2}}{\Phi^{'}}
-\frac{\pi\Phi^{'''}}{4}\Big)},
\end{equation}
which depends only on shape of the transfer function $\Phi$.  For the parameter values 
used in the phase diagram Fig.~\ref{fig:diagram} this yields a line with slope
 close to one half.  
Thus TW and OU are expected to be bistable in the wedge between $\Delta J_{1}=\Delta J_{0}/2$ and 
$\Delta J_{1}=\Delta J_{0}$.  In the case of SW, the slope of the stability line is a 
complicated function of the shape of $\Phi$ and the second Fourier coefficient $J_{2}$.  
For the parameter values used in the phase diagram Fig.~\ref{fig:diagram} the slope is 
close to 0.6.  Therefore the OU and SW states are bistable in the wedge between 
$\Delta J_{1}=0.6\Delta J_{0}$ and 
$\Delta J_{1}=\Delta J_{0}$.

\subsubsection{Network simulations}

Given that network simulations robustly reveal standing wave patterns, we would 
expect to find either mixed-mode SW-OU or bistability between SW and OU.  As we have 
shown previously, e.g. \cite{roxin05}, there is a region of bistability between SW and OU 
in network simulations for strongly modulated inhibitory connectivity.  Here we show 
additional network simulations that suggest this bistable region is in the vicinity of 
the codimension two point, i.e. it is a bistability between the OU and SW states arising 
via primary bifurcations of the unpatterned state.
\begin{figure}
\center
\includegraphics[scale=0.4]{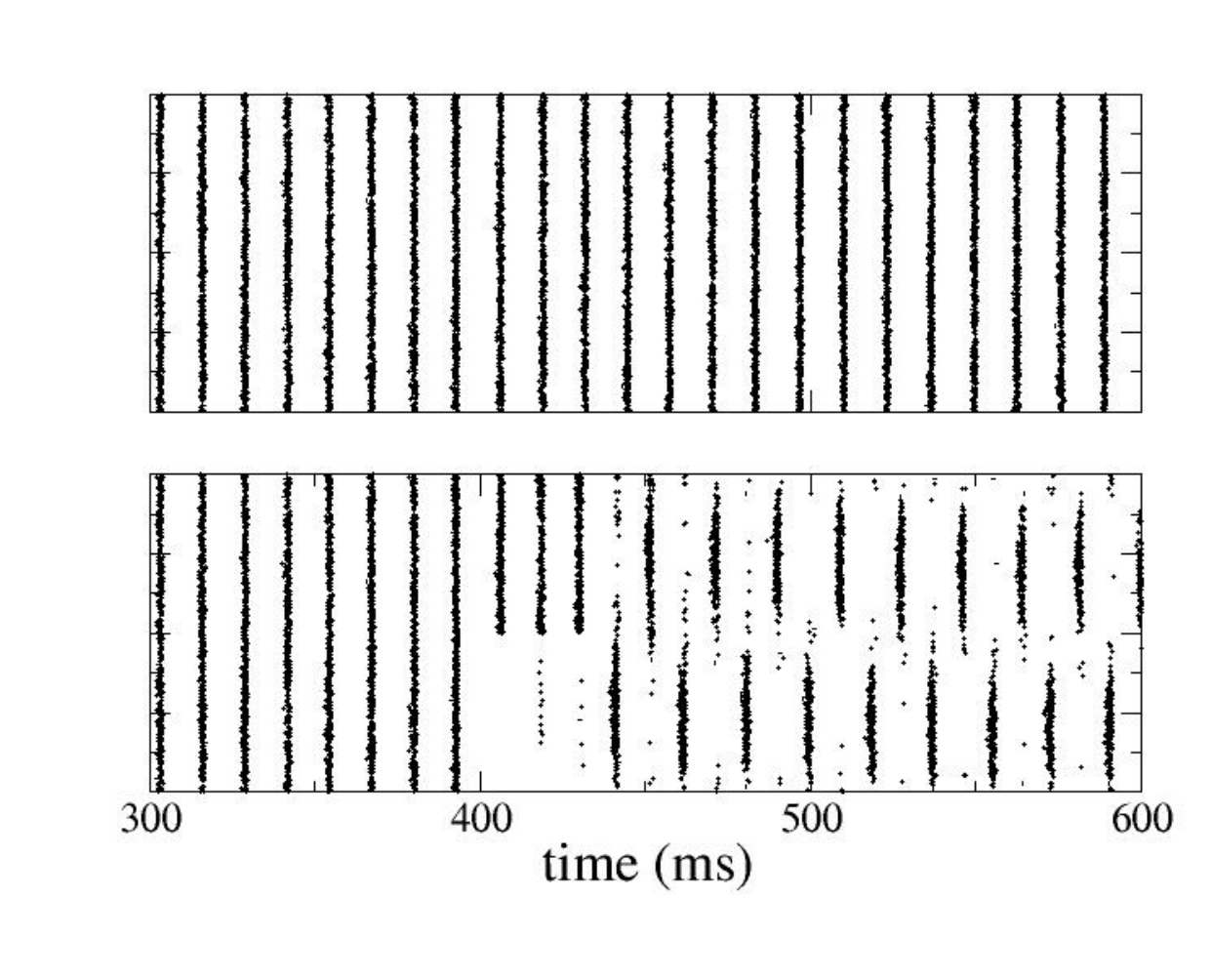}
\caption{Bistability between SW and OU states in an inhibitory network with $p_{I}(r)=0.4+0.2\cos{r}$, where $r$ is the distance between neurons, $\nu_{ext}=4500$Hz, and $g_{I}=0.1$mS$\cdot$ms/cm$^{2}$. This is,
the  only network simulation for which an explicit delay has been added of $\delta = 0.5$ms.  
Removing the 
explicit delay for these parameter values eliminates the bistability. 
Parameter values are identical for both simulations.  
In the simulation shown in the bottom raster a 
hyperpolarizing current of $I_{app}=-5.0 
\mu$A/cm$^{2}$ was injected into cells 1-1000 for 30ms at time $t=400$ms, switching the state from 
OU to SW.}\label{fig:bistab_rasters}
\end{figure}

Figure \ref{fig:bistab_rasters} shows two rasters from simulations of a purely inhibitory network with 
strongly spatially modulated connectivity. 
The top raster shows 300 milliseconds of activity in which 
homogeneous oscillations are clearly observable.  
In the bottom raster, the network is started from the 
precisely the same initial condition, but a hyperpolarizing input current is 
applied to neurons 1 to 1000 from t=400 to t=430ms.  The network activity clearly 
switches to a SW state in response to this input.  The SW state persists for as long as simulations 
were carried out (10sec).  The network thus exhibits bistability between the OU and SW states.  
\begin{figure}
\centerline{\includegraphics[scale=0.35]{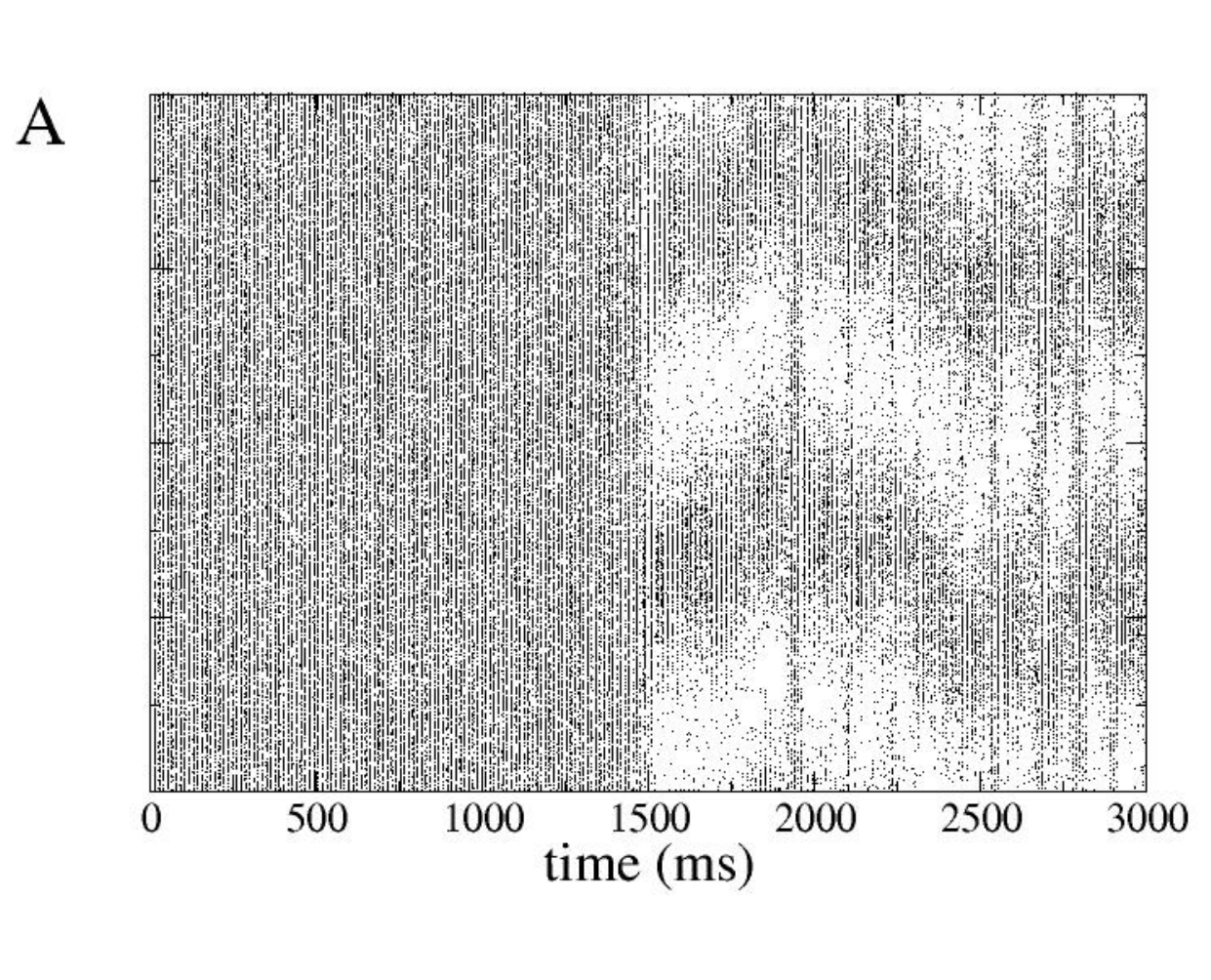}
\includegraphics[scale=0.35]{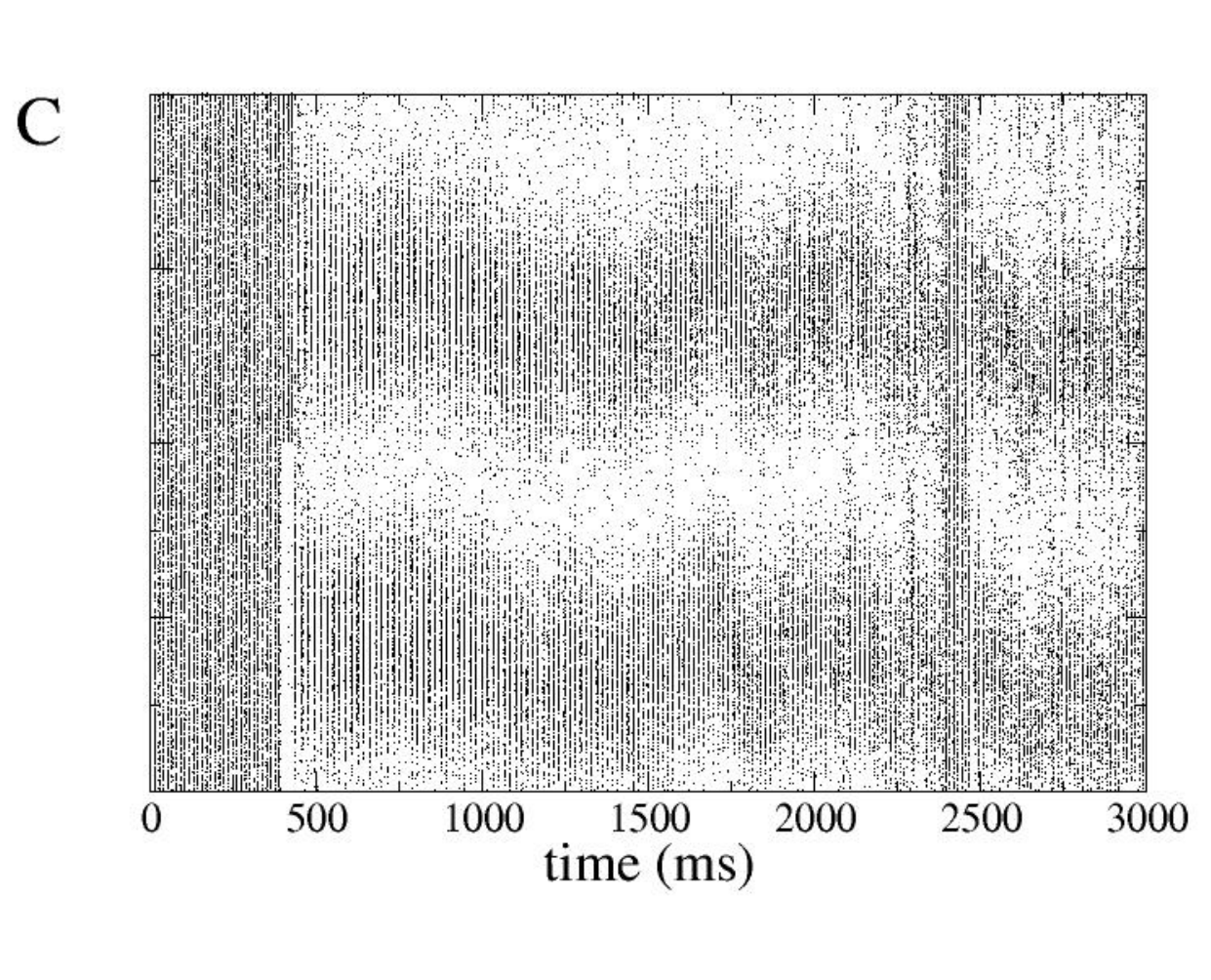}}
\centerline{\includegraphics[scale=0.35]{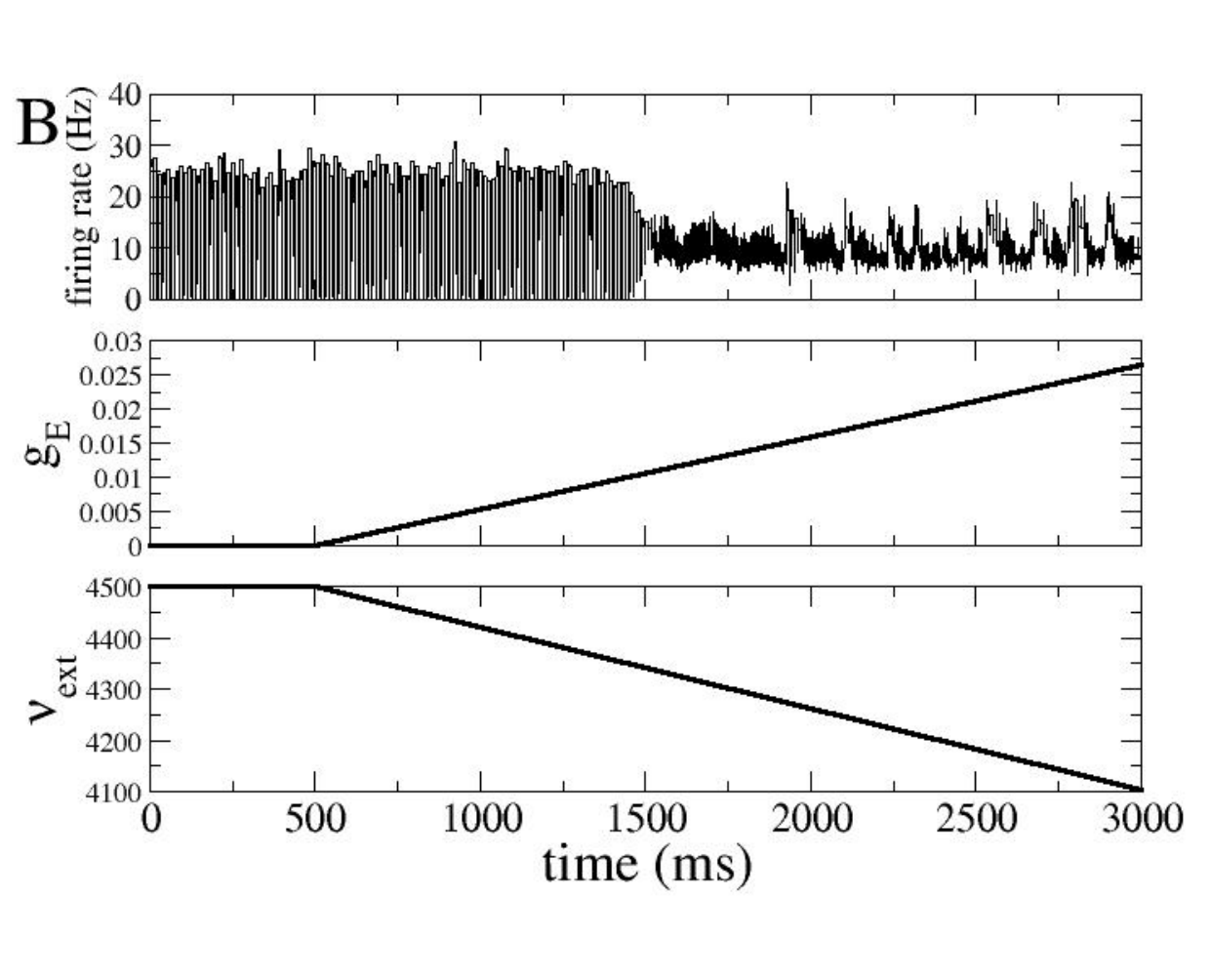}
\includegraphics[scale=0.35]{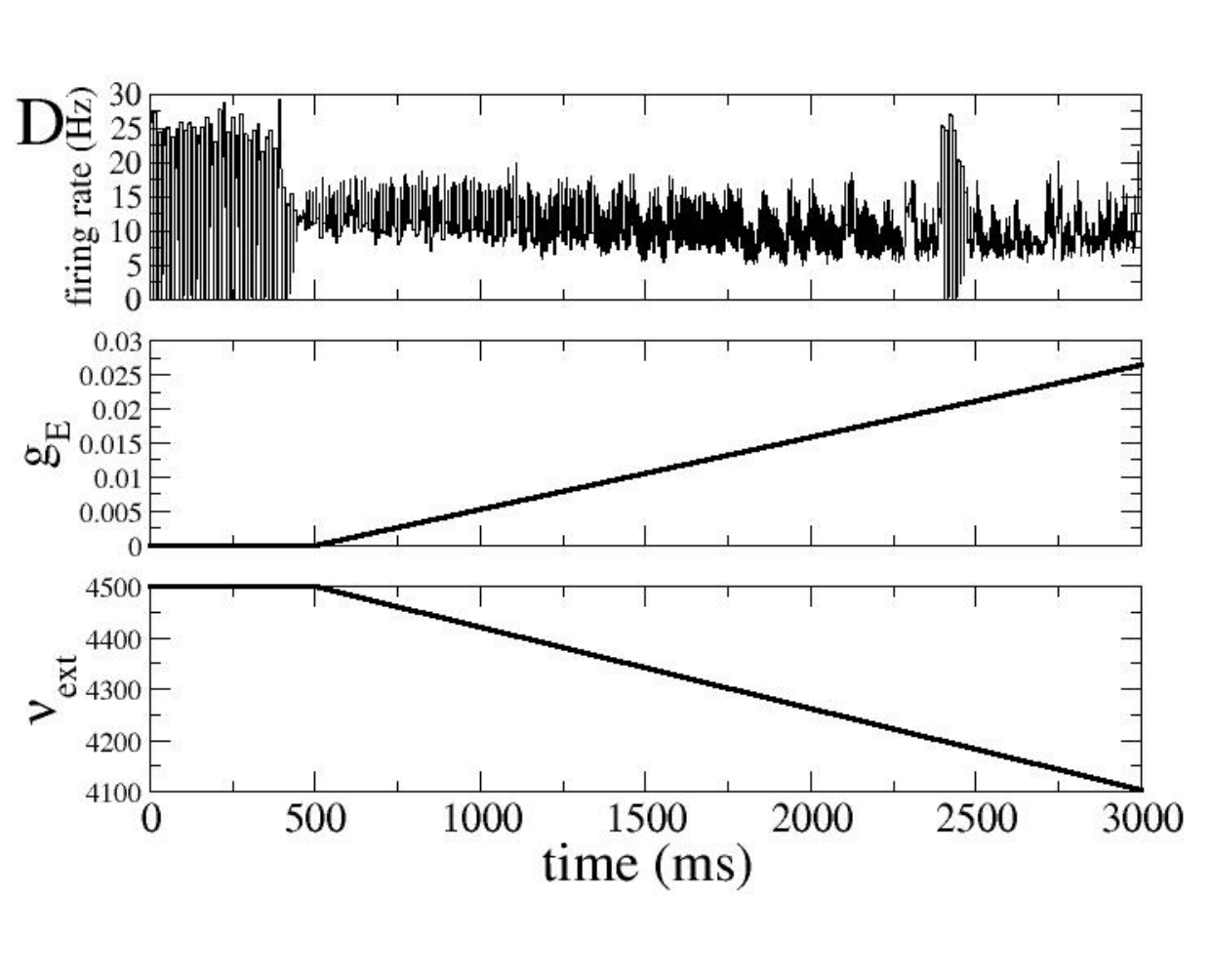}}
\caption{Increasing the strength of recurrent excitatory connections drives the system out of the bistable regime 
and into a SW state.  A: A raster plot of 3 seconds of simulation.  For the first 500ms the simulation is 
identical to that shown in the top panel of Fig.\ref{fig:bistab_rasters}.  Thereafter the weights of the recurrent 
excitatory synaptic are slowly increased while the external drive is slowly decreased so as to maintain stable mean 
firing rates.  There is a spontaneous destabilization of the OU state to a SW state around $t=1500$ms.  B: 
The network firing rate for the simulation shown in A (top), the time course of the excitatory synaptic 
weights (middle) and external drive (bottom).  C: For the first 500ms the simulation is identical to that shown in the 
bottom panel of Fig.\ref{fig:bistab_rasters}, i.e. there is a hyperpolarizing current injected at time $t=400$ms 
which switches the state from OU to SW.  Thereafter the weights of recurrent excitatory synapses are slowly 
increased while the external drive is slowly decreased so as to maintain stable mean firing rates.  D: The network 
firing rate and time courses of the excitatory synaptic weights and external drive, as in B.}\label{fig:bistab_fr}
\end{figure}

In order to determine if this region of bistability is related to the codimension 2 point, we 
adiabatically increased the recurrent excitatory connectivity in the network, thereby mimicking an 
increase in $J_{0}$ in the rate model.  This is done by generating a network of both inhibitory and 
excitatory neurons with $p_{I}(r)=0.4+0.2\cos{r}$ and $p_{E}=0.2$.  The weights of the inhibitory 
synapses are taken as $g_{I}=0.1$ while excitatory weights are allowed to vary.  Specifically, $g_{E}=0$ 
for the first 500ms of the simulation and are then slowly increased according to $g_{E}(t)=1.0526\cdot 10^{-5}(t-500)$.  
At the same time we slowly decrease the external drive in order to maintain mean firing rates.  Thus 
$\nu_{ext}=4500$ for the first 500ms and is then varied according to $\nu_{ext}=4500-0.158(t-500)$ thereafter.  
This particular functional form was determined empirically to keep the mean firing rates steady.  
Therefore, for the first 500ms of simulation time the network is equivalent to that shown in Fig.
\ref{fig:bistab_rasters} while at later times the network is no longer purely inhibitory.

Carrying out such an adiabatic increase in the recurrent excitation should 
cause the network to cross the line of instability 
of the OU state, leading to stable SW.  Thus, if we begin simulations in the OU state, they should 
destabilize at some point to SW while if we begin in the SW state they should persist.  This is 
precisely what occurs.  Fig.\ref{fig:bistab_fr}A shows a raster plot of three seconds of simulation 
time begining in the OU state.  A transition to the SW occurs around 1500ms.  Fig.\ref{fig:bistab_fr}B 
shows the network firing rate during these three seconds (top) as well as the time course of the 
excitatory synaptic weights and external drive (middle and bottom respectively).  Note that the 
mean firing rate is relatively steady in the OU state indicating that the co-variation of the synaptic 
weights and external drive balance one another.  Fig.\ref{fig:bistab_fr}C and D analogously show 
the raster and firing rate of a simulation in which the network is switched into a SW state at $t=400$ms.  
Note that the SW state persists over the whole 3 sec. simulation.

\subsection{\label{subsect:H-T}Hopf and Turing bifurcations}

We consider the co-occurrence of two instabilities: a spatially
homogeneous oscillation and a  spatially inhomogeneous steady
solution.  This occurs when the zeroth spatial Fourier mode of  the
connectivity satisfies the relation, Eq.(\ref{eq:disp3}) and the
$k_{th}$ spatial Fourier mode  satisfies $J_{1}=1/\Phi^{'}$, while  we
assume that all other Fourier modes are sufficiently below their
critical values to avoid additional instabilities.  Without loss of generality we take $k=1$ for the Turing instability.   

We expand the parameters $J_0$, $J_1$, $I$ and $r$ as in
Eqs.~(\ref{eq:Jkexp},\ref{eq:Iexp},\ref{eq:rexp}), and define the slow
time~(\ref{eq:T}).
The linear solution consists of homogeneous, global oscillations and
stationary, spatially periodic bumps with  amplitudes which  we allow
to vary slowly in time, i.e. $r_{1} = H(T)e^{i\omega
t}+A(T)e^{ix}+c.c.$.  Carrying out a weakly nonlinear analysis to
third order in $\epsilon$ leads to the coupled amplitude equations
\begin{subequations}
\begin{eqnarray}
\partial_{T}H = (\mu+i\Omega)\Delta J_{0}H+(\alpha +i\beta)|H|^{2}H
+(\kappa +i\Lambda )|A|^{2}H, \label{eq:ampeq_THH} 
\end{eqnarray}
\begin{equation}
\partial_{T}A =
\bar{\eta}\Delta J_{1}A+\Gamma |A|^{2}A+\sigma |H|^{2}A,
\label{eq:ampeq_HTA}
\end{equation}
\end{subequations}
where $\mu +i\Omega$, $\alpha +i\beta$, $\bar{\eta}$, $\Gamma$, $\kappa +i\Lambda$ 
and $\sigma $ are
given by Eqs.~(\ref{eq:hopf_lc},\ref{eq:hopf_cc},\ref{eq:turing_lc},\ref{eq:turing_cc}, 
\ref{eq:kappa},\ref{eq:sigma}), respectively.\\

\subsubsection{Solution types and their stability}

\vspace{0.1in}

Steady state solutions to Eqs.~(\ref{eq:ampeq_THH},\ref{eq:ampeq_HTA}) 
include pure mode OU,  pure
mode SB and mixed mode solutions (OU-SB).  We look
at the stability of the OU and SB solutions in turn for the general case  and
then look specifically at the case of small delay in Eq.(\ref{eq:rate}).
Since the coupling in Eqs.~(\ref{eq:ampeq_THH}-\ref{eq:ampeq_HTA}) is
only through the amplitudes we can simplify the equations by taking
$(H,A)=(\mathcal{H}e^{i\theta}, \mathcal{A}e^{i\phi})$ which yields
\begin{subequations}
\begin{equation}
\dot{\mathcal{H}} = \mu\Delta
J_{0}\mathcal{H}+\alpha\mathcal{H}^{3}+\kappa\mathcal{A}^{2}\mathcal{H},
\label{eq:ampeq_HTnrH}
\end{equation}
\begin{equation}
\dot{\mathcal{A}} = \eta\Delta
J_{1}\mathcal{A}+\Gamma\mathcal{A}^{3}+\sigma\mathcal{H}^{2}\mathcal{A}. \label{eq:ampeq_HTnrA}
\end{equation}
\end{subequations}

\vspace{0.1in}

\noindent\textit{Oscillatory Uniform (OU)}:  The uniform oscillations have the form $({\mathcal{H^*}},0)$ where  
\begin{equation}
\mathcal{H}=\sqrt{-\frac{\mu\Delta J_{0}}{\alpha}}.
\nonumber
\end{equation}
The linear stability of this solution can be calculated  with the ansatz
\begin{equation}
(\mathcal{H},\mathcal{A}) = ({\mathcal{H^*}}+\delta\mathcal{H}e^{\lambda
t},\delta\mathcal{A}e^{\lambda t}),\nonumber
\end{equation}  
which yields the two eigenvalues
\begin{subequations}
\begin{equation}
\lambda_{H} = -2\mu\Delta J_{0},  \nonumber
\end{equation}
\begin{equation}
\lambda_{A} = \eta\Delta
J_{1}+\frac{\mu\sigma}{\alpha}\Delta J_{0}. \nonumber
\end{equation}
\end{subequations}
If we assume a supercritical uniform oscillatory state then the first
eigenvalue is always negative, while  the second becomes positive
along the line 
\begin{equation}
\Delta J_{1} = \frac{\mu\sigma}{\eta\alpha}\Delta J_{0},
\label{stab:TH_OU}
\end{equation}
indicating  the growth of a bump solution.\\

For Eq.(\ref{eq:rate}) with the parameters used to generate the
phase diagram  Fig.~\ref{fig:diagram}, we find from Eq.(\ref{stab:TH_OU}) that the OU state destabilizes along the line $\Delta J_{1}\sim -0.026\Delta J_{0}$.\\

\noindent\textit{Stationary Bump (SB)}: The stationary bump solution has 
the form $(0,{\mathcal{A^*}})$ where  
\begin{equation}
\mathcal{A}=\sqrt{-\frac{\eta\Delta J_{1}}{\Gamma}}.
\nonumber
\end{equation}
The linear stability of this solution can be calculated  with the ansatz
\begin{equation}
(\mathcal{H},\mathcal{A}) = (\delta\mathcal{H}e^{\lambda
t},{\mathcal{A^*}}+\delta\mathcal{A}e^{\lambda t}), \nonumber
\end{equation}  
which yields the two eigenvalues
\begin{subequations}
\begin{equation}
\lambda_{H} = \mu\Delta J_{0}-\frac{\eta\kappa}{\Gamma}\Delta J_{1},
\nonumber
\end{equation}
\begin{equation} 
\lambda_{A} = -2\eta\Delta J_{1}.
\nonumber
\end{equation}
\end{subequations}
If we assume a supercritical stationary bump state then the second
eigenvalue is always negative, while  the first becomes positive along
the line 
\begin{equation}
\Delta J_{1} = \frac{\mu\Gamma}{\eta\kappa}\Delta J_{0},
\label{stab:TH_SB}
\end{equation}
indicating  the growth of uniform oscillations.

For Eq.(\ref{eq:rate}) with the parameters used to generate the
phase diagram~\ref{fig:diagram}, we find from Eq.(\ref{stab:TH_SB}) that the SB state destabilizes along the line $\Delta J_{1}\sim -0.144\Delta J_{0}$.\\

\textit{Mixed Mode (OU-SB)}: 
The mixed-mode solution satisfies the following matrix equation
\begin{equation}
\left(\begin{array}{cc} \alpha & \kappa \\ \sigma & \Gamma
\end{array}\right)
\left(\begin{array}{c} \mathcal{H}^{2} \\ \mathcal{A}^{2}
\end{array}\right) = -
\left(\begin{array}{c} \mu\Delta J_{0} \\ \eta\Delta J_{1}
\end{array}\right),
\nonumber
\end{equation}
which yields
\begin{subequations}
\begin{equation}
\mathcal{H}^{2} = \frac{-\mu\Gamma\Delta J_{0}+\eta\kappa\Delta
J_{1}}{\alpha\Gamma-\sigma\kappa},  
\nonumber
\end{equation}
\begin{equation}
\mathcal{A}^{2} =\frac{\mu\sigma\Delta J_{0}-\eta\alpha\Delta
J_{1}}{\alpha\Gamma-\sigma\kappa}.
\nonumber
\end{equation}
\end{subequations}
We do not study the stability of the mixed-mode solution here.

\begin{figure}
\center
\includegraphics[scale=0.27]{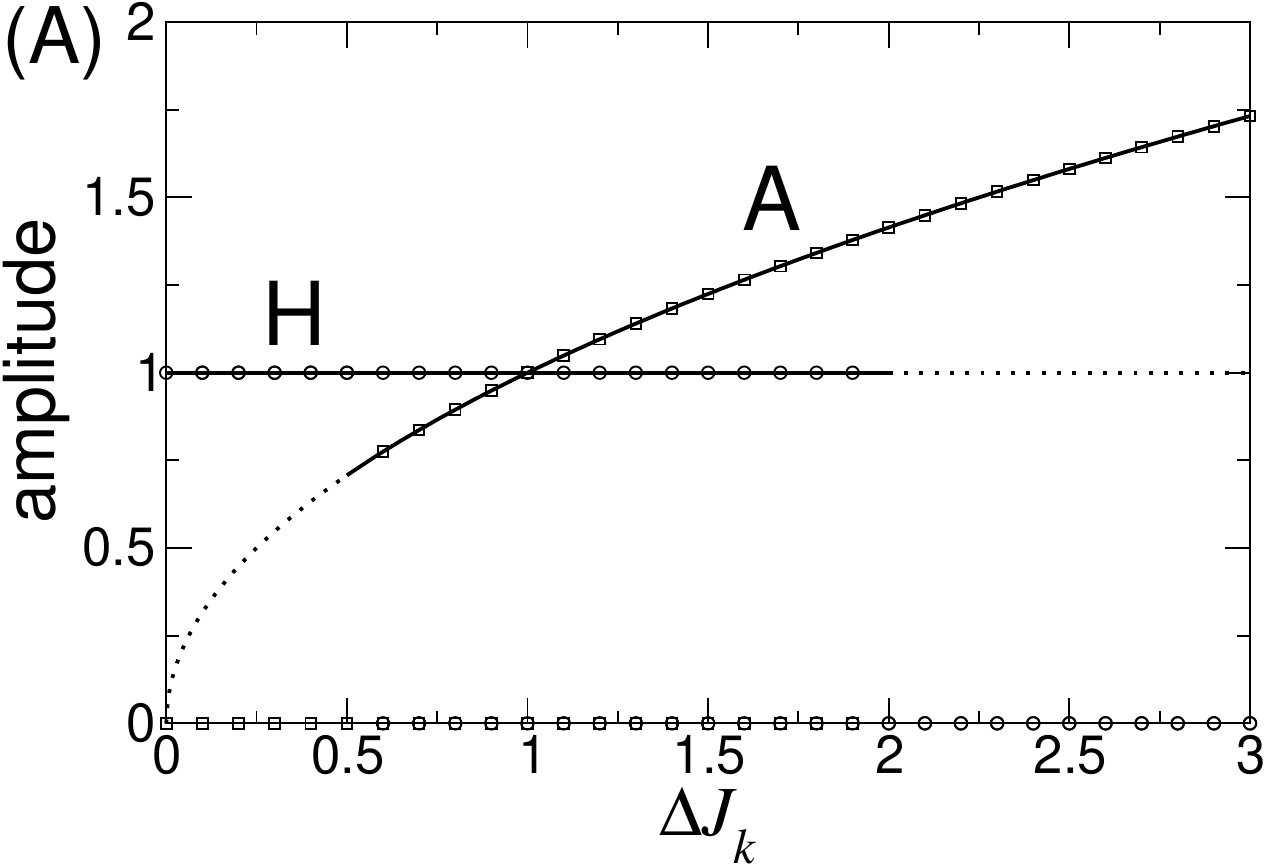}
\includegraphics[scale=0.27]{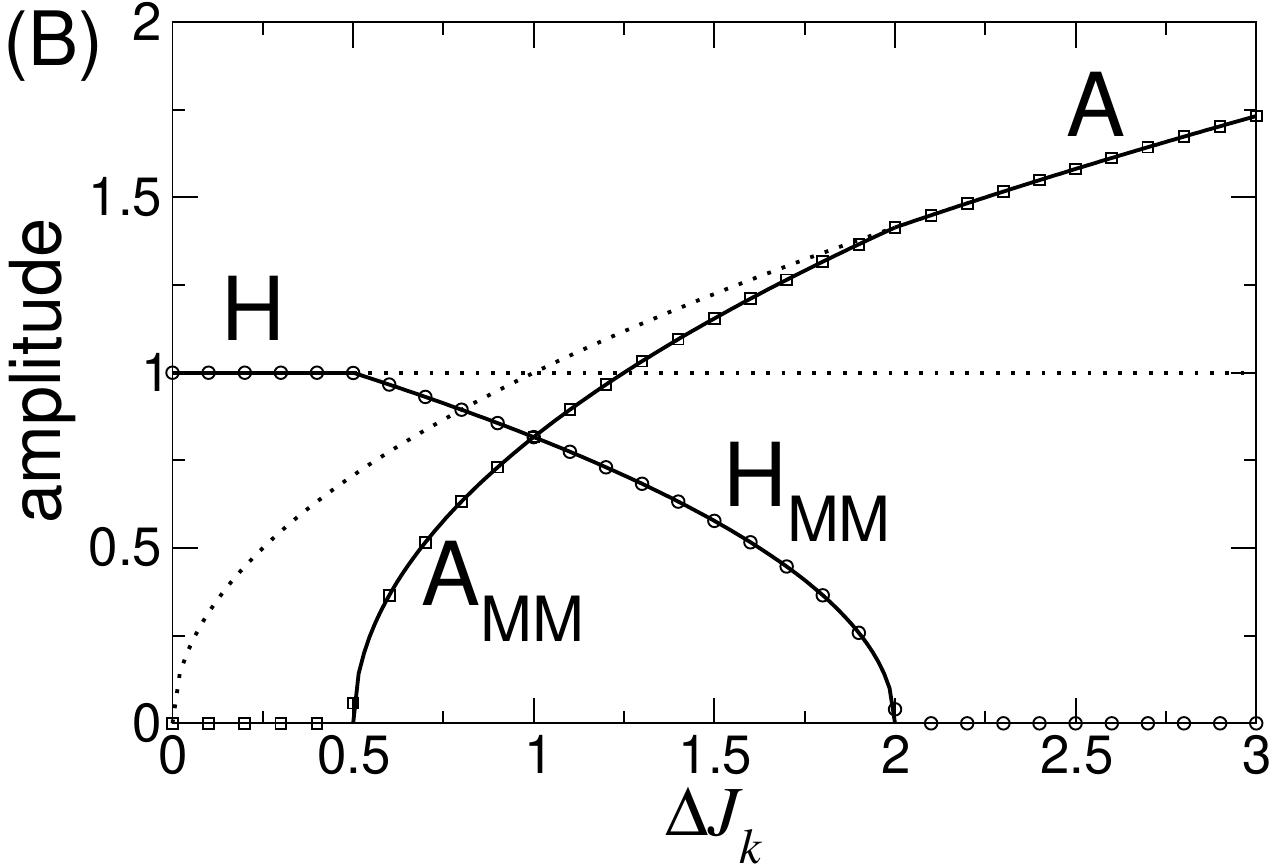}
\caption{Typical bifurcation diagrams for the competition between bumps
and global oscillations.   $\mu=-1$, $\Delta J_{0}=-1$, $\alpha=-1$,
$\eta=1$, $\Gamma=-1$.  A: $\kappa=-2$, $\sigma=-2$.  The limit cycle
and bump solutions  are bistable in the range $1/2<\Delta J_{1}<2$.
B: A mixed-mode solution is stable in the range $1/2<\Delta J_{1}<2$.
$\kappa=-0.5$, $\sigma=-0.5$.   Symbols are from simulation of the
amplitude equations  Eqs.~(\ref{eq:ampeq_HTnrH}-\ref{eq:ampeq_HTnrA})
while lines are the analytical expressions.}
\label{fig:HT_example1}
\end{figure}

\subsubsection{A simple example}
We once again illustrate the scenarios of bistability and mixed-mode
solutions with a simple example.   \\

\textit{i. Bistability~\footnote{
$\mu =\alpha =\Gamma =\Delta J_{0}=-1$, $\sigma = \kappa = -2$}}:
In this case,  the limit cycle has an amplitude $\mathcal{H} =1$ and
undergoes an instability at $\Delta J_{1}=2$.   The bump solution has
an amplitude $\mathcal{A}=\sqrt{\Delta J_{1}}$ and becomes unstable at
$\Delta J_{1}=1/2$.   The oscillatory and bump solutions are therefore
bistable in the range $1/2<\Delta J_{1}<2$.  This is borne  out in
numerical simulations of
Eqs.~(\ref{eq:ampeq_HTnrH}-\ref{eq:ampeq_HTnrA}) [see
Fig.~\ref{fig:HT_example1}A]. Symbols are from numerical simulation
(circles:limit cycle, squares:bump), while lines are analytical
solutions. \\  

\textit{ii.Mixed-mode~\footnote{$\mu =\alpha =\Gamma
=\Delta J_{0}=-1$, $\sigma = \kappa = -1/2$}}: 
In this case, the
limit cycle has an amplitude $\mathcal{H} =1$ and undergoes an
instability at $\Delta J_{1}=1/2$.   The bump solution has an
amplitude $\mathcal{A}=\sqrt{\Delta J_{1}}$ and becomes unstable at
$\Delta J_{1}=2$.   The mixed-mode solution is stable in the range
$1/2<\Delta J_{1}<2$ and has amplitudes
$\mathcal{H}_{MM}=2\sqrt{(1-\Delta J_{1}/2)/3}$  and
$\mathcal{A}=2\sqrt{(\Delta J_{1}-1/2)3}$.  The corresponding
bifurcation diagram is shown in Fig.~\ref{fig:HT_example1}B  where
symbols are from simulation of
Eqs.~(\ref{eq:ampeq_HTnrH}-\ref{eq:ampeq_HTnrA}) and lines are the
analytical results.

\subsubsection{Summary}

\begin{table*}
\begin{tabular}{lll}
Codim.-2 bifurcations & Solution types calculated & Instability boundaries \\ \hline
Hopf and Turing-Hopf & Oscillatory Uniform & $\Delta J_1=\Delta J_0$ \\
 & Travelling Waves & $\Delta J_1=a/(2 \alpha) \Delta J_0$ \\
 & Standing Waves (osc.) & $\Delta J_1=(a+c)/(4 \alpha) \Delta J_0$ \\
 & Standing Waves (stat.) &  $\Delta J_1=\Psi \Delta J_0$ \\ 
 & Mixed-Mode & Not calculated \\ 
Hopf and Turing & Oscillatory Uniform & $\Delta J_1=(\mu \sigma)/(\eta \alpha) \Delta J_0$ \\
& Stationary Bump & $\Delta J_1=(\mu \Gamma)/(\eta \kappa) \Delta J_0$\\
 & Mixed-Mode (OU-SB)  & Not calculated \\ 
\end{tabular}
\caption{Some existing dynamical states that are present close to the
  codimension-2 bifurcations, and their corresponding instability
  boundaries Eqs.~(\ref{eq:OU_steady_inst},\ref{eq:TW_inst},\ref{eq:HTH_SW_osc},\ref{eq:HTH_SW_stead},\ref{stab:TH_OU},\ref{stab:TH_SB}), 
except for the Mixed Mode solutions.}\label{table1}
\end{table*} 

The interaction between the SB and OU states can lead to one of two scenarios.  Either 
there is a region of bistability between bumps and oscillations, or there is a mixed-mode 
solution which, near the codimension-2 point at least, will consist of bumps whose amplitude 
oscillates in time, i.e. oscillating bumps (OB).  

In the limit of small $D$ the instability lines for the OU and SB states in the 
vicinity of the codimension 2 point are given by the equations
\begin{eqnarray}
\Delta J_{1} &=& -D\frac{\Big(\frac{(\Phi^{''})^{2}}{\Phi^{'}}-\Phi^{'''}\Big)}{\Big(\frac{(11\pi
-4)}{20}\frac{(\Phi^{''})^{2}}{\Phi^{'}}-\pi\frac{\Phi^{'''}}{4}\Big)}\Delta J_{0},\\
\Delta J_{1} &=& -\frac{2}{\pi}D\frac{\Big(\frac{(\Phi^{''})^{2}}{\Phi^{'}}-\frac{\Phi^{'''}}{2}
-\frac{J_{2}(\Phi^{''})^{2}}{2(1-J_{2}\Phi^{'})}\Big)}{\Big(\frac{(\Phi^{''})^{2}}{\Phi^{'}}
-\Phi^{'''}\Big)}\Delta J_{0}.
\end{eqnarray}
respectively.  The slope of both of the stability lines is proportional to $D$, indicating that 
in the small $D$ limit any region of bistability or mixed mode solution will be limited to 
a narrow wedge close to the $J_{0}$ axis.  Which scenario will be observed (bistability or 
mixed-mode) depends on the particular choice of $\Phi^{'}$ and the value of the second 
spatial Fourier mode of the connectivity $J_{2}$.  For the parameters values used to 
generate the phase diagram  Fig.~\ref{fig:diagram}
the slopes are $\sim -0.026$ and $\sim -0.144$ for the OU and SB 
stability lines respectively, indicating a mixed mode solution.

\subsubsection{Network simulations}

\begin{figure}
\center
\includegraphics[scale=0.45]{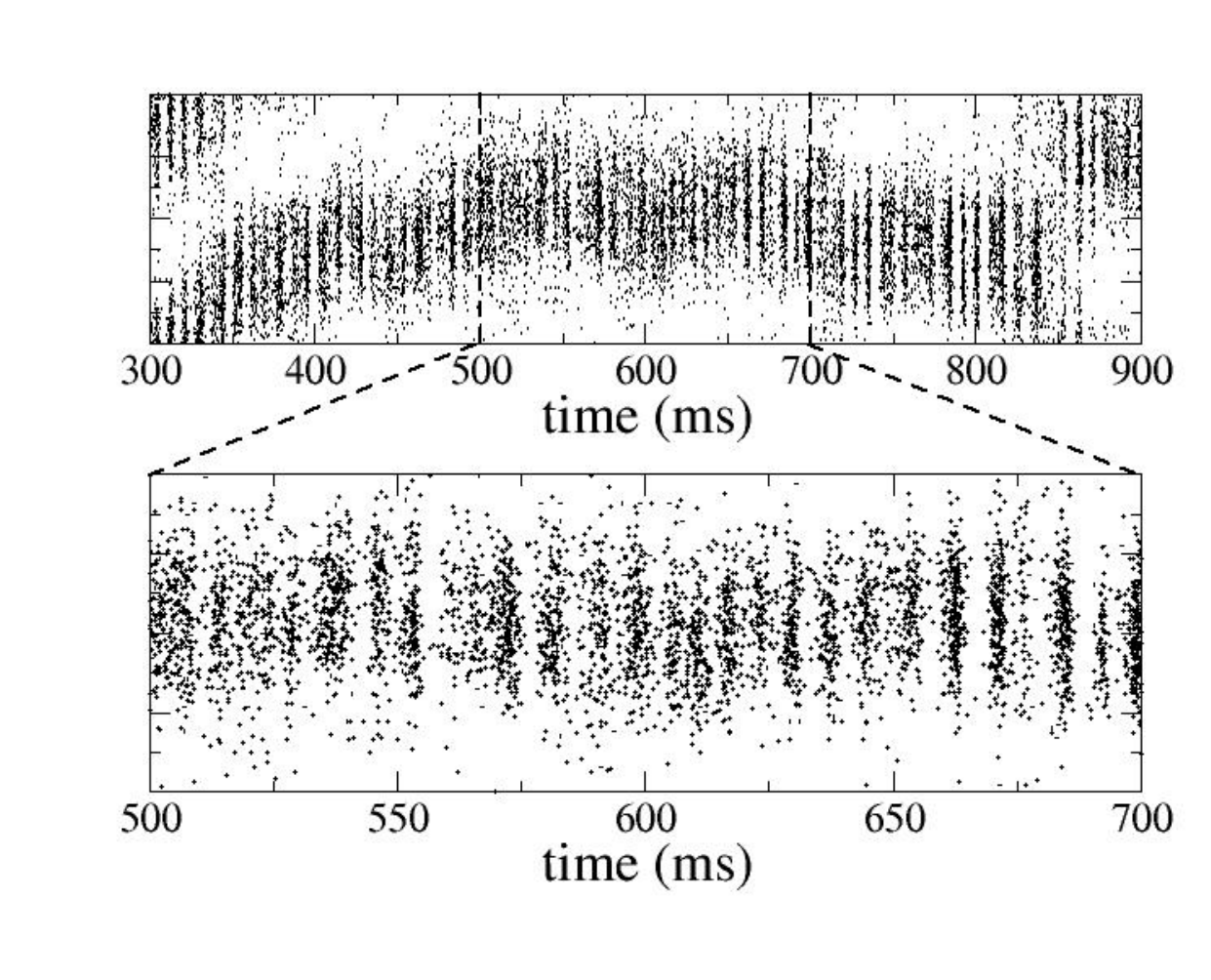}
\caption{A mixed mode (OU/SB) solution in a network of spiking neurons with $p_{E}=0.2+0.1\cos{r}$, $p_{I}=0.2$, $g_{E}=0.1$, $g_{I}=0.28$, $\nu_{ext}=2000$ and $g_{ext}=0.01$.  See text for details.}\label{fig:OB}
\end{figure}

As shown in \cite{roxin05}, oscillating bump solutions can be found in networks of spiking neurons 
with Mexican-hat connectivity and strong inhibition.  Here we have identified such oscillating bumps 
as arising via a bifurcation to mixed-mode OU and SB in the vicinity of the codimension-two point for 
homogeneous oscillations and Turing patterns.  Fig.\ref{fig:OB} shows a sample raster from a 
numerical simulatiom simulation. 
The top panel shows a mixed-mode solution which drifts in time. The bottom panel is a blow-up of the 
raster for $500$ms$ <t< 700$ms where the fast oscillations are clearly visible.  

\section{Conclusions}

Our main objective in this paper was to understand in greater detail
the  dynamical states which arise in networks of spiking neurons.
Specifically, we are interested in large networks of irregularly
spiking  neurons for which a reduced, phenomenological description in
terms of  mean firing rates is a reasonable approximation.  This is
the case, for  example, if the connectivity is sparse and
cross-correlations of the input  currents to different cells are
therefore weak.  The particular form of the  rate equation description
is motivated by the observation of emergent  fast oscillations in
simulations  of networks with dominant inhibition,
e.g. \cite{brunel99}.  The origin of these  oscillations has been well
studied in model networks and is due to an effective  delay in
neuronal interactions generated by the synaptic kinetics and
single-cell  dynamics \cite{brunel03,geisler05}. The dynamics of the
mean firing rate of the  resulting oscillatory states can be captured
by a rate equation with an  explicit time delay \cite{roxin05}.

However, the rate equation considered previously  did not agree
entirely with network simulations.  Specifically, it predicted  a
large region of traveling wave solutions while in the network
simulations  only standing waves were found to be stable.  Our
analysis here has shown that  given a more realistic transfer
function, i.e. sigmoidal or expansive power-law  nonlinearity, the
primary inhomogeneous oscillatory instability given a  cosine
connectivity will be to standing waves.  This suggests that the
standing wave states robustly observed in network simulations were due
to  the nonlinearity in the single cell fI curve, and that the rate
model with  threshold linear transfer function studied in
\cite{roxin05} was not able to  capture this effect.  

We predicted
further that altering the connectivity could  stabilize the traveling
wave state.  Specifically, more realistic patterns of  connectivity,
such as Gaussian, affect the competition between traveling and
standing waves through the second spatial Fourier mode.  In the case
of  delay-driven waves, where the primary instability occurs only for
strong  inhibition, we show that the sign of the second mode will be
negative and that  in the limit of small delay this will always lead
to traveling waves.  This  prediction is borne out in network
simulations, see Fig.\ref{fig:rasters}.  

We furthermore show that  the
bifuration to homogeneous oscillations is supercritical for  standard
choices of transfer functions, again sigmoidal and expansive
power-law.  This agrees with network simulations of conductance-based
neurons we have conducted, and with the amplitude equation derived for
a network of integrate-and-fire neurons \cite{brunel99}.  

It is more
difficult to draw clear-cut conclusions regarding the dynamical states
seen in the vicinity of the two codimension-two points we have
studied:  Hopf/Turing-Hopf and Hopf/Turing.  In general there will
either be a region of  bistability between the two states which
bifurcate via primary instabilities in the  vicinity of the
codimension-two point, or there will be one or several mixed-mode
solutions.   In network simulations we have shown examples of both:
bistable OU/SW and mixed mode OU/SB.  Many more complex dynamical
states can be  observed both in the rate model and the network
simulations further from the  primary bifurcations.  Here we have
chosen an analytical approach which is  powerful enough to allow for
arbitrary transfer functions and connectivity, but  which is limited
to the parameter space in the vicinity of the primary bifurcations.
This is a complementary approach to that taken in \cite{roxin05},
where a  specific choice of transfer function and connectivity allowed
for an  analysis of several nonlinear states, even far from the
primary bifurcations.

The approach we have taken in this paper is similar in spirit to that
of Curtu and Ermentrout \cite{curtu04}.  In that work, they study an
extension of a rate model with adaptation proposed by Hansel and
Sompolinsky \cite{hansel98}.   As in our work here, they allow for a
generic transfer function and generic connectivity  and derive
amplitude equations for the primary Turing and Turing-Hopf
bifurcations which  occur.  Thus both adaptation and the effective
delay in  neuronal interactions can lead to waves.  However, the waves
in these two cases  arise via distinct physiological mechanisms and
can exhibit very different  propagation velocities.  

In the case of
adaptation, waves arise given  patterns of synaptic connectivity
which, in the absence of adaptation would  lead to the emergence of
bump states, i.e. a Turing instability.  These tend  to be Mexican-hat
like connectivities.  In the presence of adaptation, stationary  bump
states may destabilize since the peak of activity is preferentially
suppressed compared to activity at the edge of the bump.  As a result
the  bump begins to move via a symmetry-breaking bifurcation,
generally  leading to traveling waves.  This only occurs if the
adaptation is significantly  strong, i.e. above a critical threshold.
In a two-population model with  adaptation studied in \cite{curtu04},
the frequency of the waves was shown to  be proportional to the square
root of the difference between the strength of adaptation  and a
critical value below which no bifurcation is possible.  Thus the
resulting  waves can be arbitrarily slow depending on the strength of
adaptation.  

In the case of  rate equations with delay, a Turing-Hopf
instability occurs only for connectivities which are  strongly
inhibitory locally, i.e. inverted Mexican-hat connectivities.  A
transient  perturbation which increases the firing rate locally will,
after a delay, strongly self-inhibit  while increasing the firing rate
of more distant neurons.  The process then repeats, leading to a
propagation of activity.  The frequency of such waves is clearly
related to the delay, which  itself depends on the synaptic time
constants as well as the spike generation currents
\cite{brunel03,geisler05}.

Finally, we have tried to emphasize the importance of fixed delays in
shaping the dynamics  described by Eq.(\ref{eq:rate}) and by extension
in networks of spiking neurons.  Nonetheless  both fixed and
conduction delays are present in neuronal systems and are roughly of
the  same order of magnitude in a small patch of cortex of $\sim 1$mm
in extent.  It remains  to be studied how these delays interact to
shape patterns of spontaneous activity.



\appendix

\section{\label{app:network}The network of spiking neurons}

In this appendix we describe the network of conductance-based neurons used in 
simulations.  The single cell model is taken from \cite{wang96}.  
The network consists of two populations of neurons: one excitatory and 
one inhibitory.  The number of neurons in each population is $N_{E}$ and $N_{I}$ 
respectively.  The membrane voltage of the $i$th neuron in the excitatory population evolves 
according to the differential equation
\begin{equation}
C_{m}\dot{V}_{i}=-I_{Na,i}-I_{K,i}-I_{L,i}-I_{syn,i}+I_{app,i},
\nonumber
\end{equation}
where the membrane capacitance $C_{m} = 1\mu F/cm^{2}$ and the applied current $I_{app,i}$ has the 
units $\mu A/cm^{2}$.  

The leak current is 
\begin{equation}
I_{L,i}=g_{L}(V_{i}-E_{L}),
\nonumber
\end{equation} 
where $g_{L}=0.1 mS/cm^{2}$.  Action potential generation is dependent on a sodium and a 
potassium current.  

The sodium current is 
\begin{equation}
I_{Na,i}=g_{Na}m_{\infty,i}^{3}h_{i}(V_{i}-E_{Na}), 
\nonumber
\end{equation}
where $g_{Na}=35 mS/cm^{2}$, $E_{Na}=55mV$ and 
the activation variable $m$ is assumed fast and therefore taken at its equilibrium 
value $m_{\infty,i}=\alpha_{m,i}/(\alpha_{m,i}+\beta_{m,i})$, where $\alpha_{m,i}=-0.1(V_{i}+35)/
(\mathrm{exp}(-0.1(V_{i}+35))-1)$ and $\beta_{m,i}=4\mathrm{exp}(-(V_{i}+60)/18)$.  The 
inactivation variable $h_{i}$ follows the first order kinetics
\begin{equation}
\dot{h}_{i}=\phi\Big(\alpha_{h,i}(1-h_{i})-\beta_{h,i}h_{i}\Big),
\nonumber
\end{equation}
where $\phi=5$, $\alpha_{h,i}=0.07\mathrm{exp}(-(V_{i}+58)/20)$ and $\beta_{h,i}
=(\mathrm{exp}(-0.1(V_{i}+28))+1)^{-1}$.

The potassium current is
\begin{equation}
I_{K,i}=g_{K}n_{i}^{4}(V_{i}-E_{k}),
\nonumber
\end{equation}
where $g_{K}=9 mS/cm^{2}$ and $E_{K}=-90mV$.  The activation variable $n_{i}$ follows the first 
order kinetics
\begin{equation}
\dot{n}_{i}=\phi\Big(\alpha_{n,i}(1-n_{i})-\beta_{n,i}n_{i}\Big)
\nonumber
\end{equation}
where $\phi=5$, $\alpha_{n,i}=-0.01(V_{i}+34)/(\mathrm{exp}(-0.1(V_{i}+34))-1)$ and 
$\beta_{n,i}=0.125\mathrm{exp}(-(V_{i}+44)/80)$.

The synaptic current is
\begin{equation}
I_{syn,i}=g_{EE}s_{E,i}(t)(V_{i}-E_{syn,E})+g_{EI}s_{I,i}(t)(V_{i}-E_{syn,I})+g_{ext}s_{ext,i}(t)
(V_{i}-E_{syn,E}),
\nonumber
\end{equation}
where the reversal potentials for excitatory and inhibitory synapses are $E_{syn,E}=0$ and 
$E_{syn,I}=-80$ respectively.  The conductance change from the activation of recurrent excitatory 
connections is given by $g_{EE}s_{E,i}(t)$ where 
\begin{eqnarray}
\tau_{E,2}\dot{s}_{E,i}&=&-s_{E,i}+x_{E,i},\nonumber\\
\tau_{E,1}\dot{x}_{E,i}&=&-x_{E,i}+\sum_{j=1}^{N_{E}}w_{ij}\sum_{k}\delta (t-t_{j}^{k}), \label{eq:x}
\end{eqnarray}
where the $w_{ij}s\in \{0,1\}$ indicate the presence or absence of a synaptic contact from cell 
$j$ to cell $i$.  The double sum in Eq.(\ref{eq:x}) is over all neurons in the excitatory population 
and over all spikes, i.e. $t_{j}^{k}$ is the time of the $k$th spike of neuron $j$ which will cause 
a jump of amplitude $1/\tau_{E,1}$ in the variable $x_{i}$ of neuron $i$ if $w_{ij}=1$.  The resulting 
post-synaptic conductance change in cell $i$ from a single presynaptic spike at time $t^{*}$ is 
given by 
\begin{equation}
g_{EE}s_{E,i}(t) = \frac{g_{EE}}{\tau_{E,2}-\tau_{E,1}}\Big(e^{-(t-t^{*})/\tau_{E,2}}-e^{-(t-t^{*})/\tau_{E,1}}\Big),
\nonumber
\end{equation}
which has units of $mS/cm^{2}$.  The time course is therefore a difference of exponentials with a 
rise time given by $\tau_{E,1}$ and a decay time $\tau_{E,2}$.  Note that the time integral of 
the response $s_{E,i}$ from $t=t_{*}$ to $t=\infty $ has been normalized to 1 and so $g_{EE}$ has units 
of $mS\cdot ms/cm^{2}$.  The synaptic current from inhibitory connections is analogous, with 
time constants $\tau_{I,1}$ and $\tau_{I,2}$.  

Finally, external inputs have the same functional 
form as the recurrent excitatory inputs.  External presynaptic spikes to excitatory cells 
are modeled as a Poisson process with rate $\nu_{E,ext}$.  The Poisson process is independent from 
cell to cell. Unless otherwise noted, all external inputs have weight $g_{ext}=0.0019$mS$\cdot$ms/cm$^{2}$, and synpatic time constants are taken to be $\tau_{1}=1$ms and 
$\tau_{2}=3$ms.

Inhibitory neurons are modeled analogously to excitatory ones.  For this work, we take all 
single cell parameters to be identical to the excitatory cells.  Synaptic time constants 
are taken to be the same, i.e. $\tau_{E,1}=\tau_{I,1}=\tau_{1}$ and $\tau_{E,2}=\tau_{I,2}=\tau_{2}$.  
In addition we take $g_{EE}=g_{IE}=g_{E}$ and $g_{II}=g_{EI}=g_{I}$.  Thus excitatory and inhibitory 
synapses have identical time courses but may have different strengths.

\subsection{Connectivity}

We choose a prescription for choosing $w_{ij}$s which leads to sparse, random connectivity which 
is spatially modulated.  We do this by defining a probability for a connection to be made from cell 
$j$ in population $\beta\in\{ E,I \}$ to a cell $i$ in population $\alpha\in\{ E,I \}$ of the form
\begin{equation}
p^{\alpha\beta }(i,j)=p_{0}^{\alpha\beta}+p_{1}^{\alpha\beta}\cos{r}+p_{2}^{\alpha\beta}\cos{2r},
\label{conn}
\end{equation}
where $r$ is the distance between cells $i$ and $j$ which are situated on a ring , 
normalized such that $r\in\{-\pi, \pi\}$.
In order to compare with the rate model we choose $p_{l}^{EE}=p^{IE}=p^{E}$ and 
$p_{l}^{II}=p_{l}^{EI}=p_{l}^{I}$ where $l\in\{0,1,2\}$.

\section{\label{app:ampeq}Amplitude Equations}

In this Appendix we outline the calculation of the amplitude equations
which  describe the slow temporal evolution of the various
instabilities near their  respective bifurcations.

\subsection{General framework for the weakly nonlinear calculation: Codimension 1 bifurcations}

Here we briefly describe the general framework for the weakly
nonlinear calculation for the Turing, Hopf and  Turing-Hopf
bifurcations.  We use the standard multiple-scales approach which takes advantage
of  the fact there is a near-zero eigenvalue in the vicinity of a bifurcation which
is responsible  for the slow temporal evolution of the critical
eigenmode (see e.g. \cite{holmes95}).

For simplicity we first rewrite Eq.(\ref{eq:rate}) as
\begin{equation}
\dot{r} = -r+\Phi\Big(\langle Jr \rangle +I \Big), \label{eq:rate_app}
\end{equation}
where $\langle fg \rangle\equiv
\frac{1}{2\pi}\int_{-\pi}^{\pi}dyf(y-x)g(y,t-D)$.  We study the
stability of the steady state solution $R=\Phi\Big( J_{0}R+I\Big)$,
where $J(x)=J_{0}+2\sum_{n=1}^{\infty}J_{n}\cos{nx}$.  We expand the rates,
the connectivity and the input current as
\begin{eqnarray}
r(x,t) &=& R+\epsilon r_{1}(x,t,T)+\epsilon r_{2}(x,t,T)+\dots ,\nonumber \\
J(x) &=& \bar{J}(x)+\epsilon^{2}\Delta J(x), \nonumber \\ 
I &=& \bar{I}+\epsilon^{2}\Delta I,\nonumber
\end{eqnarray}
where the small parameter $\epsilon $ is defined by the distance from the critical value of the connectivity, given by Eqs.~(\ref{eq:disp2},\ref{eq:disp3}). 
Plugging these expansions into Eq.(\ref{eq:rate_app}) yields
\begin{equation}
(\mathcal{L}+\epsilon^{2}\mathcal{L}_{2})(\epsilon
r_{1}+\epsilon^{2}r_{2}+...) = \epsilon^{2}N_{2}(r_{1})+
\epsilon^{3}N_{3}(r_{1},r_{2}),\nonumber
\end{equation}
where
\begin{eqnarray}
\mathcal{L}r &=& \partial_{t}r+r-\langle\bar{J}r\rangle,\nonumber \\
\mathcal{L}_{2}r &=& \partial_{T}\langle\bar{J}r\rangle
-\langle\Delta Jr\rangle,\nonumber \\ N_{2} &=&
\frac{\Phi^{''}}{2}\langle\bar{J}r_{1}\rangle^{2}, \nonumber \\ N_{3} &=&
\Phi^{''}\langle\bar{J}r_{1}\rangle\langle\bar{J}r_{2}\rangle+\frac{\Phi^{'''}}{6}\langle\bar{J}r_{1}\rangle^{3}.\nonumber
\nonumber
\end{eqnarray}
We now collect terms by order in $\epsilon$. At first order we have
\begin{equation}
\vartheta (\epsilon ):\quad \mathcal{L}r_{1} = 0.
\nonumber
\end{equation}
This equation gives the linear dispersion relation Eq.(\ref{eq:disp}).
The values of the connectivity and input current for which it is
satisfied are $J(x)=\bar{J}(x)$ and $I=\bar{I}$. At second order we 
obtain 
\begin{equation}
\vartheta (\epsilon^{2} ):\quad \mathcal{L}r_{2} = N_{2}(r_{1}).
\nonumber
\end{equation}
The second order solution $r_{2}$ is the particular solution of this
linear differential equation. And finally, at third order
\begin{equation}
\vartheta (\epsilon^{3} ):\quad
\mathcal{L}r_{3} = N_{3}(r_{1},r_{2})-\mathcal{L}_{2}r_{1}.
\nonumber
\end{equation}
At this order secular terms arise which have the same temporal and/or
spatial frequency as the linear solution.  In order for the  above
equation to have a solution, these terms must therefore be eliminated,
yielding the desired amplitude equation for the  instability.

\subsubsection{Steady Bifurcation: $\omega = 0$, $k=0$}

For completeness we include here the derivation of the amplitude
equation  for the transcritical bifurcation.

The $0^{th}$ spatial Fourier mode of the connectivity is given by the
critical value $\bar{J}_{0} = \frac{1}{\Phi^{'}}$, while  we assume
that all other Fourier modes are sufficiently below their critical
values to avoid additional instabilities.   We expand
\begin{eqnarray}
J_{0} &=& \bar{J}_{0}+\epsilon\Delta J_{0}, \label{eq:J0exp}\\ 
I &=& \bar{I}+\epsilon\Delta I, \nonumber 
\\ r &=& R+\epsilon
r_{1}+\epsilon^{2}r_{2}+\dots, \nonumber
\end{eqnarray}
where the small parameter $\epsilon$ is defined by Eq.(\ref{eq:J0exp}).
We define the slow time $T = \epsilon t$.   The linear solution is a
spatially homogeneous  amplitude which  we allow to vary slowly in
time, i.e. $r_{1} = A(T)$.  Carrying out a weakly nonlinear analysis
to  second order in $\epsilon$ leads to the normal form for a
transcritical bifurcation given by
\begin{eqnarray}
\partial_{T}A &=& \eta\Delta J_{0}A+\gamma A^{2}, \nonumber \\ 
\eta &=& \frac{\Phi^{'}}{1+D}, \nonumber \\ 
\gamma &=&
\frac{\Phi^{''}}{2(1+D)}\bar{J}_{0}^{2}. \label{eq:ampeq-steady}
\end{eqnarray}
\begin{figure}
\center
\includegraphics[scale=0.4]{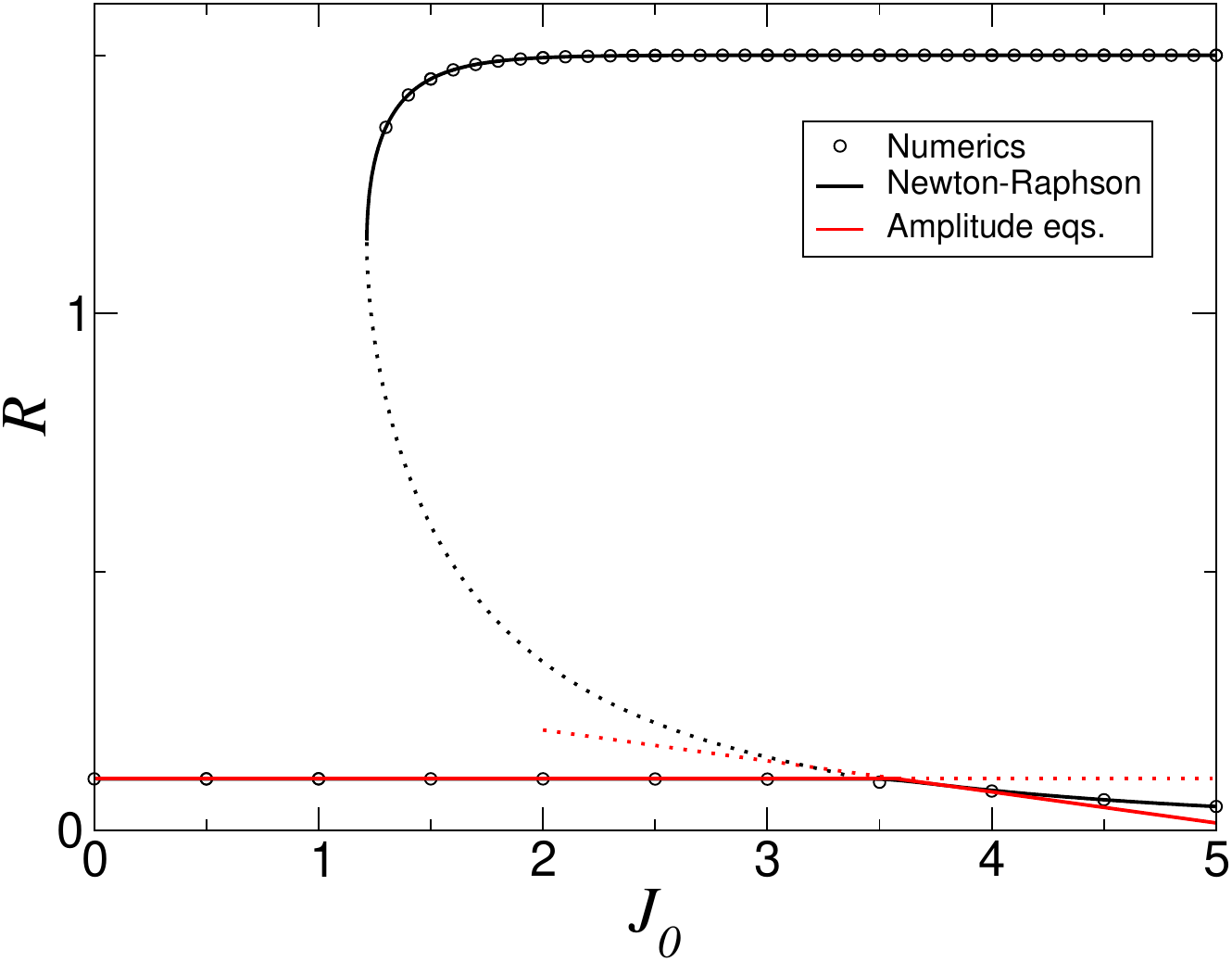}
\caption{(Color online) Bifurcation diagram for the steady instability.  Open
circles:  numerical simulation of Eq.(\ref{eq:rate}).  Red Lines:
amplitude equation solution from Eq.(\ref{eq:ampeq-steady}).   Black
lines: steady-state solution of Eq.(\ref{eq:rate}) using a
Newton-Raphson solver.  Solid lines indicate stable  solutions and
dotted lines unstable ones.  $\Phi (x) = \frac{\alpha}{1+e^{-\beta
x}}$ where $\alpha=1.5$ and $\beta = 3$.  $J(x) = J_{0}+J_{1}\cos{x}$
where $J_{1}=0$.  The input current $I$ is varied so as to keep the
uniform stationary  solution fixed at $R=0.1$.}
\end{figure}

\subsubsection{Turing bifurcation}

\noindent $\vartheta (\epsilon )$: The solution to the linear equation is spatially periodic with slowly
varying amplitude $A$,
\begin{equation}
r_{1} = A(T)e^{ix}+c.c.
\nonumber
\end{equation}

\noindent $\vartheta (\epsilon^{2} )$: The nonlinear forcing and resulting second order solution are
\begin{eqnarray}
N_{2} &=& \Phi^{''}J_{1}^{2}(A^{2}e^{2ix}+c.c.+2|A|^{2})/2, \nonumber\\ 
r_{2} &=& r_{22}e^{2ix}+c.c.+r_{20},\nonumber \\ r_{22} &=&
\frac{\Phi^{''}J_{1}^{2}}{2(1-J_{2}\Phi^{'})}A^{2},\nonumber \\ 
r_{20} &=&
\frac{\Phi^{''}J_{1}^{2}}{1-J_{0}\Phi^{'}}|A|^{2}.\nonumber
\end{eqnarray}

\noindent $\vartheta (\epsilon^{3} )$: The nonlinear forcing at cubic order is
\begin{equation}
N_{3} =
(\Phi^{''}J_{1}J_{0}Ar_{20}+\Phi^{''}J_{1}J_{2}{A^*}r_{22} +\Phi^{'''}J_{1}^{3}|A|^{2}A/2)e^{ix}+  c.c.+\dots \nonumber
\end{equation}
Eliminating all terms of periodicity $e^{ix}$ at this order yields
the amplitude equation, Eq.(\ref{eq:ampeq_turing}).
\begin{equation}
\partial_{T}A = \eta\Delta J_{1}A+\Gamma |A|^{2}A,
\nonumber 
\end{equation}
with the coefficients
\begin{eqnarray} 
\eta &=& \frac{\Phi^{'}}{1+D}, \nonumber \\ 
\Gamma &=& \frac{\bar{J}_{k}^{3}}{1+D}
\Bigg(\frac{J_{0}(\Phi^{''})^{2}}{1-J_{0}\Phi^{'}}
+\frac{J_{2}(\Phi{''})^{2}}{2(1-J_{2}\Phi^{'})}+\frac{\Phi^{'''}}{2}\Bigg).\nonumber
\end{eqnarray}

\subsubsection{Hopf bifurcation}
\vspace{0.1in}

\noindent $\vartheta (\epsilon )$: The solution to the linear equation is a time periodic function with slowly varying amplitude $H$
\begin{equation}
r_{1} = H(T)e^{i\omega t}+c.c.
\nonumber
\end{equation}

\noindent $\vartheta (\epsilon^{2} )$: The nonlinear forcing and 
resulting second order solution are
\begin{eqnarray}
N_{2} &=& \Phi^{''}J_{0}^{2}(H^{2}e^{2i\omega (t-D)}+c.c.+2|H|^{2})/2,
\nonumber \\ 
r_{2} &=& r_{22}e^{2i\omega t}+c.c.+r_{20},\nonumber \\ 
r_{22} &=&
\frac{\Phi^{''}J_{0}^{2}}{2(2i\omega +1-\Phi^{'}J_{0}e^{-2i\omega
D})}e^{-2i\omega D}H^{2}, \nonumber\\ 
r_{20} &=&\frac{\Phi^{''}J_{0}^{2}}{1-J_{0}\Phi^{'}}|H|^{2}.\nonumber
\end{eqnarray}

\noindent $\vartheta (\epsilon^{3} )$: The nonlinear forcing at cubic 
order is

\begin{eqnarray}
N_{3}& =& (\Phi^{''}J_{0}^{2}Hr_{20}+\Phi^{''}J_{0}^{2}{H^*}r_{22} 
\nonumber\\ &+&\Phi^{'''}J_{0}^{3}|A|^{2}A/2)e^{i\omega
(t-D)}+c.c.+\dots \nonumber
\end{eqnarray}
Eliminating all terms of periodicity $e^{i\omega t}$ at this order
yields the amplitude equation, Eq.(\ref{eq:hopf}).
\begin{equation}
\partial_{T}H = (\mu +i\Omega )\Delta J_{0}H +(\alpha +i\beta)|H|^{2}H,
\nonumber 
\end{equation}
with the coefficients
\begin{eqnarray}
\mu + i\Omega &=&
\frac{\Phi^{'}e^{-i\omega D}}{1+D(1+i\omega)},  \label{eq:hopf_lc}\\
\alpha +i\beta &=& \frac{e^{-i\omega D}}{1+D(1+i\omega)} 
\times \label{eq:hopf_cc}\\ 
&& \Bigg(\frac{\bar{J}_{0}^{4}(\Phi^{''})^{2}}{1-\Phi^{'}\bar{J}_{0}} +\frac{\bar{J}_{0}^{4}(\Phi^{''})^{2}e^{-2i\omega D}}{2(2i\omega
+1-\Phi^{'}\bar{J}_{0}e^{-2i\omega D})}
+\frac{\bar{J}_{0}^{3}\Phi^{'''}}{2}\Bigg).\nonumber 
\end{eqnarray}

\subsubsection{Turing-Hopf bifurcation}
\vspace{0.1in}

\noindent $\vartheta (\epsilon )$: The solution to the linear equation 
are two sets of periodic waves, one left-traveling with slowly 
varying  amplitude $A$ and the other right-traveling with slowly 
varying amplitude $B$
\begin{eqnarray}
r_{1} &=& A(T)e^{i\omega t+ix}+B(T)e^{-i\omega t+ix}+c.c. \nonumber 
\\ &=& A(T)e^{\psi}+B(T)e^{\phi}+c.c. \nonumber
\end{eqnarray}

\noindent $\vartheta (\epsilon^{2} )$: The nonlinear forcing and 
resulting second order solution are

\begin{eqnarray}
N_{2} &=& \Phi^{''}J_{1}^{2} \Big[ A^{2}e^{2\psi-2i\omega D}+2ABe^{\psi
+\phi}+2A{B^*}e^{\psi +{\phi^*}-2i\omega D} +\nonumber \\ 
&&B^{2}e^{2\phi +2i\omega D}+\text{c.c.} 
+2(|A|^{2}+|B|^{2})\Big]/2, \nonumber \\ 
r_{2} &=&
r_{2\psi}e^{2\psi}+r_{\psi\phi}e^{\psi+\phi}+r_{\psi {\phi^*}}e^{\psi+{\phi^*}}+r_{2\phi}e^{2\phi}+\text{c.c.}
+r_{20}, \nonumber \\ r_{2\psi} &=& \frac{\Phi^{''}J_{1}^{2}}{2(2i\omega
+1-\Phi^{'}J_{2}e^{-2i\omega D})}e^{-2i\omega D}A^{2},\nonumber \\
r_{\psi\phi} &=& \frac{\Phi^{''}J_{1}^{2}}{1-\Phi^{'}J_{2}}AB,\nonumber \\
r_{\psi {\phi^*}} &=& \frac{\Phi^{''}J_{1}^{2}}{2i\omega
+1-J_{0}\Phi^{'}e^{-2i\omega D}}e^{-2i\omega D}A B^*,\nonumber \\ r_{2\phi}
&=& \frac{\Phi^{''}J_{1}^{2}}{2(-2i\omega
D+1-J_{2}\Phi^{'}e^{2i\omega D})}e^{2i\omega D}B^{2},\nonumber \\ r_{20} &=&
\frac{\Phi^{''}J_{1}^{2}(|A|^{2}+|B|^{2})}{1-J_{0}\Phi^{'}}.\nonumber
\end{eqnarray}

\noindent $\vartheta (\epsilon^{3} )$: The nonlinear forcing at cubic 
order is

\begin{eqnarray}
N_{3} &=&
[\Phi^{''}J_{1}(J_{0}Ar_{20}+J_{2} {A^*}r_{2\psi}+J_{0}Br_{\psi {\phi^*}}
+J_{2}{B^*}r_{\psi\phi})+\nonumber\\
&& \Phi^{'''}(|A|^{2}A/2+|B|^{2}A)]e^{\psi-i\omega
D}+\nonumber\\
&&[\Phi^{''}J_{1}(J_{0}A {r^*}_{\psi {\phi^*}}+J_{2}{A^*}r_{\psi\phi}+J_{0}Br_{20}+J_{2} {B^*}r_{2\phi})+\nonumber\\
&&\Phi^{'''}(|B|^{2}B/2+|A|^{2}B)]e^{\phi
+i\omega D}+ \dots \nonumber
\end{eqnarray}
Eliminating all terms with dependencies $e^{i\psi}$ and $e^{i\phi}$
yields the two coupled amplitude equations~(\ref{eq:ampeq_THA},\ref{eq:ampeq_THB}):
\begin{eqnarray}
\partial_{T}A &=& (\mu+i\Omega)\Delta
J_{1}A+(a+ib)|A|^{2}A +(c+id)|B|^{2}A, \nonumber \\
\partial_{T}B &=& (\mu-i\Omega)\Delta
J_{1}B+(a-ib)|B|^{2}B+(c-id)|A|^{2}B, \nonumber
\end{eqnarray}
with the coefficients 
\begin{eqnarray}
a+ib &=& \frac{\bar{J}_{k}^{3}e^{-i\omega
D}}{1+D\Phi^{'}\bar{J}_{k}e^{-i\omega
D}}\times \nonumber\\
&&\Bigg(\frac{J_{0}(\Phi^{''})^{2}}
{1-\Phi^{'}J_{0}}+\frac{J_{2}(\Phi^{''})^{2}e^{-2i\omega
D}}{2(2i\omega +1-\Phi^{'}J_{2}e^{-2i\omega D})}
+\frac{\Phi^{'''}}{2} \Bigg), \label{eq:TH_cc1}\\ 
c+id &=&
\frac{\bar{J}_{k}^{3}e^{-i\omega D}}{1+D\Phi^{'}\bar{J}_{k}e^{-i\omega
D}}\times \nonumber\\ 
&&\Bigg(\frac{J_{0}(\Phi^{''})^{2}}{1-\Phi^{'}J_{0}} +\frac{J_{0}(\Phi^{''})^{2}e^{-2i\omega D}}{2i\omega
+1-\Phi^{'}J_{0}e^{-2i\omega D}}+
\frac{J_{2}(\Phi^{''})^{2}}{1-\Phi^{'}J_{2}}+\Phi^{'''}\Bigg).
\label{eq:TH_ccc1}
\end{eqnarray}

\subsection{Codimension 2 bifurcations }
\vspace{0.1in}

\subsubsection{Double zero eigenvalue: Hopf, Turing-Hopf}
\vspace{0.1in}

\noindent $\vartheta (\epsilon )$: The solution to the linear equation 
are periodic oscillations and traveling waves

\begin{equation}
r_{1} =  H(T)e^{i\omega t}+A(T)e^{\psi}+B(T)e^{\phi}+c.c. \nonumber
\end{equation}
where we have defined $\psi\equiv i(k x+\omega t)$ and
$\phi\equiv i(k x-\omega t)$.\\

\noindent $\vartheta (\epsilon^{2} )$: The nonlinear forcing and 
resulting second order solution are

\begin{eqnarray}
N_{2} &=& \Phi^{''}J_{0}^{2}(H^{2}e^{2i\omega (t-D)}+c.c.+2|H|^{2})/2+
\Phi^{''}J_{1}^{2} [ A^{2}e^{2\psi-2i\omega D}+2ABe^{\psi
+\phi}\nonumber \\  &&+2A {B^*}e^{\psi + {\phi^*}-2i\omega D}
+B^{2}e^{2\phi +2i\omega D}+c.c.+2(|A|^{2}+|B|^{2})]/2\nonumber \\
&&+\Phi^{''}J_{0}J_{1}(HAe^{2i\omega (t-D)+ix}+ {H^*}Be^{-2i\omega
(t-D)+ix}+A{H^*}e^{ix}+HBe^{ix}+c.c.), \nonumber \\ 
r_{2} &=& r_{22}e^{2i\omega t}+r_{2\psi}e^{2\psi}+r_{\psi\phi}
e^{\psi+\phi}+r_{\psi{\phi^*}}e^{\psi+{\phi^*}}+
r_{2\phi}e^{2\phi} +r_{HA}e^{2i\omega t+ix}+
\nonumber \\
&&r_{B{H^*}}e^{-2i\omega t+ix}+r_{A{H^*}}e^{ix}+r_{HB}e^{ix}+c.c.+r_{20},\nonumber \\ r_{HA} &=&
\frac{\Phi^{''}J_{0}J_{1}}{2i\omega +1-\Phi^{'}J_{1}e^{-2i\omega
D}}e^{-2i\omega D}HA, \nonumber \\  
r_{B{H^*}} &=&
\frac{\Phi^{''}J_{0}J_{1}}{-2i\omega +1-\Phi^{'}J_{1}e^{2i\omega
D}}e^{2i\omega D}B{H^*}, \nonumber \\ r_{A{H^*}} &=&
\frac{\Phi^{''}J_{0}J_{1}}{1-\Phi^{'}J_{1}} A{H^*}, \nonumber\\ r_{HB} &=&
\frac{\Phi^{''}J_{0}J_{1}}{1-\Phi^{'}J_{1}}HB,\nonumber \\ r_{20} &=&
\frac{\Phi^{''}(J_{0}^{2}|H|^{2}+J_{1}^{2}|A|^{2}+J_{1}^{2}|B|^{2})}{1-J_{0}\Phi^{'}}.\nonumber
\end{eqnarray}

\noindent $\vartheta (\epsilon^{3} )$: The nonlinear forcing at cubic 
order is
\begin{eqnarray}
N_{3} &=&
[\Phi^{''}J_{0}^{2}(Hr_{20}+{H^*}r_{22})
+\Phi^{''}J_{1}^{2}(A{r^*}_{A{H^*}}+{A^*}r_{HA}+
{B^*}r_{A{H^*}}+\nonumber
\\  && B{r^*}_{B{H^*}} +A{r^*}_{HB}+{B^*}r_{HB})+
\Phi^{'''}J_{0}^{3}|H|^{2}H/2]e^{i\omega
(t-D)}+\nonumber\\
&&[\Phi^{''}J_{1}(J_{0}Ar_{20}+J_{2}{A^*}r_{2\psi}+J_{0}Br_{\psi{\phi^*}}+J_{2}{B^*}r_{\psi\phi} +
\nonumber
\\  && J_{0}(Hr_{A{H^*}}+Hr_{HB}+{H^*}r_{HA}))+
\Phi^{'''}(|A|^{2}A/2+|B|^{2}A)]e^{\psi-i\omega
D}+ \nonumber\\
&&[\Phi^{''}J_{1}(J_{0}A{r^*}_{\psi{\phi^*}}+J_{2}{A^*}r_{\psi\phi}+J_{0}Br_{20}+J_{2}{B^*}r_{2\phi}]+\nonumber \\
&& J_{0}(Hr_{B{H^*}}+{H^*}r_{A{H^*}}+{H^*}r_{HB}))
+\Phi^{'''}(|B|^{2}B/2 +|A|^{2}B)]e^{\phi+i\omega D} +\dots \nonumber
\end{eqnarray}
Eliminating terms with dependencies $e^{i\omega t}$, $e^{i\psi}$ and
$e^{i\phi}$ yields the three coupled  amplitude equations~(\ref{eq:ampeq_HTHH}-\ref{eq:ampeq_HTHB}):
\begin{eqnarray}
\partial_{T}H &=& (\mu+i\Omega)\Delta J_{0}H 
+2(\alpha+i\beta)\Big[(\frac{|H|^{2}}{2}+|A|^{2}+|B|^{2})H  
+A{H^*}{B^*}\Big],\nonumber \\ 
\partial_{T}A &=&
(\mu+i\Omega)\Delta J_{1}A+(a+ib)|A|^{2}A+(c+id)|B|^{2}A+ \nonumber \\ 
&&(\alpha+i\beta ) (2 |H|^{2}A +H^{2}B),
\nonumber \\ 
\partial_{T}B &=& (\mu-i\Omega)\Delta
J_{1}B+(a-ib)|B|^{2}B+(c-id)|A|^{2}B + \nonumber \\
&&(\alpha -i\beta) (2 |H|^{2}B +{H^*}^{2}A),
\nonumber
\end{eqnarray}
where $\alpha +i\beta$, $a+ib$ and $c+id$ are given by
Eqs.~(\ref{eq:hopf_cc}, \ref{eq:TH_cc1},\ref{eq:TH_ccc1}), respectively.\\

In the small delay limit ($D\to 0$) we can use the asymptotic values given 
by Eqs.~(\ref{eq:D0}) to obtain, to leading order,
\begin{eqnarray}
a +ib &=&
-\frac{\chi(\frac{\pi}{2}+i)}{(D \Phi)^{3}}\Bigg(
\frac{(\Phi^{''})^{2}}{\Phi^{'}}-\frac{\Phi^{'''}}{2}\Bigg),
\label{eq:HTH_cc2}\\ 
c+id &=&
-\frac{\chi}{(D \Phi)^3}\Bigg(
\frac{(\Phi^{''})^{2}}{\Phi^{'}}\frac{(3\pi
-2)}{5}
-\frac{\pi}{2}\frac{J_{2}(\Phi^{''})^{2}}
{1-\Phi^{'}J_{2}} -\frac{\pi\Phi^{'''}}{2}+ \nonumber \\
&& i\Bigg[\frac{(\Phi^{''})^{2}}{\Phi^{'}} \frac{(6+\pi
)}{5}-\frac{J_{2}(\Phi^{''})^{2}}{1-\Phi^{'}J_{2}}-\Phi^{'''}\Bigg]\Bigg). \label{eq:HTH_ccc2}
\end{eqnarray}
where  $\chi \equiv \pi^{3}/(8+2 \pi^{2})$ and $\alpha +i\beta$ is given by Eq.(\ref{eq:hopf_cc2}).
 
\begin{figure}
\center
\includegraphics[scale=0.35]{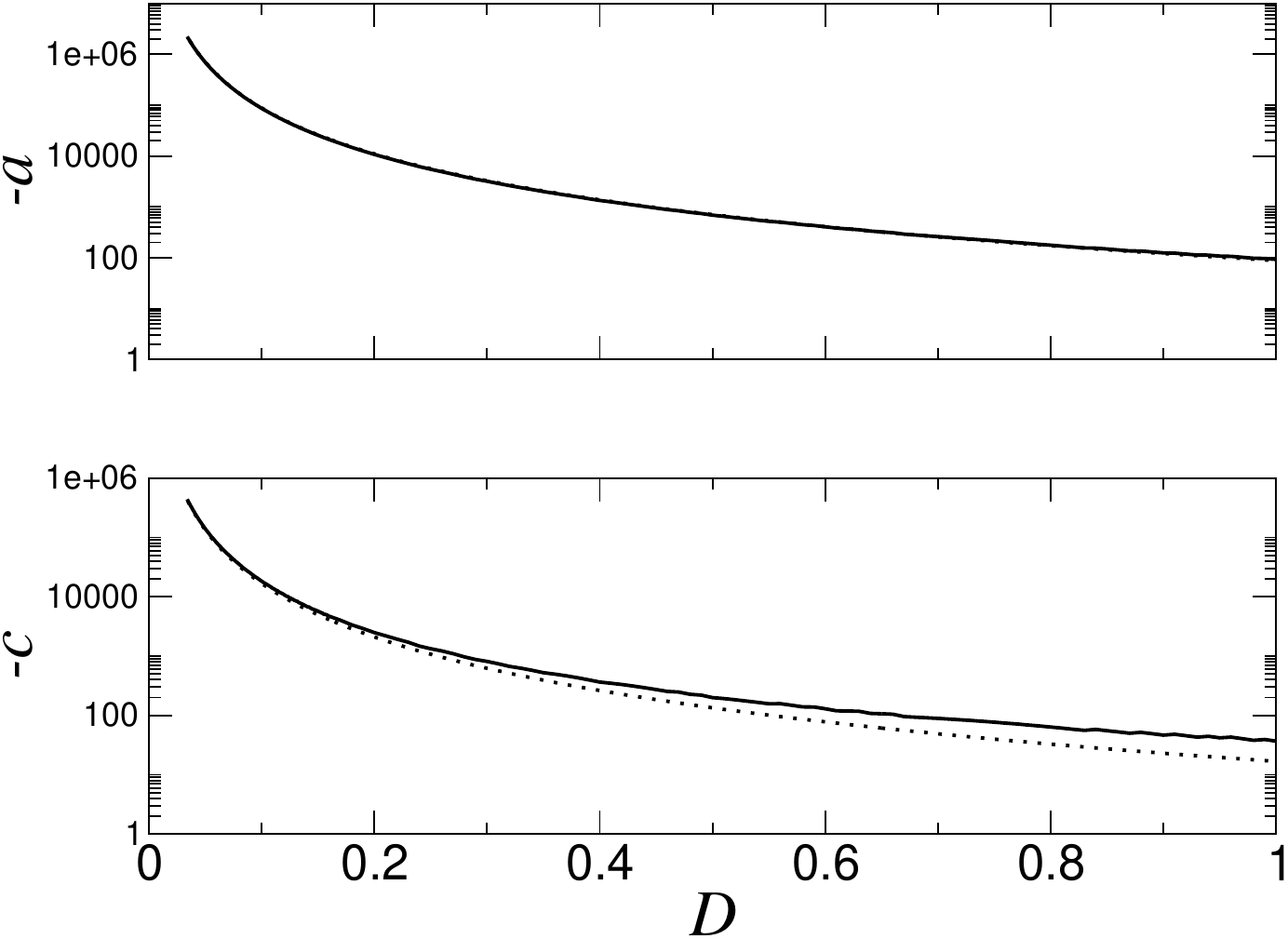}
\caption{Top: The real part of the cubic coefficient at the
codimension-2 point.  The solid line  is the full expression,
Eq.(\ref{eq:TH_cc1}) and the dotted line is the asymptotic result in the
$D\to 0$ limit,  Eq.(\ref{eq:HTH_cc2}).  Bottom: The real part of the
cross-coupling coefficient at the codimension-2 point.  The  solid
line in the full expression, Eq.(\ref{eq:TH_ccc1}) and the dotted line
is the asymptotic results in the  $D\to 0$ limit,
Eq.(\ref{eq:HTH_ccc2}).}
\end{figure}

 \vspace{0.1in}

\noindent\textbf{Solutions and their stability}
\vspace{0.1in}

\noindent\textit{Oscillatory uniform OU:}
\vspace{0.1in}

This solution can be expressed as $(H,A,B)=(\mathcal{H}e^{i\omega t},0,0)$,
where
\begin{eqnarray}
\mathcal{H}&=&\sqrt{\frac{-\mu\Delta J_{0}}{\alpha}},\nonumber \\
\omega &=&\Big(\Omega-\frac{\beta\mu}{\alpha}\Big)\Delta J_{0}.\nonumber
\end{eqnarray}
The stability of this solution can be studied with the ansatz
\begin{eqnarray}
(H,A,B) &=& ( M  \mathcal{H}e^{i\omega t}(1+\delta H_{+}e^{\lambda
t}+{\delta H^*}_{-}e^{{\lambda^*}t}), e^{i\omega t}(\delta A_{+}e^{\lambda t}+{\delta A^*}_{-}e^{{\lambda^*}t}),  \nonumber \\
& & e^{-i\omega t}(\delta B_{+}e^{\lambda t}+{\delta B^*}_{-}e^{{\lambda^*}t})), \nonumber
\end{eqnarray}
which leads to three pairs of coupled linear equations which determine
the six eigenvalues $\lambda$. The first pair is restricted to the
linear subspace of the small amplitude limit cycle and results in the
standard stability problem which yields one stable eigenvalue 
$\lambda = -\mu\Delta J_{0}$ and one zero eigenvalue corresponding 
to  a shift in the phase of the
oscillation. The other two pairs, which span the subspaces of $(\delta A_{+},
\delta B_{+})$ and  $(\delta A_{-}, \delta B_{-})$ respectively, give

\begin{eqnarray}
&&\left( \begin{array}{cc} M &  -(\alpha +i\beta )\mathcal{H}^{2}
\\ \nonumber -(\alpha -i\beta )\mathcal{H}^{2} & M^*
\end{array}\right) \left( \begin{array}{c} \delta A_{+} \\ \delta B_{+} 
\end{array}
\right) = 0, \nonumber
\end{eqnarray}
where $M=\lambda -\mu\Delta
J_{1}-2\alpha\mathcal{H}^{2}+i(\Omega(\Delta J_{0}-\Delta
J_{1})-\beta\mathcal{H}^{2})$, 
and the complex conjugate matrix spanning $(\delta A_{-}, \delta B_{-})$.   Setting the determinant equal to  zero yields the characteristic equation
\begin{eqnarray}
0&=&\lambda^{2}-2\lambda (\Delta J_{1}-2\Delta J_{0})\mu+\mu^{2}(\Delta J_{1}
-4\Delta J_{1}\Delta J_{0}+3\Delta J_{0})+ \nonumber \\
&& \Omega^{2}(\Delta J_{1}-\Delta J_{0})^{2}
-2\frac{\beta}{\alpha}\mu\Omega \Delta J_{0}(\Delta J_{1}-\Delta J_{0}).
\nonumber
\end{eqnarray}
We find an oscillatory instability for $\Delta J_{1}=2\Delta J_{2}$ while a steady instability occurs for $\Delta J_{1}=\Delta J_{0}$.  
The steady instability therefore always precedes the oscillatory one.
\vspace{0.1in}

\noindent\textit{Traveling waves (TW):}
\vspace{0.1in}

This solution can be expressed as $(H,A,B)=(0,\mathcal{A}_{TW}e^{i\omega t},0)$, where
\begin{eqnarray}
\mathcal{A}_{TW}&=&\sqrt{\frac{-\mu\Delta J_{1}}{a}},\nonumber \\
\omega &=&\Big(\Omega-\frac{b\mu}{a}\Big)\Delta J_{1}. \nonumber
\end{eqnarray}
The stability of this solution can be studied with the ansatz
\begin{eqnarray}
(H,A,B) & = & (e^{i\omega t}(\delta H_{+}e^{\lambda t}+{\delta
H^*}_{-}e^{{\lambda^*}t}), 
\mathcal{A}_{TW}e^{i\omega t}(1+\delta A_{+}e^{\lambda t}+{\delta A^*}_{-}e^{{\lambda^*}t}),  \nonumber \\ &&
e^{-i\omega t}(\delta B_{+}e^{\lambda t}+{\delta B^*}_{-}e^{{\lambda^*}t})),\nonumber
\end{eqnarray}
which results in four coupled linear equations corresponding to the
stability problem for TW in the competition  between SW and TW (see
section D, Turing-Hopf Bifurcation).  Here we assume that the TW
solution is supercritical  and stable.  We then turn our attention to
the remaining two linear equations which describe the growth  of the
oscillatory uniform mode.  These equations are uncoupled and yield the 
complex conjugate eigenvalues
\begin{equation}
\lambda = -\mu \Big(2\frac{\alpha }{a}\Delta J_{1}-\Delta J_{0}\Big)\pm i\Big(\Omega 
(1-\frac{a}{2\alpha})+(b-2\beta )\frac{\mu}{2\alpha }\Big)\Delta J_{0}, \nonumber
\end{equation}
from which it is easy to see that an instability occurs for $\Delta J_{1}=\frac{a}{2\alpha }\Delta J_{0}$.   This instability will generically 
occur with non-zero frequency.

\vspace{0.1in}

\noindent\textit{Standing waves (SW):}
\vspace{0.1in}

This solution can be expressed as $(H,A,B)=(0,\mathcal{A}_{SW}e^{i\omega t},\mathcal{A}_{SW}e^{-i\omega t})$, 
where $\mathcal{A}_{SW}$ and $\omega $ are given by 
Eqs.~(\ref{eq:HTH_SW_amp}, \ref{eq:HTH_SW_freq}) 
\begin{eqnarray}
\mathcal{A}_{SW}&=&\sqrt{\frac{-\mu\Delta J_{1}}{(a+c)}},\nonumber \\
\omega  &=& \Big(\Omega-\frac{(b+d)}{(a+c)}\mu \Big)\Delta J_{1}. 
\nonumber
\end{eqnarray} 
The stability of this solution can be studied with the ansatz
\begin{eqnarray}
(H,A,B) &=& (e^{i\omega t}(\delta H_{+}e^{\lambda t}+{\delta
H^*}_{-}e^{{\lambda^*}t}), 
\mathcal{A}_{SW}e^{i\omega t}(1+\delta
A_{+}e^{\lambda t}+{\delta A^*}_{-}e^{{\lambda^*}t}),\nonumber \\&&
\mathcal{A}_{SW}e^{-i\omega t}(1+\delta B_{+}e^{\lambda
t}+{\delta B^*}_{-}e^{{\lambda^*}t})).\nonumber
\end{eqnarray}
This ansatz results in four coupled equations for the stability of SW
in the competition between SW and TW.  Here we assume that the  SW
solution is supercritical and stable.  The remaining two equations
describe the growth of the oscillatory uniform mode.   

\begin{eqnarray}
&&\left( \begin{array}{cc} N & -2\mathcal{A}_{SW}(\alpha
    +i\beta) \\ -2\mathcal{A}_{SW}(\alpha -i\beta) &  N^*
\end{array}\right) \left( \begin{array}{c} \delta H_{+} \\ 
\delta H_{-} \end{array}\right) = 0.
\end{eqnarray}
where $N=\lambda +i\omega -(\mu +i\Omega )\Delta J_{0}-
4\mathcal{A}_{SW}(\alpha+i\beta)$.
Setting the determinant to zero yields the characteristic equation for
the eigenvalues
\begin{eqnarray}
0 &=& 
\mu^{2}\left(\Delta J_{0}-\frac{4 \alpha \Delta J_{1}} {(a+c)} \right)^{2}+
\Big(\Delta J_{1}\Big[\Omega -\frac{(b+d)\mu}{(a+c)}
  +\frac{4\beta \mu}{(a+c)} \Big] -\Omega\Delta
  J_{0}\Big)^{2}-
\nonumber\\ 
&&\frac{4\mu^{2} \Delta J_{1}^{2}(\alpha^{2}+\beta^{2})}{(a+c)^{2}}
-2\mu\lambda \Big[\Delta J_{0}-\frac{4\alpha \Delta
  J_{1}}{(a+c)}\Big]+\lambda^{2}.
\end{eqnarray}
The conditions for oscillatory and steady instabilities~
(\ref{eq:HTH_SW_osc},\ref{eq:HTH_SW_stead}),  are found by
setting $\lambda $ equal to  $i\bar{\omega}$ and $0$ respectively.
\vspace{0.1in}

\noindent\textit{Mixed Mode:}
\vspace{0.1in}

Mixed mode solutions are found by applying the ansatz Eq.(\ref{eq:HTH_mmansatz}) to 
Eqs.~(\ref{eq:ampeq_HTHH}-\ref{eq:ampeq_HTHB}).  This gives
\begin{eqnarray}
\dot{\mathcal{H}}&=&\mu\Delta J_{0}\mathcal{H}+\alpha
(\mathcal{H}^{2}+2\mathcal{A}^{2}+2\mathcal{B}^{2})\mathcal{H}
+2\mathcal{HAB}(\alpha\cos{\phi}-\beta\sin{\phi}), \nonumber \\
\dot{\mathcal{A}}&=&\mu\Delta
J_{1}\mathcal{A}+a\mathcal{A}^{3}+c\mathcal{B}^{2}\mathcal{A}
+2\alpha\mathcal{H}^{2}\mathcal{A}+\mathcal{H}^{2}\mathcal{B}(\alpha\cos{\phi}+\beta\sin{\phi}), \nonumber \\ 
\dot{\mathcal{B}}&=&\mu\Delta
J_{1}\mathcal{B}+a\mathcal{B}^{3}+c\mathcal{A}^{2}\mathcal{B}
+2\alpha\mathcal{H}^{2}\mathcal{B}+\mathcal{H}^{2}\mathcal{A}(\alpha\cos{\phi}+\beta\sin{\phi}), \nonumber \\ 
\dot{\phi}&=&2\Omega (\Delta J_{1}-\Delta
J_{0})+2\beta\mathcal{H}^{2}+(b+d-4\beta)(\mathcal{A}^{2}
+\mathcal{B}^{2})-\nonumber\\
&&\alpha\sin{\phi}\Big(\frac{\mathcal{H}^{2}\mathcal{B}}{\mathcal{A}}
+\frac{\mathcal{H}^{2}\mathcal{A}}{\mathcal{B}}+4\mathcal{AB}\Big)+\beta\cos{\phi} \Big(\frac{\mathcal{H}^{2}\mathcal{B}}{\mathcal{A}}
+\frac{\mathcal{H}^{2}\mathcal{A}}{\mathcal{B}}-4\mathcal{AB}\Big),
\label{eq:mm_phi}
\end{eqnarray}
where $\phi = \psi_{A}-\psi_{B}-2\theta $. One steady state solution of these equations takes the form
$(\mathcal{H},\mathcal{A},\mathcal{B},\phi) =
(\mathcal{\hat{H}},\mathcal{\hat{A}},-\mathcal{\hat{A}},\hat{\phi})$,
where
\begin{eqnarray}
\mathcal{\hat{H}}^{2} &=& \frac{-\mu\Delta J_{0}(a+c)-2\mu\Delta
  J_{1}\big( -2\alpha +\alpha\cos{\hat{\phi}}
  -\beta\sin{\hat{\phi}}\big)}{\alpha (a+c)-\big[
    4\alpha-2(\alpha\cos{\hat{\phi}}
    -\beta\sin{\hat{\phi}})\big]\big[2\alpha-\alpha\cos{\hat{\phi}}+\beta\sin{\hat{\phi}}\big]},
\label{eq:mmh2} \\ \mathcal{\hat{A}}^{2} &=& \frac{-\mu\Delta
  J_{0}(-2\alpha+\alpha\cos{\hat{\phi}}-\beta\sin{\hat{\phi}})-\mu\Delta
  J_{1} \alpha}{\alpha (a+c)-\big[ 4\alpha-2(\alpha\cos{\hat{\phi}}
    -\beta\sin{\hat{\phi}})\big]\big[2\alpha-\alpha\cos{\hat{\phi}}+\beta\sin{\hat{\phi}}\big]},
\label{eq:mma2}
\end{eqnarray}
and $\hat{\phi}$ is found by plugging Eqs.~(\ref{eq:mmh2},\ref{eq:mma2}) 
into Eq.(\ref{eq:mm_phi}) and setting the  left hand side
equal to zero.  We do not study the stability of the mixed-mode state here.\\

\subsubsection{Double zero eigenvalue: Turing, Hopf}

$\vartheta (\epsilon )$: The solutions to the linear equation are time 
periodic oscillations and spatially periodic functions
\begin{equation}
r_{1} = H(T)e^{i\omega t}+A(T)e^{ix}+c.c. \nonumber
\end{equation}

$\vartheta (\epsilon^{2} )$: The nonlinear forcing and resulting second order solution are
\begin{eqnarray}
N_{2} &=& \frac{\Phi^{''}}{2}J_{0}^{2}(H^{2}e^{2i\omega
(t-D)}+c.c.+2|H|^{2})+\frac{\Phi^{''}}{2}J_{1}^{2}(A^{2}e^{2ix}+c.c.+2|A|^{2})\nonumber
\\  && \Phi^{''}J_{0}J_{1}HAe^{ix+i\omega
t}+\Phi^{''}J_{0}J_{1}A{H^*}e^{-i\omega t+ix}+c.c., \nonumber \\ 
r_{2} &=&
r_{2H}e^{2i\omega t}+r_{2A}e^{2ix}+r_{AH}e^{i\omega
t+ix}+r_{A{H^*}}e^{-i\omega t+ix}+c.c.+r_{20}, \nonumber \\ r_{AH} &=&
\frac{\Phi^{''}J_{0}J_{1}}{i\omega +1-\Phi^{'}J_{1}e^{-i\omega
D})}e^{-i\omega D}AH,\nonumber \\  
r_{A{H^*}} &=&\frac{\Phi^{''}J_{0}J_{1}}{-i\omega +1-\Phi^{'}J_{1}e^{i\omega
D}}e^{i\omega D}A{H^*}, \nonumber \\ 
r_{20} &=&\frac{\Phi^{''}(J_{0}^{2}|H|^{2}+J_{1}^{2}|A|^{2})}{1-J_{0}\Phi^{'}}.
\nonumber
\end{eqnarray}
$\vartheta (\epsilon^{3} )$: The nonlinear forcing at cubic order is
\begin{eqnarray}
N_{3} &=&
\Big[\Phi^{''}(J_{0}^{2}Hr_{20}+J_{0}^{2}{H^*}r_{2H}+J_{1}^{2}A{r^*}_{A\hat{H}}+J_{1}^{2}{A^*}r_{AH})
+ \nonumber \\
&&\Phi^{'''}(J_{0}^{2}J_{1}|H|^{2}A+J_{1}^{3}|A|^{2}A/2)\Big]e^{i\omega
t} + \nonumber\\  &&
\Big[\Phi^{''}(J_{0}J_{1}Hr_{A{H^*}}+J_{0}J_{1}{H^*}r_{AH}+J_{0}J_{1}Ar_{20}+J_{1}J_{2}{A^*}r_{2A})+ \nonumber\\  &&
\Phi^{'''} (J_{0}^{2}J_{1}|H|^{2}A
+J_{1}^{3}|A|^{2}A/2)\Big]e^{i  x}.\nonumber
\end{eqnarray}
Eliminating terms with dependencies $e^{i\omega t}$, $e^{ix}$ yields
the two coupled  amplitude equations,
Eqs.~(\ref{eq:ampeq_THH}-\ref{eq:ampeq_HTA}),
\begin{eqnarray}
\partial_{T}H &=& (\mu+i\Omega)\Delta J_{0}H+(\alpha +i\beta)|H|^{2}H
+(\kappa +i\Lambda )|A|^{2}H,\nonumber\\ 
\partial_{T}A &=&
\bar{\eta}\Delta J_{1}A+\Gamma |A|^{2}A+\sigma |H|^{2}A,
\nonumber
\end{eqnarray}
where $\mu +i\Omega$, $\alpha +i\beta$, $\bar{\eta}$ and $\Gamma$ are
given by Eqs.~(\ref{eq:hopf_lc}, \ref{eq:hopf_cc},\ref{eq:turing_lc},
\ref{eq:turing_cc}) respectively, and
\begin{eqnarray}
\kappa +i\Lambda &=& \frac{e^{-i\omega D}}{1+D\Phi^{'}\bar{J}_{0}e^{-i\omega D}}\Bigg(
\frac{\bar{J}_{0}^{2}\bar{J}_{k}^{2}}{1-\Phi^{'}\bar{J}_{0}}+2\frac{\bar{J}_{1}^{3}\bar{J}_{0}
(\Phi^{''})^{2}e^{-i\omega D}}{i\omega +1-\Phi^{'}\bar{J}_{1}e^{-i\omega D}}+\nonumber\\ 
&&\Phi^{'''}\bar{J}_{0}\bar{J}_{1}^{2}\Bigg) \label{eq:kappa}, \\
\sigma &=& \frac{1}{1+D\Phi^{'}\bar{J}_{1}}\Bigg( 
\frac{\bar{J}_{0}^{3}\bar{J}_{1}(\Phi^{''})^{2}}{1-\Phi^{'}\bar{J}_{0}}+\Phi^{'''}\bar{J}_{0}^{2}\bar{J}_{1}+
\bar{J}_{0}^{2}\bar{J}_{1}^{2} (\Phi^{''})^{2}\times\nonumber\\
&&
\Big(\frac{e^{i\omega D}}{-i\omega +1-\Phi^{'}\bar{J}_{1}e^{i\omega D}}+\frac{e^{-i\omega D}}{i\omega +1-\Phi^{'}\bar{J}_{1}e^{-i\omega D}}\Big)\Bigg). \label{eq:sigma}
\end{eqnarray}

In the small delay limit ($D\to 0$) we can use the asymptotic values~(\ref{eq:D0}) to obtain, to leading order,
\begin{eqnarray}
\kappa +i\Lambda &=&
-\frac{\pi}{2D(\Phi^{'})^{3}}\frac{(\pi/2+i)}{(1+\pi^{2}/4)}\Bigg(\frac{(\Phi^{''})^{2}}{\Phi^{'}}
-\Phi^{'''}\Bigg)+\vartheta (1), \nonumber \\ 
\eta &=& \Phi^{'}+\vartheta (D), \nonumber
\\ 
\Gamma &=&
\frac{1}{(\Phi^{'})^{3}}\Bigg(-\frac{(\Phi^{''})^{2}}{\Phi^{'}}+\frac{\Phi^{'''}}{2} +\frac{J_{2}(\Phi^{''})}{2(1-J_{2}\Phi^{'})}\Bigg)+\vartheta (D), 
\nonumber \\
\sigma &=&
-\frac{\pi^{2}}{4D^{2}(\Phi^{'})^{3}}\Bigg(\frac{(\Phi^{''})^{2}}{\Phi^{'}}-\Phi^{'''}\Bigg)+\vartheta(1/D),\nonumber
\end{eqnarray}
and $\mu+i\Omega$ and $\alpha +i\beta$ are given by
Eqs.~(\ref{eq:hopf_lc2},\ref{eq:hopf_cc2}) respectively.\\

\bibliographystyle{elsarticle-num}








\end{document}